\documentclass[twocolumn]{aastex62}

\usepackage{amsmath}
\usepackage{wasysym}
\usepackage{savesym}
\savesymbol{tablenum}
\usepackage{siunitx}
\restoresymbol{SIX}{tablenum}
\usepackage{booktabs}
\usepackage{rotating}
\usepackage[T1]{fontenc}
\usepackage{blindtext}

\DeclareSIUnit\Mearth{M_\oplus}
\DeclareSIUnit\Mjupiter{M_J}
\DeclareSIUnit\Rearth{R_\oplus}
\DeclareSIUnit\Rjupiter{R_J}
\DeclareSIUnit\Mmoon{M_\textrm{\leftmoon}}
\DeclareSIUnit\JEM{J_{EM}}

\graphicspath{{./}{figures/}}

\received{31.03.2025}
\revised{19.05.2025}
\accepted{20.05.2025}
\submitjournal{ApJ}

\shorttitle{Origin of Jupiter's fuzzy core}
\shortauthors{Meier et al.}

\begin{document}
\title{On the origin of Jupiter's fuzzy core:\\ constraints from N-body, impact, and evolution simulations}

\correspondingauthor{Thomas Meier}
\email{thomas.meier5@uzh.ch}

\author[0000-0001-9682-8563]{Thomas Meier}
\affiliation{Department of Astrophysics, University of Zurich, Winterthurerstrasse 190, CH-8057 Zurich, Switzerland}

\author[0000-0002-4535-3956]{Christian Reinhardt}
\affiliation{Department of Astrophysics, University of Zurich, Winterthurerstrasse 190, CH-8057 Zurich, Switzerland}
\affiliation{Physics Institute, Space Research and Planetary Sciences, Center for Space and Habitability, University of Bern, Sidlerstrasse 5, CH-3012 Bern, Switzerland}

\author[0000-0002-5418-6336]{Sho Shibata}
\affiliation{Department of Astrophysics, University of Zurich, Winterthurerstrasse 190, CH-8057 Zurich, Switzerland}
\affiliation{Department of Earth, Environmental and Planetary Sciences, 6100 Main MS 126, Rice University, Houston, TX 77005, USA}

\author[0000-0002-8278-8377]{Simon Müller}
\affiliation{Department of Astrophysics, University of Zurich, Winterthurerstrasse 190, CH-8057 Zurich, Switzerland}

\author[0000-0001-7565-8622]{Joachim Stadel}
\affiliation{Department of Astrophysics, University of Zurich, Winterthurerstrasse 190, CH-8057 Zurich, Switzerland}

\author[0000-0001-5555-2652]{Ravit Helled}
\affiliation{Department of Astrophysics, University of Zurich, Winterthurerstrasse 190, CH-8057 Zurich, Switzerland}



\begin{abstract}
It has been suggested that Jupiter's fuzzy core could be a result of a giant impact. Here, we investigate the expected impact conditions from N-body simulations. We then use state-of-the-art SPH simulations to investigate the results of impacts with different conditions including various impactor masses and composition, different formation stages in Jupiter's growth, and different resolutions. We next simulate the long-term thermal evolution of Jupiter post-impact. We find that 3D N-body simulations predict rather oblique impacts, and that head-on collisions are rare. Moreover, our results show that even under a head-on collision, Jupiter's fuzzy core cannot be formed. We next simulated Jupiter's thermal evolution and showed that unless post-impact temperatures are extremely low, a giant impact would not lead to an extended dilute core as inferred by interior models. We conclude that Jupiter's fuzzy core is not caused by an impact and is likely to be an outcome of its formation process. 
\end{abstract}



\section{Introduction}\label{sec:Introduction}
The Juno mission provided accurate measurements of Jupiter's gravitational field, providing tighter constraints for interior models (e.g. \citealt{folknerJupiterGravityField2017,iessMeasurementJupiterAsymmetric2018,duranteJupiterGravityField2020}). Jupiter structure models that fit Juno data find that the planet is inhomogeneous in composition and consists of a fuzzy (dilute) core (e.g. \citealt{wahlComparingJupiterInterior2017,vazanJupiterEvolutionPrimordial2018,debrasNewModelsJupiter2019,nettelmannTheoryFiguresSeventh2021,militzerJunoSpacecraftMeasurements2022,miguelJupiterInhomogeneousEnvelope2022}). 

Several scenarios for the formation of Jupiter's fuzzy core were proposed. The first is related to Jupiter's formation process and the concurrent accretion of heavy elements and H-He gas \citep{vallettaGiantPlanetFormation2020,venturiniJupiterHeavyelementEnrichment2020}. The second scenario suggests that Jupiter's primordial compact core was eroded by large-scale or double-diffusive convection \citep{guillotInteriorJupiter2004,mollDoublediffusiveErosionCore2017}. Finally, the third scenario argues that a giant impact mixed the primordial heavy-element core, distributing the heavy elements beyond the central region \citep{liEMBRYOIMPACTSGAS2010}. A comprehensive discussion of these aspects can be found in \citet{helledRevelationsJupiterFormation2022}.

While the possibility of forming Jupiter's fuzzy core via a giant impact has been studied in prior work, further investigations of this scenario are required, as each study was limited in several aspects. \citet{liuFormationJupitersDiluted2019} identified a scenario where a \SI{10}{\Mearth} body collides head-on with a nearly fully formed proto-Jupiter of \SI{306.7}{\Mearth} which sufficiently disrupts the compact core to create a dilute core that is stable over time, also when accounting for Jupiter's post-impact thermal evolution. However, their N-body simulations were only 2-dimensional, skewing the impact distribution towards head-on impacts. In addition, it remains unclear to what extent the mixing between the core and envelope is caused by the diffusive nature of the Eulerian grid method used in their study \citep{springelPurSiMuove2010,robertsonComputationalEulerianHydrodynamics2010}. In addition, \citet{liuFormationJupitersDiluted2019} used rather simplified equations of state and treatment of self-gravity. 

Recently, \citet{sandnesNoDiluteCore2024} used a newly developed extension to the Lagrangian hydrodynamics method SPH called REMIX \citep{sandnesREMIXSPHImproving2025}. This method does not suffer from the spurious mixing that large bulk motion introduces in Eulerian methods with improved mixing compared to traditional SPH. Using this improved method, it was shown that although impacts can lead to mixing in the deep interior, settling takes place quickly, leading to a layered structure. It was therefore concluded that Jupiter's fuzzy core was not created by a giant impact. However, unlike in \citet{liuFormationJupitersDiluted2019}, \citet{sandnesREMIXSPHImproving2025} have not considered the post-impact thermal evolution which is critical for assessing the outcome of the impact on longer timescales. 

Currently, it remains unclear whether a giant impact can indeed lead to the formation of Jupiter's fuzzy core. In this study, we re-assess the giant impact scenario while considering the constraints imposed by using 3D N-body formation simulations with tidal damping, high resolution smoothed particle hydrodynamics (SPH) impact simulations with state-of-the-art equations of state and proper treatment of self-gravity and post-impact thermal evolution simulations. Our paper is structured as follows. In Section~\ref{sec:N-body_results} we show the results of the N-body simulations and identify the most likely impact conditions. In Section~\ref{sec:Impact_Simulations} we present the SPH simulations of the impacts on Jupiter. In Section~\ref{sec:Post-impact_evolution} we show the results of the long-term evolution simulations of the post-impact states. In Section~\ref{sec:Discussion} we discuss how our results compare to prior studies, collisions on early Jupiter and challenges for modeling mixing in SPH. Finally, in Section~\ref{sec:Conclusions} we present our conclusions. In Appendix~\ref{sec:appendix:Mixing_metrics} we propose several mixing metrics to measure mixing in SPH and discuss the challenges. In Appendix~\ref{sec:appendix:Modelling_mixing_in_SPH} we explore the challenges in modeling mixing in SPH. In Appendix~\ref{sec:appendix:Extended_data_tables} the parameters of all SPH simulations are listed.

\section{Expected impacts onto Jupiter}\label{sec:N-body_results}
\citet{liuFormationJupitersDiluted2019} investigated the impact conditions of four \SI{10}{\Mearth} embryos during the growth of proto-Jupiter using N-body simulations. They found that in \SIrange{40}{60}{\percent} of their simulations a head-on collision (with an impact parameter $\theta \lesssim \SI{30}{\degree}$) is the most probable collision outcome. However, in these simulations the embryos were distributed on the co-planar orbit of proto-Jupiter (2D N-body simulation). This assumption is rather unrealistic because mutual scattering between embryos inclines their mutual orbits. In the case of embryos with inclined orbits, the collision probability decreases and the impact angle increases. We therefore re-assess the expected impact conditions between proto-Jupiter and large embryos using 3D N-body simulations. 

\subsection{Numerical setup}
\begin{figure}[ht!]
\centering
\includegraphics[width=\linewidth]{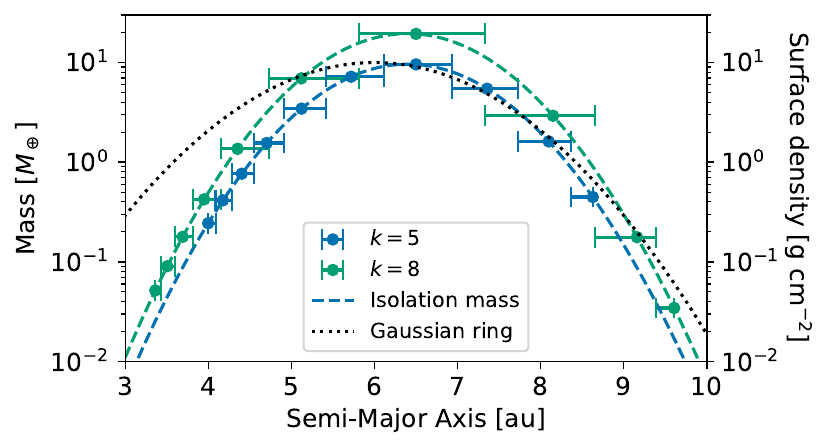}
\caption{Initial mass distribution of embryos with initial orbital separation between the embryos of $k=5$ (blue) and $k=8$ (green) mutual Hill radii. The dashed lines show the planetesimal isolation mass calculated with the surface density. The error bars on the dots show $k$ mutual Hill radii on each side. The dotted line shows the surface density of planetesimals which has a Gaussian profile with $\mu=\SI{6}{\astronomicalunit}$ and $\sigma=\SI{1.13}{\astronomicalunit}$. The total mass of the planetesimal ring is \SI{40}{\Mearth}. 
}
\label{fig:disk}
\end{figure}

We use the N-body simulation code developed in \citet{shibataFateRemnantSolid2024}, which uses the IAS15 integrator algorithm \citep{reinIAS15FastAdaptive2015}. We distribute 10 embryos around Jupiter's formation location at $\sim\SI{6}{\astronomicalunit}$. The mass of each embryo is set to the planetesimal isolation mass, assuming a Gaussian profile for the planetesimal ring \citep[e.g.][]{batyginFormationRockySuperearths2023, wooTerrestrialPlanetFormation2023}. The total mass of the solid and the highest surface density are set to \SI{40}{\Mearth} and \SI{10}{\gram\per\centi\meter\tothe2}, respectively. Our Gaussian profile has the mean location at $\mu=\SI{6}{\astronomicalunit}$ and the standard deviation of the distribution $\sigma=\SI{1.13}{\astronomicalunit}$. Figure~\ref{fig:disk} shows the surface density of planetesimals (dotted line). We assume that proto-Jupiter forms in the location where the isolation mass is the highest. The other embryos are distributed assuming an orbital separation of $k$ mutual Hill radii to the next embryos, and their masses are set to the local isolation mass. In the case of planetesimal accretion, the embryos reach the oligarchic phase and deplete the surrounding planetesimals. Following this picture, we set the width of the feeding zone to the same as the orbital separation. The planetesimal isolation mass is presented with the dashed lines in Figure~\ref{fig:disk}. At the beginning of the simulations, proto-Jupiter's mass is $10 M_\oplus$ and $20 M_\oplus$ for $k=5$ and $k=8$, respectively. Due to the wider feeding zone, the embryos' mass is larger for $k=8$. Note that since the isolation mass scales with $\Sigma^{3/2} r^3$, where $\Sigma$ is the surface density of planetesimals, the shape of the isolation mass differs from that of the surface density of planetesimals. The orbital separation factor $k$, and the initial orbital inclination of the embryos $\sin i_0$ are input parameters. We perform 520 simulations for each parameter set, changing the initial orbital angles of the embryos. 

Unlike \citet{liuFormationJupitersDiluted2019}, we include the tidal damping of embryos using type-I and type-II prescriptions \citep{cresswellThreedimensionalSimulationsMultiple2008, kanagawaRadialMigrationGapopening2018}. At the beginning of the N-body simulations, proto-Jupiter enters the runaway gas accretion phase. The mass of proto-Jupiter increases via the rapid accretion of hydrogen-helium (H-He) gas. The gaseous disk model and the growth model of proto-Jupiter are taken from \citet{shibataHeavyelementAccretionProtoJupiter2023}. The radius of proto-Jupiter is calculated assuming a fixed bulk density of \SI{0.125}{\gram\per\centi\meter\tothe3}. 

\subsection{Impact condition in 3D}
\begin{figure}[ht!]
\centering
\includegraphics[width=\linewidth]{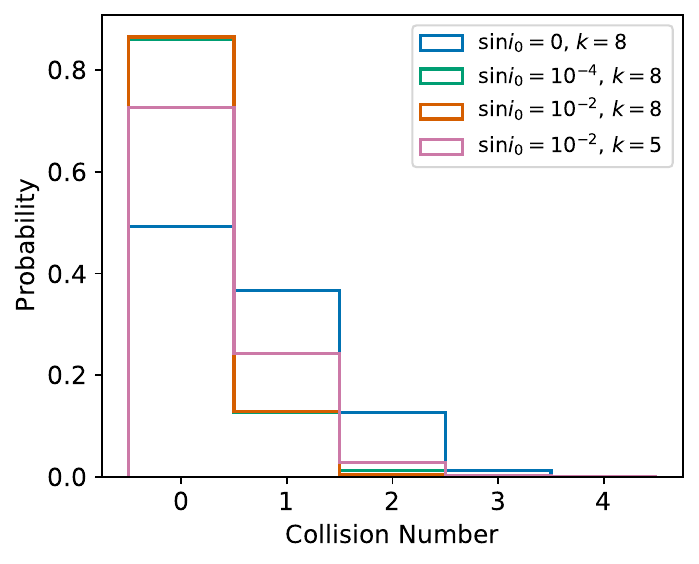}
\caption{Probability distribution of collisions obtained in each simulation. $i_0$ and $k$ are the initial inclination and orbital separation factor of the colliding embryos.}
\label{fig:Collision_Number}
\end{figure}

\begin{figure*}[ht!]
\centering
\includegraphics[width=\linewidth]{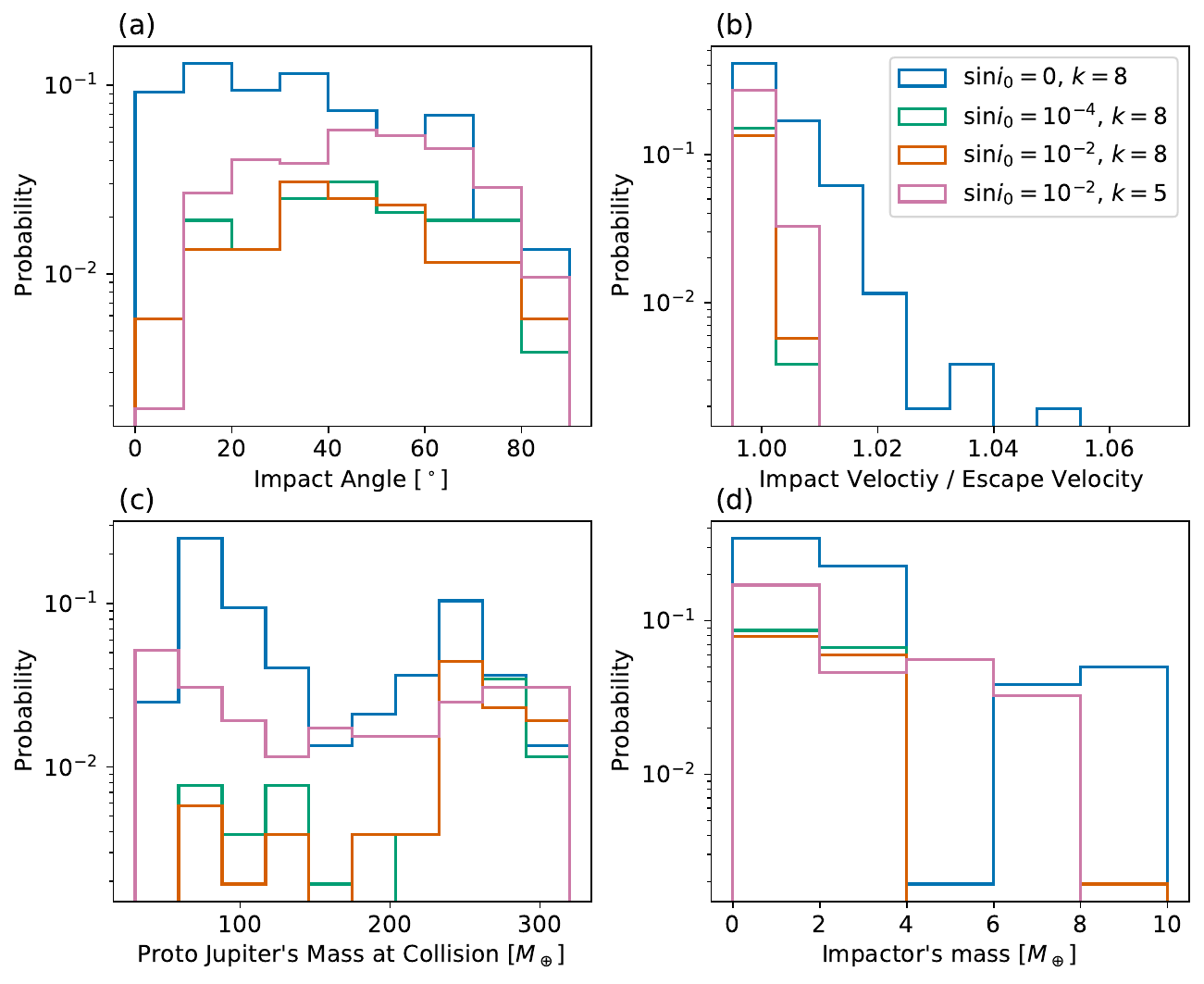}
\caption{Impact conditions of embryos onto a growing proto-Jupiter. The different colors correspond to the results when assuming different initial inclinations of embryos. (a) Impact angles, (b) impact velocity divided by the escape velocity, (c) proto-Jupiter's mass at the collision, and (d) impactor's mass. The probability shows the occurrence rate in one simulation.} 
\label{fig:Impact_Conditions}
\end{figure*}

Figure~\ref{fig:Collision_Number} shows the inferred number of collisions between proto-Jupiter and the embryos. We only count the impacts where the embryo's mass is larger than \SI{0.5}{\Mearth}. The blue lines show the results with 2D simulations. We find that \SI{50}{\percent} include collisions with proto-Jupiter, a result that is consistent with \citet{liuFormationJupitersDiluted2019}. However, the collision probability is only \SIrange{10}{20}{\percent} in the 3D simulations.

Figure~\ref{fig:Impact_Conditions} shows the impact conditions inferred by our N-body simulations. \citet{liuFormationJupitersDiluted2019} found that the head-on collision is the dominant collision in the 2D case. Our results suggest that the most probable collision is around $\theta\sim\SI{45}{\degree}$ in the 3D cases. We also find that the initial inclination rarely affects the impact condition in the 3D cases.

Most of the collisions in our simulations occur before Jupiter reaches its final mass. If the orbital separation is as small as $k=5$, \SI{10}{\percent} of our simulations include collisions before the growing Jupiter reached a mass of \SI{100}{\Mearth}. The relative size of the primordial core is larger when the envelope mass is smaller, which may assist in mixing the primordial core and forming a fuzzy core. The impact velocities are close to the escape velocity (from $10$ to \SI{40}{\kilo\meter\per\second}) due to proto-Jupiter's strong gravity (panel-b). Our results also show that the impact velocities are higher in the 2D case than in the 3D case. Impactors with impact velocities higher than the escape velocity have high eccentricities ($e\gtrsim0.1$). In the 3D cases, both the inclination and the eccentricity are high due to the energy equi-partition. Since the collision probability decreases with the embryo's inclination, such high-velocity impacts are rare in the 3D case.

In the 2D cases, \SI{10}{\percent} of the simulations had impactors larger than \SI{4}{\Mearth}. However, such giant impacts become rare in the 3D simulations (panel-d). When the orbital separation is as wide as $k~=~8$, only impactors smaller than \SI{4}{\Mearth} collide with the growing Jupiter. Collisions of more massive embryos are possible only when the orbital separation is $k=5$. In this case, \SI{10}{\percent} of the simulations have impactors with masses of \SIrange{4}
{8}{\Mearth}. 

Our results suggest that head-on collisions are actually uncommon and that the mixing of the primordial core after Jupiter reached its final mass is, in fact, rather unlikely. However, giant impacts could occur in the early gas accretion phase of proto-Jupiter, which may dilute the primordial heavy-element core (see Section~\ref{sec:Discussion_Collisions_on_early_Jupiter} for a discussion).

\section{Impact Simulations}\label{sec:Impact_Simulations} 
The simulations above identify the expected impact parameters between proto-Jupiter and the surrounding embryos. We next determine the outcome of these collisions. We model the collisions using the smoothed particle hydrodynamics (SPH) method \citep{lucyNumericalApproachTesting1977,monaghanSmoothedParticleHydrodynamics1992,springelSmoothedParticleHydrodynamics2010} which has been extensively used to model giant impacts (\citealt{canupSimulationsLateLunarforming2004,asphaugMercuryOtherIronrich2014,emsenhuberSPHCalculationsMarsscale2018,reinhardtFormingIronrichPlanets2022,kegerreisImmediateOriginMoon2022,timpeSystematicSurveyMoonforming2023,ballantyneInvestigatingFeasibilityImpactinduced2023,meierSystematicSurveyMoonforming2024,ballantyneSputnikPlanitiaImpactor2024}). We use a new SPH implementation (T. Meier et al. 2025, in preparation) built upon the gravity code \texttt{pkdgrav3}, derived from the Lagrangian \citep{springelCosmologicalSmoothedParticle2002,priceSmoothedParticleHydrodynamics2012}, that accounts for corrections due to a variable smoothing length and contains kernels that do not suffer from the pairing instability and allow accurate sampling with 400 neighbors \citep{dehnenImprovingConvergenceSmoothed2012}. The code includes state-of-the-art improvements for modeling giant impacts such as an interface/free-surface correction \citep{reinhardtNumericalAspectsGiant2017,reinhardtBifurcationHistoryUranus2020,ruiz-bonillaDealingDensityDiscontinuities2022} and a generalized EOS interface \citep{meierEOSResolutionConspiracy2021,meierEOSlib2021} which simplifies the use of different equations of state and allows the use of multiple EOS in the same simulation. The code shows excellent scaling on HPC systems with either CPU based or hybrid CPU/GPU architectures enabling the use of several billion particles in a simulation.

All the collision simulations begin with two distinct bodies: the \textit{target} (proto-Jupiter) and the \textit{impactor}, whereby the target is the more massive of the two bodies. In all the simulations proto/young Jupiter is assumed to have a compact core of heavy elements. We consider different compositions and therefore use various equations of state (EOS). For the hydrogen-helium mixture (H-He), we construct an EOS table from the REOS.3 tables for pure hydrogen (H) and pure helium (He) \citep{beckerINITIOEQUATIONSSTATE2014,wooDidUranusRegular2022} using the additive volume law (\SI{72.5}{\percent} H and \SI{27.5}{\percent} He, corresponding to the solar abundance, e.g. \citet{loddersSolarSystemAbundances2010}). In some simulations, instead of REOS.3, an extended version of the SCvH EOS \citep{saumonEquationStateLowMass1995,vazanEffectCompositionEvolution2013,matzkevichOutcomeCollisionsGaseous2024} is used for the H-He mixture. To model water, we use the ice \citep{mordasiniPlanetaryEvolutionAtmospheric2020} EOS constructed from ANEOS (ANalytic Equation Of State) \citep{thompsonImprovementsCHARTRadiationhydrodynamic1974}, while for iron and rock, we use the iron \citep{stewartEquationStateModel2020a} and forsterite \citep{stewartEquationStateModel2019} EOS constructed from M-ANEOS \citep{thompsonImprovementsCHARTRadiationhydrodynamic1974, meloshHydrocodeEquationState2007, thompsonMANEOS2019}.

The particle representation of the colliding bodies are created with the \texttt{ballic} code \citep{reinhardtNumericalAspectsGiant2017}, including improvements for multi-component models as described in \citet{chauFormingMercuryGiant2018} and \citet{reinhardtBifurcationHistoryUranus2020}. The number of particles per simulation ranges between \SI{e7}{} and \SI{2.1e9}{}. Details of the models, e.g., their masses, composition and resolution, as well as the impact conditions and simulation time of all SPH simulations are given in Appendix~\ref{sec:appendix:Extended_data_tables}.

In order to quantify mixing in the SPH simulations, we employ a set of mixing metrics that we derive and benchmark in Appendix~\ref{sec:appendix:Mixing_metrics}. From these metrics, we choose $M_{mix,\beta}$, $M_{mix,\delta,0.2}$ and $M_{mix,\beta\gamma}$ to apply on the SPH results.

\subsection{Oblique Impact}\label{sec:Oblique_Impact}
\begin{figure*}[ht!]
\centering
\includegraphics[width=\linewidth]{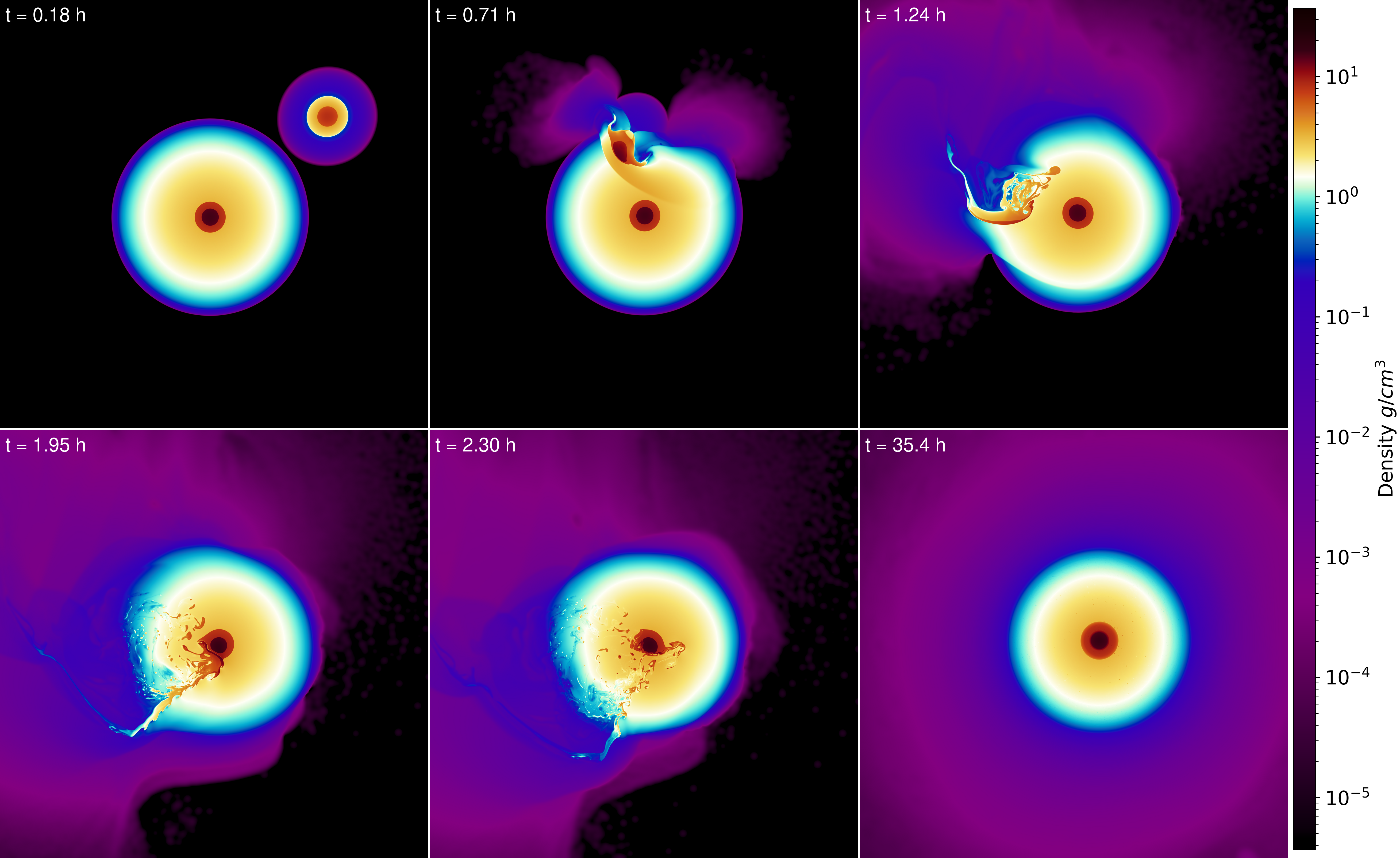}
\caption{Density slices through the oblique collision with \SI{e9}{} particles (run 5.1.2) at different times. The panels have a side length of \SI{50}{\Rearth}. The density is calculated by distributing the mass of each particle to each pixel according to the intersection between the pixel area and the particle's kernel using the SPH kernel function. Jupiter's core is not disrupted but the core of the impactor is completely sheared apart, with parts of it violently mixed into the envelope, but it accumulates quickly onto the existing solid core of Jupiter; no dilute core is formed. Animations of this impact at resolutions of \SI{e8}{} and \SI{e9}{} particles can be found at \href{https://youtu.be/QZ-CCR2OMlY}{youtu.be/QZ-CCR2OMlY} and \href{https://youtu.be/81AuWQP59qc}{youtu.be/81AuWQP59qc}.}
\label{fig:Oblique_Collision}
\end{figure*}

\begin{figure*}[ht!]
\centering
\includegraphics[width=\linewidth]{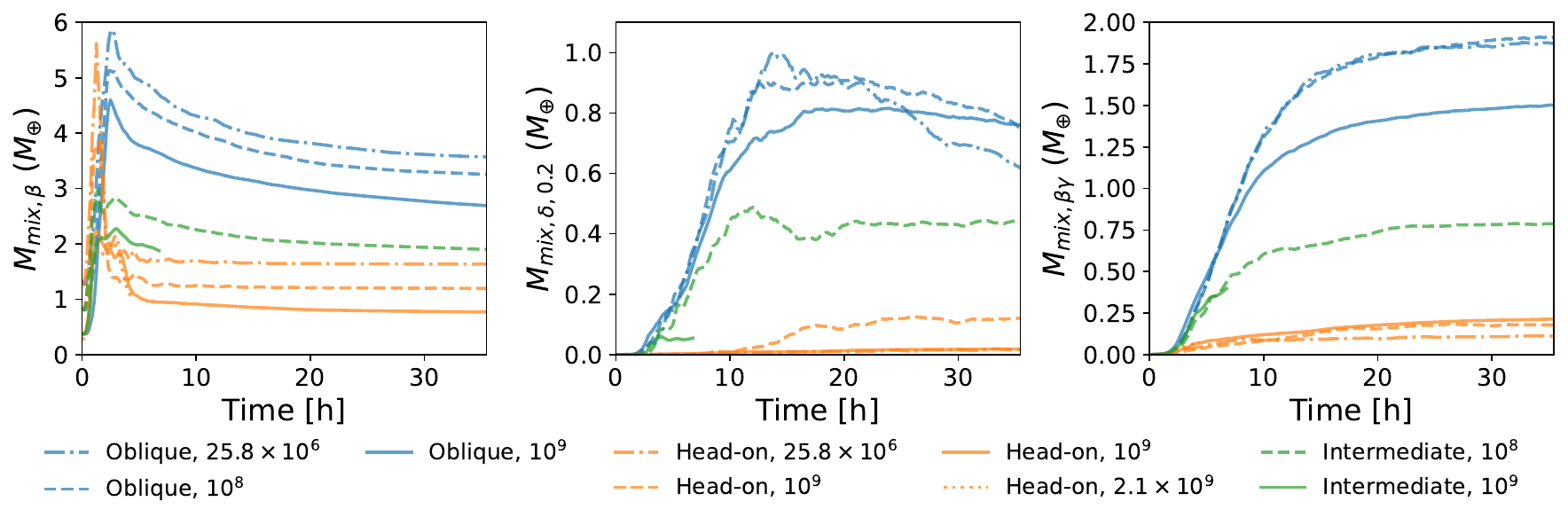}
\caption{Timeseries of the mixing metrics as defined in Appendix~\ref{sec:appendix:Mixing_metrics} applied to the simulation results of the oblique impact, the head-on impact and the intermediate impact discussed in Sections~\ref{sec:Oblique_Impact},~\ref{sec:Head-on_Impact} and~\ref{sec:Intermediate_Impact} respectively. In each case, the amount of mixing measured by the three metrics depends on the resolution.}
\label{fig:jupiter_impacts_mixing_timeseries}
\end{figure*}

First, we investigate the most likely impact scenario identified in Section~\ref{sec:N-body_results}: an oblique impact at an impact angle of $\theta=\SI{45}{\degree}$ close to the mutual escape velocity of \SI{46}{\kilo\meter\per\second} onto a nearly fully formed Jupiter with a mass of \SI{306.7}{\Mearth} and a heavy-element core of \SI{10}{\Mearth}. Our N-body simulations show that such giant impacts are rare. However, in order to maximize the mixing potential of the impact, we use an impactor with a mass of \SI{10}{\Mearth}. The impact is simulated using three different resolutions (\SI{25.8e6}{}, \SI{e8}{} and \SI{e9}{} particles, corresponding to runs 3.1.2, 5.1.4 and 5.1.2, respectively, see Table~\ref{tab:Jupiter_impact_parameters_late} for details). This impact was also considered by \citet{liuFormationJupitersDiluted2019} who found that such an impact does not change the core-envelope structure of Jupiter. Figure~\ref{fig:Oblique_Collision} shows snapshots of this collision simulation at a resolution of \SI{e9}{} particles. The impactor misses Jupiter's core and gets sheared apart in the envelope, violently mixing parts of the impactor's core with Jupiter's H-He envelope while the rest of the heavy elements settle onto the core. This happens rapidly, within a few hours, increasing the mass of Jupiter's compact core from \SI{10}{\Mearth} to $\sim\SI{18}{\Mearth}$. However, some of the heavy-element particles remain in the H-He envelope. Figure~\ref{fig:jupiter_impacts_mixing_timeseries} shows the time series of the three mixing metrics (blue lines, see Appendix~\ref{sec:appendix:Mixing_metrics}), the values of which all depend on the resolution. $M_{mix,\beta}$ is between \SI{3}{\Mearth} and \SI{4}{\Mearth}, while $M_{mix,\delta,0.2}$ is between \SI{0.6}{\Mearth} and \SI{0.8}{\Mearth} and $M_{mix,\beta\gamma}$ is between \SI{1.5}{\Mearth} and \SI{1.8}{\Mearth}. This suggests that such an impact does not lead to the formation of a dilute core. However, some heavy material is mixed into the envelope.

\subsection{Head-on Impact}\label{sec:Head-on_Impact}
\begin{figure*}[ht!]
\centering
\includegraphics[width=\linewidth]{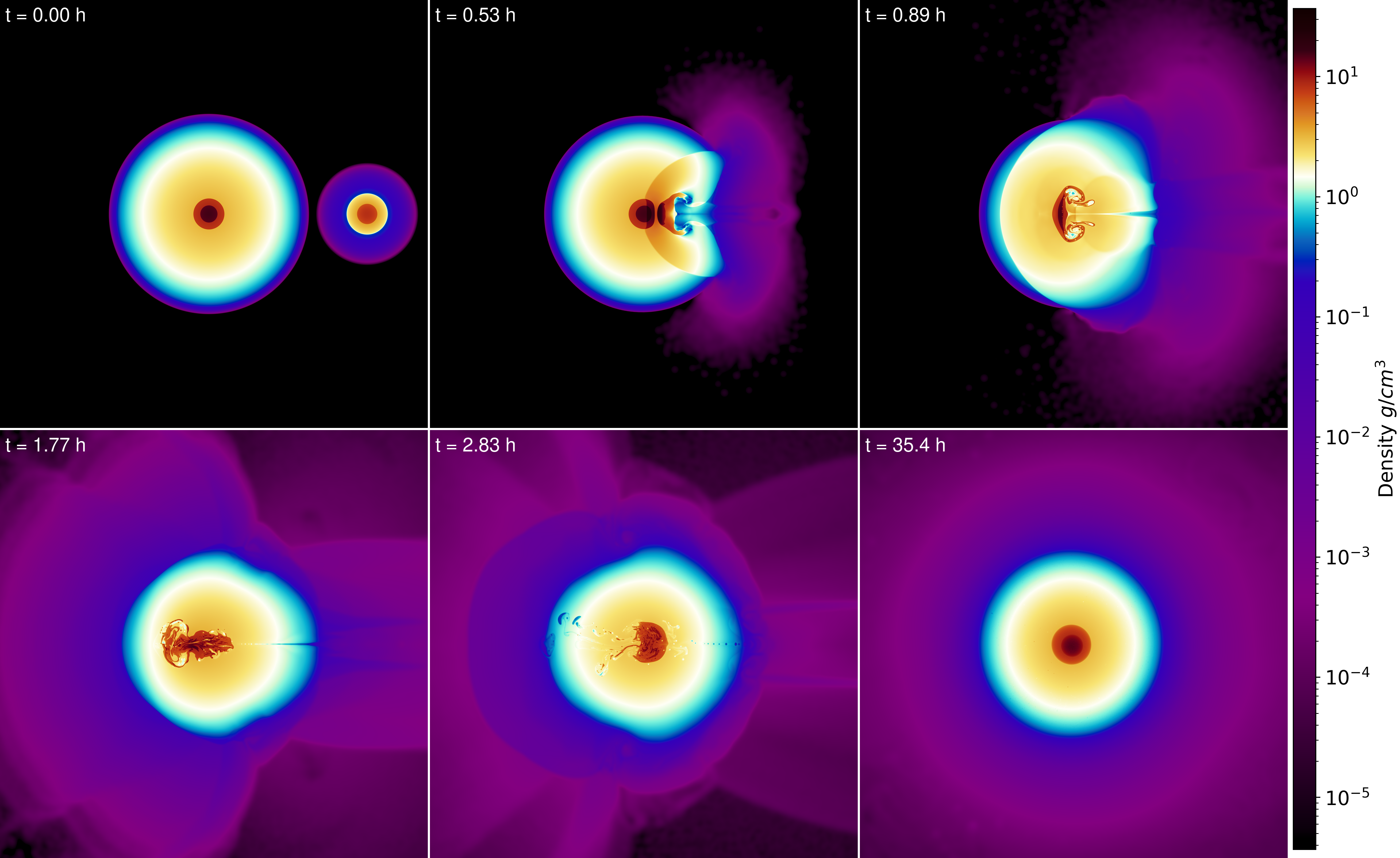}
\caption{Density slices through the head-on collision with \SI{e9}{} particles (run 5.1.1) at different times. Jupiter's core is completely disrupted but all the heavy elements settle quickly into a heavy-element core, so no dilute core is formed. Animations of this impact at resolutions of \SI{e8}{} and \SI{e9}{} particles can be found at \href{https://youtu.be/Q6eRPfo4isE}{youtu.be/Q6eRPfo4isE} and \href{https://youtu.be/0OpGIeFuERI}{youtu.be/0OpGIeFuERI}, while animations of the early stages of this impact at a resolution of \SI{2.1e9}{} particles can be found at \href{https://youtu.be/KnWIBrK4XQQ}{youtu.be/KnWIBrK4XQQ} (slices as in this figure) and \href{https://youtu.be/9iIU-Zz_Ar0}{youtu.be/9iIU-Zz\_Ar0} (different styles of volumetric rendering).}
\label{fig:Headon_Collision}
\end{figure*}

Next, we investigate the scenario identified by \citet{liuFormationJupitersDiluted2019}, a head-on collision between the same bodies as in the oblique impact above. We simulate this impact at four different resolutions (\SI{25.8e6}{}, \SI{e8}{}, \SI{e9}{} and \SI{2.1e9}{} particles, corresponding to runs 3.1.1, 5.1.3, 5.1.1 and 5.1.10, respectively, see Table~\ref{tab:Jupiter_impact_parameters_late} for details). Figure~\ref{fig:Headon_Collision} presents snapshots of this collision simulation at a resolution of \SI{e9}{} particles. During the collision, Jupiter's core is completely disrupted by the impactor's core and is stretched more than halfway towards Jupiter's surface. However, in this case, most of the heavy elements also rapidly settle to the center, forming a compact core. Figure~\ref{fig:jupiter_impacts_mixing_timeseries} shows the time series of the three mixing metrics as orange lines. Depending on the resolution, this impact results in values for $M_{mix,\beta}$ between \SI{1}{\Mearth} and \SI{2}{\Mearth} which is only marginally higher than the value for the initial condition. The two other metrics, $M_{mix,\delta,0.2}$ and $M_{mix,\beta\gamma}$ show values below $\SI{0.1}{\Mearth}$ and $\SI{0.25}{\Mearth}$ respectively. This indicates that no real mixing occurs and that most of the contribution towards $M_{mix,\beta}$ comes from the interface between the core and the envelope.

\subsection{Intermediate Impact}\label{sec:Intermediate_Impact}
\begin{figure*}[ht!]
\centering
\includegraphics[width=\linewidth]{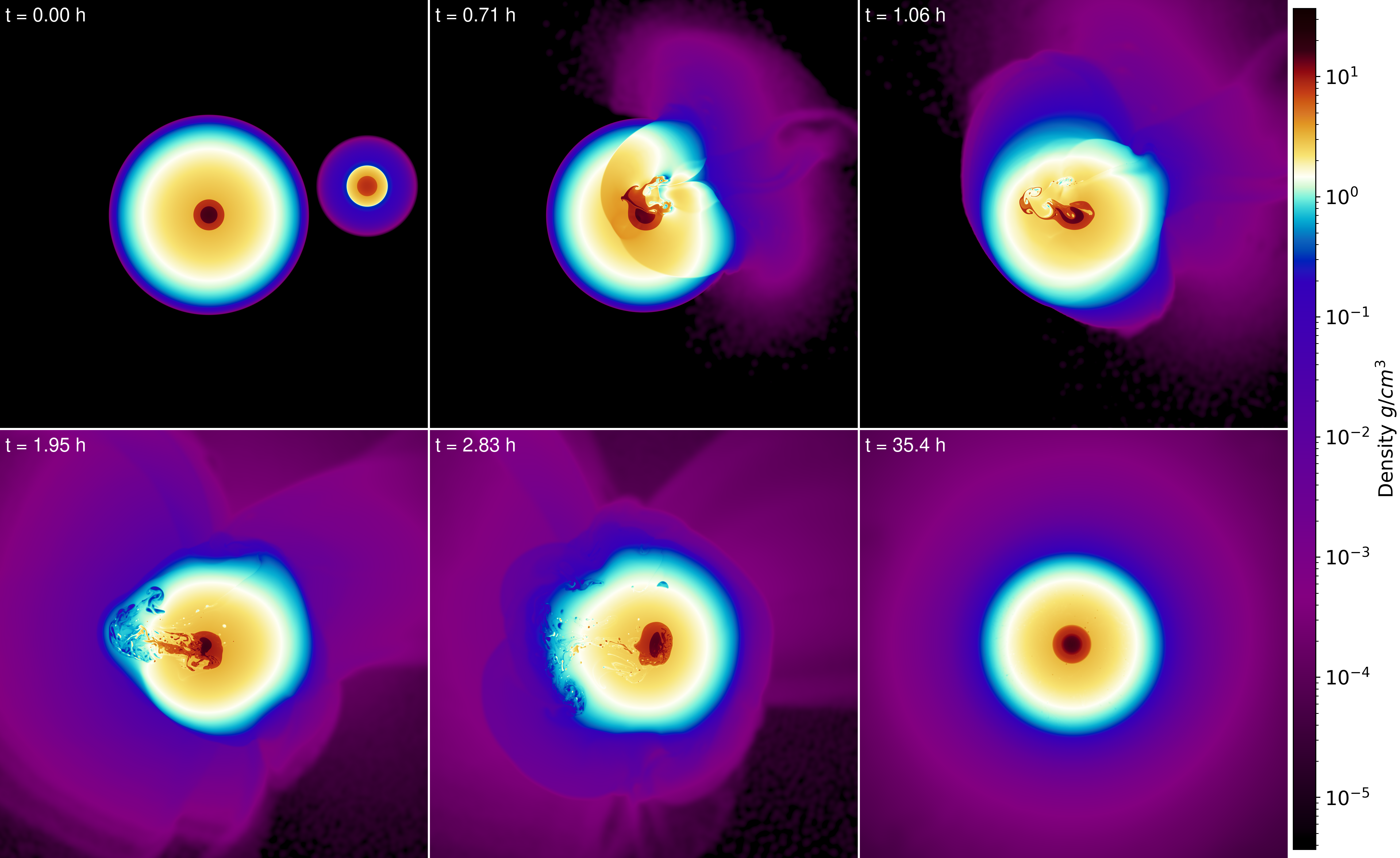}
\caption{Density slices through the intermediate impact simulation with \SI{e9}{} particles (run 5.1.9) at different times, with the last frame being from the simulation with \SI{e8}{} particles (run 5.1.8). The impact partially disrupts Jupiter's core, but most of the heavy elements quickly settle into a compact heavy-element core, i.e., no dilute core is formed. An animation of this impact at a resolution of \SI{e8}{} can be found at \href{https://youtu.be/tbopyIyjZuI}{youtu.be/tbopyIyjZuI} and of the initial stages of this impact at a resolution of \SI{e9} at \href{https://youtu.be/Ng9nWHMaUMI}{youtu.be/Ng9nWHMaUMI}.}
\label{fig:Core_Hit_Collision}
\end{figure*}

We also consider an intermediate situation between the oblique impact in Section~\ref{sec:Oblique_Impact} and the head-on impact in Section~\ref{sec:Head-on_Impact}. We use a value of $b=0.2$ for the impact parameter, resulting in an impact angle of $\theta=\SI{11.54}{\degree}$ in which case the core of the impactor makes contact with Jupiter's core during the collision, but it is still oblique so some material can be mixed into the envelope. We simulate this impact at two different resolutions (\SI{e8}{} and \SI{e9}{} particles, corresponding to runs 5.1.8 and 5.1.9 respectively, see Table~\ref{tab:Jupiter_impact_parameters_late} for details). Figure~\ref{fig:Core_Hit_Collision} shows snapshots of this collision simulation at a resolution of \SI{e9}{} particles (but the last snapshot is from the \SI{e8}{} particle simulation). We find that this impact significantly (but not fully) disturbs Jupiter's core. As expected, it leads to a compact core with slightly more mixed material than the head-on collision, but less than the oblique collision at $\theta=\SI{45}{\degree}$. Figure~\ref{fig:jupiter_impacts_mixing_timeseries} presents the time series of the three mixing metrics as green lines. $M_{mix,\beta}$, $M_{mix,\delta,0.2}$ and $M_{mix,\beta\gamma}$ are found to be \SI{2}{\Mearth}, \SI{0.4}{\Mearth} and \SI{0.75}{\Mearth}, respectively.

\subsection{Other Impacts}\label{sec:Other_Impacts}
We also explored several different modifications to the cases presented above. First, a giant impact may occur shortly after Jupiter's formation where it is expected to be significantly hotter and more extended. We therefore increased the surface temperature of both Jupiter and the impactor from \SI{300}{\kelvin} to \SI{1200}{\kelvin}. This leads to an increase in Jupiter's central temperature from \SI{26100}{\kelvin} to \SI{46800}{\kelvin}. In the head-on collision (run 5.1.5), the qualitative results are unchanged as we still get a compact core. However, the values of the mixing metrics increase significantly, $M_{mix,\beta}$ increases from \SI{1.12}{\Mearth} to \SI{2.51}{\Mearth}, $M_{mix,\delta,0.2}$ increases from \SI{0.121}{\Mearth} to \SI{1.23}{\Mearth} and $M_{mix,\beta\gamma}$ increases from \SI{0.180}{\Mearth} to \SI{0.900}{\Mearth}. The collision with $\theta=\SI{45}{\degree}$ (run 5.1.6) results in a hit-and-run: due to the higher surface temperature the impactor is so expanded that its core does not hit Jupiter. The collision with $b=0.2$ (run 5.1.7) leads to an interaction between the two cores, although it is less direct than with the colder bodies. In this case the qualitative results are also similar, but the values of the mixing metrics change, $M_{mix,\beta}$ increases from \SI{1.90}{\Mearth} to \SI{3.95}{\Mearth}, $M_{mix,\delta,0.2}$ increases from \SI{0.437}{\Mearth} to \SI{1.69}{\Mearth} and $M_{mix,\beta\gamma}$ increases from \SI{0.787}{\Mearth} to \SI{1.92}{\Mearth}.

As a second modification, we changed the H-He EOS equation from REOS.3 \citep{beckerINITIOEQUATIONSSTATE2014} to SCvH \citep{saumonEquationStateLowMass1995,vazanEffectCompositionEvolution2013,matzkevichOutcomeCollisionsGaseous2024}. Interestingly, this leads to a significantly smaller central temperature (\SI{17900}{\kelvin} vs. \SI{26100}{\kelvin} in the REOS.3 case). The qualitative aspects of the impact are unchanged. However, the values of the mixing metrics are generally smaller than in the REOS.3 simulations. We therefore suggest that there is a correlation between the amount of mixing and the temperature in the central region of Jupiter and we hope to explore this in future research. 

Third, we replaced the differentiated \SI{10}{\Mearth} impactor with a \SI{1}{\Mearth} pure rock body and simulated the oblique and head-on impacts at \SI{25e6}{} particles (runs 3.2.1 and 3.2.2, see Table~\ref{tab:Jupiter_impact_parameters_late}). However, such a small impactor does not disrupt Jupiter's core, instead, it is simply added to the core.

Finally, Table~\ref{tab:Jupiter_simplified_impact_parameters} lists a set of simulations that use simplified Jupiter and impactor models where Jupiter consists only of two layers (core and hydrogen-helium envelope) while the impactor consists of only one material. We combine three different materials for both the core and the impactor (rock, ice and pure helium). We find that using a simpler structure for Jupiter and the impactor (as used by \citet{sandnesNoDiluteCore2024}) does not significantly change the simulation's outcome.

\section{Post-impact evolution}\label{sec:Post-impact_evolution}
To simulate Jupiter's post-impact evolution, we used a modified version \citep{mullerChallengeFormingFuzzy2020,mullerTheoreticalObservationalUncertainties2020,mullerCanJupitersAtmospheric2024} of the Modules for Experiments in Stellar Astrophysics code (MESA; \citet{paxtonMODULESEXPERIMENTSStelLAR2010,paxtonMODULESEXPERIMENTSStelLAR2013,paxtonMODULESEXPERIMENTSStelLAR2015, paxtonModulesExperimentsStellar2018,paxtonModulesExperimentsStellar2019,jermynModulesExperimentsStellar2023}). The relevant modifications compared to the official version concerned the EOS, which was altered to handle arbitrary mixtures of hydrogen, helium, and a heavy element. For H-He, we used the \citet{chabrierNewEquationState2021} EOS, and the heavy elements were modeled as a 50-50 water-rock mixture with QEOS \citep{moreNewQuotidianEquation1988}. We note that for the thermal evolution calculations, we did not use the REOS.3 H-He EOS because the official version does not provide the entropy, and there are known issues with the thermodynamic integration \citep{xieAccurateThermodynamicallyConsistent2025}. Previous work comparing the SCvH and \citet{chabrierNewEquationState2019} EOSs found only small differences in the inferred outcomes of convective mixing \citep{mullerChallengeFormingFuzzy2020,knierimConvectiveMixingGas2024}. Therefore, we don't expect this inconsistency to affect our conclusions. MESA uses the standard mixing-length theory approach to treat convective mixing as a diffusive process \citep[e.g.,][]{kippenhahnStellarStructureEvolution2013}, and the regions unstable to large-scale convection were determined by the Ledoux criterion \citep{ledouxStellarModelsConvection1947} in our simulations. The default treatment of the energy equation in the version of MESA used in this work neglects the change in internal energy coming purely from a change in composition \citep[see, e.g.,][for a discussion]{paxtonModulesExperimentsStellar2018,surAPPLEEvolutionCode2024}. We therefore implemented a time-centered finite-difference calculation of this extra term to include all relevant effects in the energy equation.

\begin{figure}[ht!]
\centering
\includegraphics[width=\linewidth]{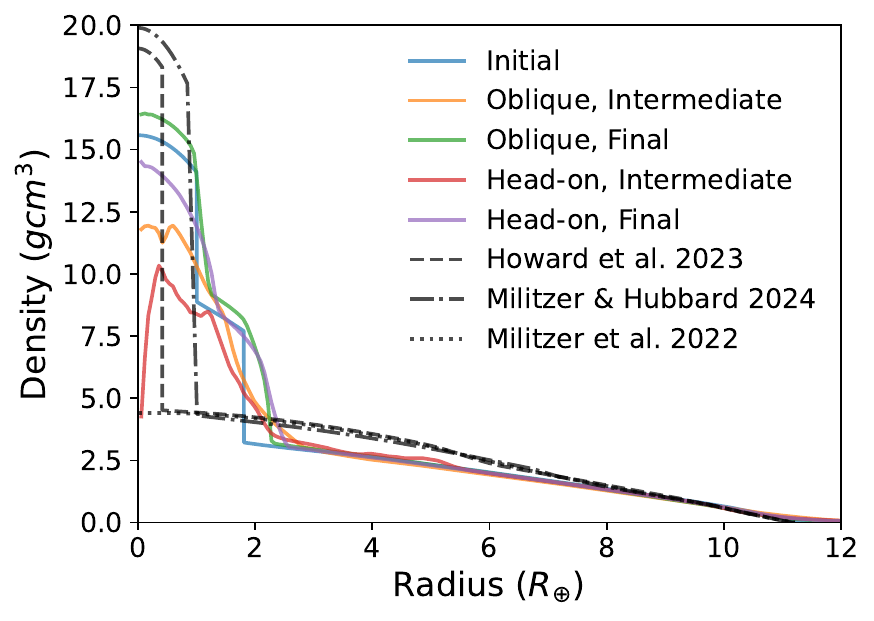}
\caption{Density profiles of proto-Jupiter for the intermediate and the final states after the oblique and head-on impacts (bottom left and bottom right frame of Figures~\ref{fig:Oblique_Collision} and~\ref{fig:Headon_Collision}). The dashed black line shows a density profile of Jupiter with a dilute core and a very small compact center \citep{howardJupitersInteriorJuno2023} while the dot-dashed line shows a profile with a more massive compact core \citep{militzerStudyJupitersInterior2024}. The dotted line corresponds to a dilute core without a pure heavy-element central region \citep{militzerJunoSpacecraftMeasurements2022}.}
\label{fig:jupiter_profiles}
\end{figure}

We calculated 1D profiles for the oblique and the head-on impacts from the results shown in the bottom left (intermediate state) and bottom right (final state) panels of Figures~\ref{fig:Oblique_Collision} and~\ref{fig:Headon_Collision}. We averaged the density, temperature, and composition in 500 spherical shells distributed equally between $R=\SI{0}{\Rearth}$ and $R=\SI{30}{\Rearth}$, centered on the connection between the center of proto-Jupiter at the start and of Jupiter at the end of the simulation. The density profiles are shown in Figure~\ref{fig:jupiter_profiles}. The intermediate-state profile represents a configuration where more heavy elements are mixed within the interior, a situation that would occur if mixing is not suppressed by SPH (see Appendix for detailed discussion). In the final state of both impacts, the top region of the rock and ice layers are extremely hot (peak temperatures up to \SI{1.6e5}{\kelvin}). This is because shock-heated impactor material gets compressed as it falls onto the core, heating it further. The target's core is also shock-heated, but only moderately. 

The EOSs used in the SPH and the thermal evolution simulations are different. In addition, mixtures in the SPH simulations are not treated properly (i.e., the thermodynamics of mixtures is not included - see Section~\ref{sec:Mixing_and_de-mixing_in_SPH} for details), and radiative cooling is also not considered. As a result, we did not use the temperature profiles inferred from the SPH simulations. To account for the uncertainties associated with the heating of Jupiter's interior with such impacts, we constructed three different initial thermal states corresponding to hot, warm, and cold post-impact interiors. The hot and warm states correspond to a post-impact Jupiter that is significantly inflated (at 2.3 and \SI{1.3}{\Rjupiter}) due to the energy deposition, while the cold state post-impact Jupiter has a radius of \SI{1}{\Rjupiter}. The models have initial central temperatures of roughly \SIrange{7}{8.5e4}{\kelvin} (hot), \SIrange{5}{6e4}{\kelvin} (warm) and \SIrange{3}{5e4}{\kelvin} (cold), depending on the composition profile. 
The hot and warm initial conditions are consistent with comprehensive formation models of Jupiter, which yield a radius of about \SI{2}{\Rjupiter} and maximum temperatures exceeding $\sim$ \SI{6e4}{\kelvin} at the end of the runaway gas accretion phase \citep[e.g.,][]{cummingPrimordialEntropyJupiter2018,stevensonMixingCondensableConstituents2022}. The cold initial condition does not match any current formation models. In general, a colder interior implies there is less energy available to destroy a dilute core created by a giant impact. Therefore, this initial condition was included to investigate the outcome of favorable initial conditions for sustaining an extended fuzzy core over billions of years. Our impact simulations suggest that Jupiter's interior is extremely hot ($>$ \SI{1e5}{\kelvin}) shortly after the impact. However, our evolution models were unable to reach such high temperatures for the initial models. Increasing the temperatures by heating the planet leads to a significantly inflation of the radius, preventing the central temperatures from rising. We note that the hot initial condition is the most realistic scenario and the one which is most consistent with the SPH simulations.

\begin{figure}
\centering
\includegraphics[width=\linewidth]{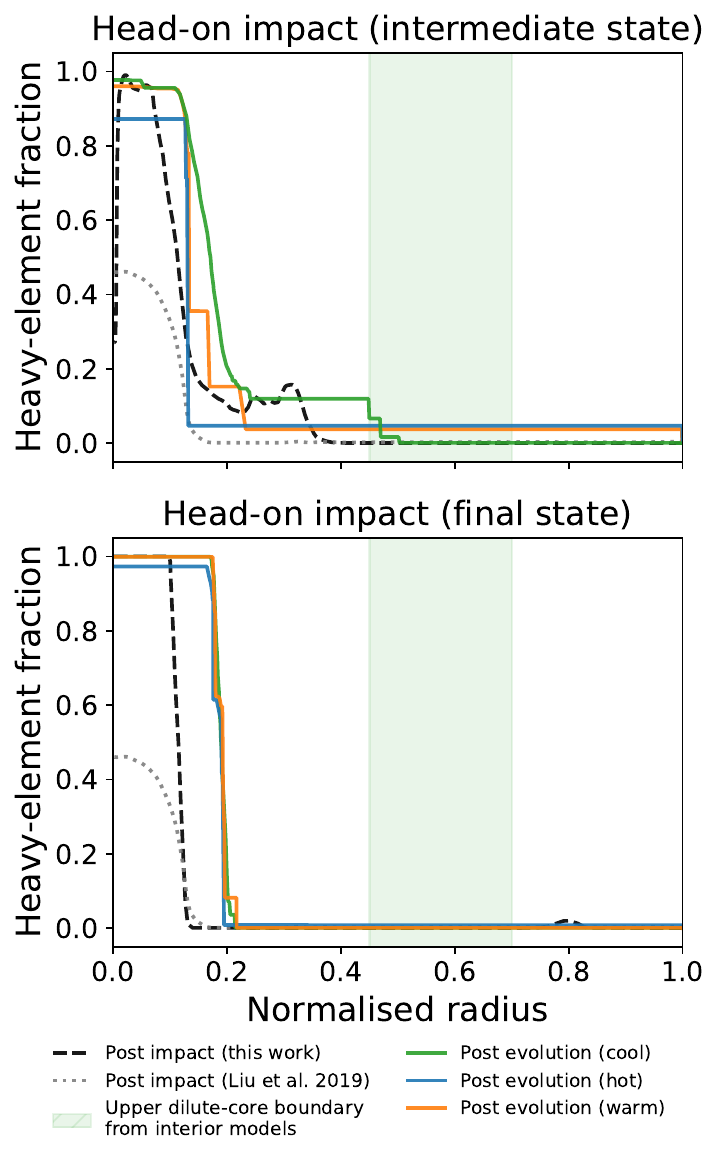}
\caption{Heavy-element fraction vs. the normalised radius for the head-on impact. The initial composition profile from the impact simulation is shown as dashed black lines. The dotted gray lines show the head-on impact profile from \citet{liuFormationJupitersDiluted2019}. The colored lines show the current-day profiles after 4.56 Gyr of cooling for three different initial thermal states. The shaded green region shows the upper boundary of Jupiter's fuzzy core as inferred by interior models. Except for the cool intermediate-state initial conditions, no models lead to an extended dilute core today (see text for details).}
\label{fig:composition_profiles_headon}
\end{figure}

\begin{figure}
\centering
\includegraphics[width=\linewidth]{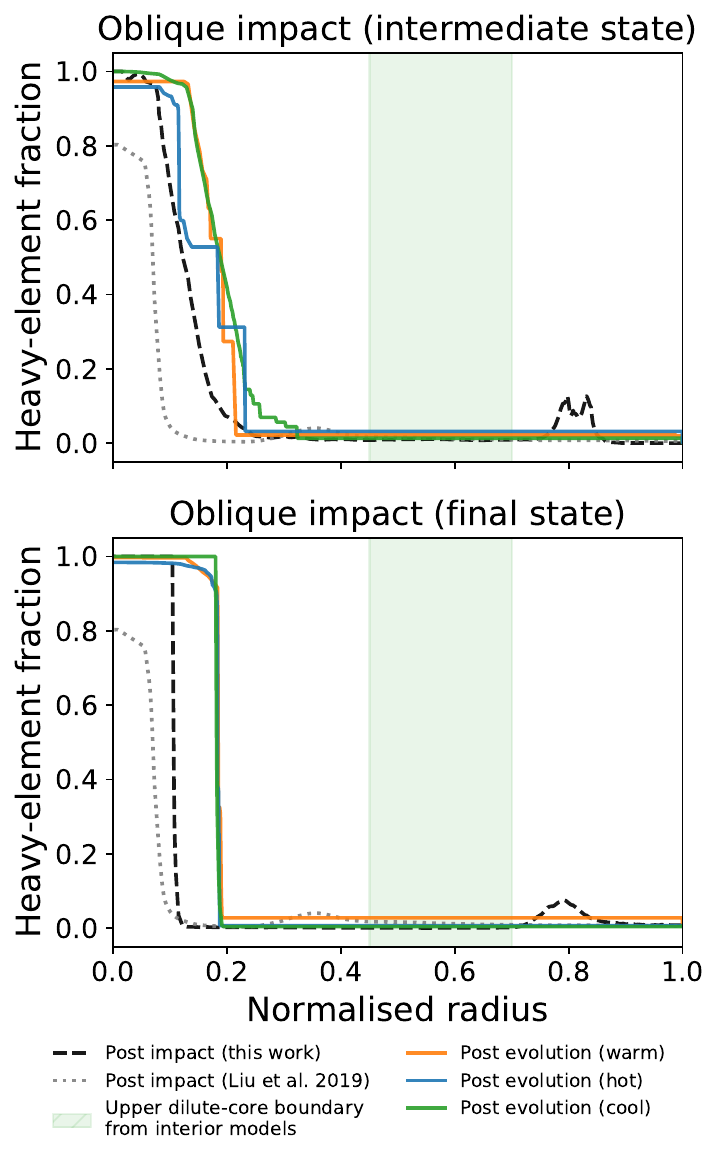}
\caption{Same as Figure~\ref{fig:composition_profiles_headon} but for the oblique impact. None of the models lead to an extended dilute core today.}
\label{fig:composition_profiles_oblique}
\end{figure}

Our approach outlined above resulted in twelve different models: Four composition profiles with each three initial thermal states. All the models were evolved for 4.56 Gyr, predicting Jupiter's current-day composition profiles. The results are shown in Figures~\ref{fig:composition_profiles_headon} and~\ref{fig:composition_profiles_oblique}.

Overall, our results are incompatible with internal structure models of Jupiter fitting its gravitational moments measured by the Juno orbiter \citep{boltonJupitersInteriorDeep2017}. These models infer an extended dilute core with an upper boundary extending to roughly \SIrange{45}{60}{\percent} of Jupiter's radius as highlighted in Figures~\ref{fig:composition_profiles_headon} and~\ref{fig:composition_profiles_oblique} \citep{wahlComparingJupiterInterior2017,debrasNewModelsJupiter2019,nettelmannTheoryFiguresSeventh2021,militzerJunoSpacecraftMeasurements2022,miguelJupiterInhomogeneousEnvelope2022,howardJupitersInteriorJuno2023}. The only model leading to a marginally consistent extended dilute core used the intermediate state from the head-on impact with likely unrealistically cold initial conditions. All the other scenarios we considered lead to a Jupiter with a homogeneous envelope and a compact core extending at most around \SI{20}{\percent} in radius. This is because the result of the giant impact is already a stable compact core and because convective mixing erodes the dilute core as Jupiter evolves.

We find that there are only minor differences between the head-on and oblique impacts; the result depends much more on which state from the SPH simulation was used for the initial composition profile (intermediate or final). When the final states of the SPH simulations are considered, there is very limited convective mixing throughout the evolution, almost independent of the initial thermal state. This is because the initial composition profile is akin to a pure heavy-element compact core, with an extremely steep and therefore stable composition gradient at the boundary around $m \approx \SI{0.05}{\Mjupiter}$. In contrast, the intermediate states have a more dilute-core-like structure where the heavy-element fraction gradually decreases from the center to the envelope. These are also more similar to the post-impact profiles from \citet{liuFormationJupitersDiluted2019}, although our impact simulations suggest a larger pure heavy-element core. In these cases, the outcome strongly depends on Jupiter's internal temperatures after the impact. The hot initial models erode the core much more efficiently compared to the cold ones. The only model that leads to an extended dilute core (with a boundary at about $m \approx \SI{0.25}{\Mjupiter}$ or $r \approx \SI{0.45}{\Rjupiter}$) today uses the intermediate state from the head-on impact and assumes that Jupiter was cold and not inflated after the impact. However, we note that such a cold initial thermal state is unlikely; even without the impact energy, Jupiter would likely have been much hotter at the time of the impact \citep[e.g.,][]{cummingPrimordialEntropyJupiter2018,stevensonMixingCondensableConstituents2022}. None of the other models we considered lead to an extended dilute-core structure today.

Note that thermal evolution simulations may overestimate the amount of convective mixing. Instead of large-scale convection, oscillatory double-diffusive convection could develop and transport the energy from the deep interior to the envelope \citep[e.g.,][]{leconteNewVisionGiant2012,woodNEWMODELMIXING2013,radkoDoubleDiffusiveRecipesPart2014,garaudDoubleDiffusiveConvectionLow2018}, leaving the dilute core intact. However, hydrodynamic simulations suggest that double diffusive layers are transient and tend to merge quickly \citep{fuentesLayerFormationStably2022,tulekeyevConstraintsLongtermExistence2024,fuentes3DSimulationsSemiconvection2025}, except under special conditions or if merging is significantly inhibited by rotation \citep{fuentesEvolutionSemiconvectiveStaircases2024}. We therefore conclude that it is unlikely that Jupiter's dilute core resulted from a giant impact.

\section{Discussion}\label{sec:Discussion}
Our study represents a step forward towards a better understanding of the impact conditions onto proto-Jupiter, the expected outcomes of different collision scenarios and their impact on the long-term evolution. However, more work is clearly required. In this section, we discuss how our results compare to previous work, the possibility that the fuzzy cores of Jupiter and Saturn are caused by the formation processes, and the probability of impacts during the early stages in Jupiter's formation. We also discuss the limitations of SPH simulations in modeling mixing and de-mixing. 

\subsection{Comparison with previous studies}
The possibility of forming Jupiter's fuzzy core by a giant impact has been suggested before. Using 2D N-body simulations, \citet{liuFormationJupitersDiluted2019} identified a head-on impact which sufficiently disrupts Jupiter's core to create a dilute core that is stable over time if proto-Jupiter is not too hot after the impact, according to their post-impact evolution modeling. However, our study, which uses an N-body analysis in 3D, shows that head-on impacts are rather unlikely. Instead, the most likely impact is found to be very oblique. Under such a geometry, the impactor misses Jupiter's core in the initial phase of the collision, and therefore, a dilute core cannot be formed. Our hydrodynamics simulations show that even a head-on impact does not lead to sufficient mixing of core material towards Jupiter's outer region. The fuzzy core found in \citet{liuFormationJupitersDiluted2019} is expected to be the result of the overly diffusive nature of the numerical method used in the study. Our conclusion that head-on collisions cannot lead to the formation of Jupiter's fuzzy core confirms the findings of \citet{sandnesNoDiluteCore2024}. 

\subsection{Fuzzy cores as a result of the formation process}
As we discussed earlier, interior models of Jupiter that fit Juno data require an extended dilute core. However, by now it became clear that this is also the case for Saturn \citep[see e.g.,][for a review]{helledFuzzyCoresJupiter2024}. In fact, for Saturn, the evidence for a fuzzy core is even stronger than for Jupiter. In addition to constraints from the gravitational moments, observations of oscillations of Saturn's rings (ring seismology) suggest that Saturn has an extended region enriched in heavy elements in its deep interior, and that this region is stably stratified \citep{fullerSaturnRingSeismology2014,mankovichDiffuseCoreSaturn2021}, meaning that there is no large-scale convection occurring. Fuzzy cores are therefore not the exception, but our current understanding of the solar system giant planets implies that they are the norm. It is therefore unlikely that low-probability giant impacts that required very specific conditions are the explanation for both Saturn's and Jupiter's dilute cores. The more probable alternative is that fuzzy cores (and composition gradients) are outcomes of the planetary formation process \citep[e.g.,][]{helledFuzzinessGiantPlanets2017,stevensonMixingCondensableConstituents2022}. However, this explanation also faces the challenge that hot primordial interiors tend to lead to large-scale convection that erodes or even destroys composition gradients \citep{vazanCONVECTIONMIXINGGIANT2015,mullerChallengeFormingFuzzy2020,knierimConvectiveMixingGas2024}. Two recent studies identified the initial conditions that lead to an extended dilute core in Jupiter today assuming that it had a primordial composition gradient \citep{tejadaarevaloJupiterEvolutionaryModels2025, surSimultaneousEvolutionaryFits2025}. They found that a fuzzy core that extends to about \SI{45}{\percent} radially today, Jupiter must have formed rather cold, with central temperatures of $\sim$ \SI{4e4}{\kelvin}. This confirms the earlier results from \citet{mullerChallengeFormingFuzzy2020}, who showed that warm to hot initial conditions lead to more compact cores.

\subsection{Collisions on early Jupiter}\label{sec:Discussion_Collisions_on_early_Jupiter}
The N-body simulations presented in Section~\ref{sec:N-body_results} clearly show that the last giant impact on Jupiter is unlikely to have happened before the onset of runaway gas accretion. However, we still investigate whether a giant impact very early in the evolution of Jupiter may lead to a fuzzy core. We run a set of low-resolution collision simulations (\SI{e6}{} particles per \SI{}{\Mearth}, see Table~\ref{tab:Jupiter_impact_parameters_early}) with targets containing a \SI{10}{\Mearth} heavy-element core (either rock or ice) and an envelope of either \SI{5}{\Mearth} or \SI{10}{\Mearth} H-He and impactors of \SI{1}{\Mearth} and \SI{10}{\Mearth} of either rock or ice. None of these simulations lead to a "dilute core", as they all lead to cores that are pure heavy-element in composition. The scenario with the most mixing is an oblique impact ($b=0.2$) of a \SI{10}{\Mearth} ice impactor on a proto-Jupiter with a \SI{10}{\Mearth} ice core and \SI{10}{\Mearth} H-He envelope at \SI{20}{\kilo\meter\per\second} (run 2.3.2) which results in $M_{mix,\beta}=\SI{2.13}{\Mearth}$, $M_{mix,\delta,0.2}=\SI{1.48}{\Mearth}$ and $M_{mix,\beta\gamma}=\SI{0.71}{\Mearth}$, while all other simulations have much smaller values, do not have all the heavy elements bound to Jupiter, or are hit-and-run encounters (see Table~\ref{tab:Jupiter_impact_results_early} in Appendix~\ref{sec:appendix:Extended_data_tables}).

\subsection{Modeling mixing in SPH}\label{sec:Mixing_and_de-mixing_in_SPH}
It is well known that traditional SPH suppresses mechanical mixing at the interface between dissimilar materials \citep{woolfsonPracticalSPHModels2007,dengPrimordialEarthMantle2019,dengEnhancedMixingGiant2019,reinhardtBifurcationHistoryUranus2020,ruiz-bonillaDealingDensityDiscontinuities2022}. It is therefore not surprising that the simulations presented in Section~\ref{sec:Impact_Simulations} do not produce large values for the mixing metrics, although we use an interface correction \citep{ruiz-bonillaDealingDensityDiscontinuities2022} which substantially improves the modeling of interfaces. Our SPH simulations also suggest that even when materials are mechanically mixed during the initial phase of the impact, they tend to separate from each other, with the denser component quickly falling into the center of Jupiter. As we discuss below, this separation is caused by the lack of a thermodynamically consistent treatment of mixtures.

\subsubsection{Mixing under various conditions}
\begin{figure}[ht!]
\centering
\includegraphics[width=\linewidth]{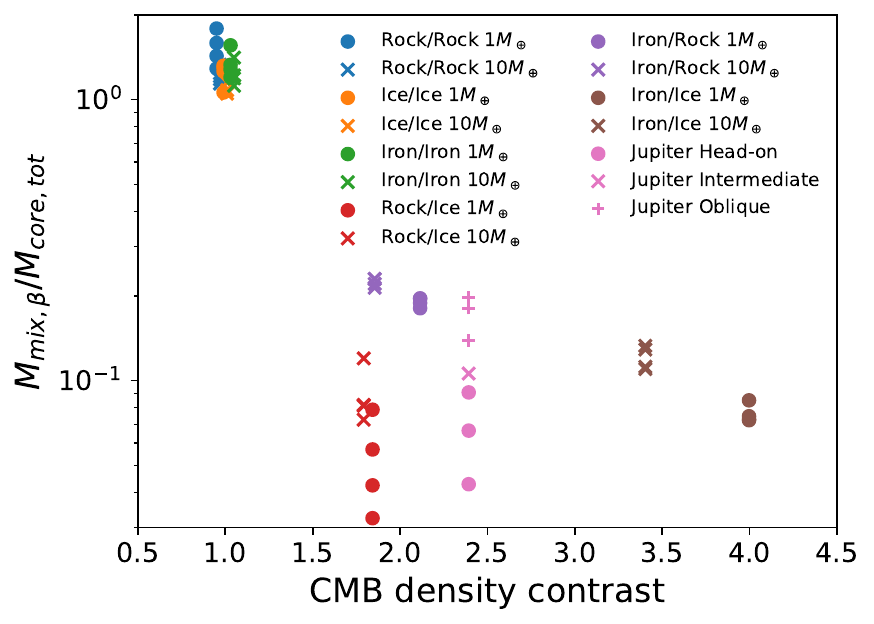}
\caption{$M_{mix,\beta}$, scaled to the total mass of core material, at the end of the simulation for the impact simulations from Tables~\ref{tab:Jupiter_impact_parameters_late},~\ref{tab:Mixing_Runs_Pure} and~\ref{tab:Mixing_Runs} that were run to \SI{35.4}{\hour} as a function of the density contrast at the core-mantle boundary (CMB). The points with a CMB density contrast of \SI{1}{} are slightly spread out horizontally for better visibility. While all values of the simulations where core and mantle are of the same material are above \SI{1}{}, the results of the impacts with iron/rock and iron/ice bodies show a clear trend that $M_{mix,\beta}$ decreases with increasing CMB density contrast. The results of the impacts between rock/ice bodies and the impacts on Jupiter do not follow this trend and they also show a strong decreasing trend with resolution.}
\label{fig:Mbeta_vs_CMB_density_contrast}
\end{figure}

In Appendix~\ref{sec:appendix:Modelling_mixing_in_SPH} we consider various simulations with different setups in order to explore the behavior of mixing in our simulations. We find that in collisions without a material interface that could suppress mixing, $M_{mix,\beta}$ increases with resolution and no de-mixing occurs. Without a collision, models that start with a pre-mixed state both with and without H-He settle back into a differentiated state relatively quickly. We find that the settling timescale decreases with increasing resolution. We can infer that if an impact shows a steady or increasing trend in $M_{mix,\beta}$ with increasing resolution, the impact very likely mixes the materials, although the true extent of the mixing is suppressed by the shortcomings of the SPH method. On the other hand, in the case of a decreasing trend, it remains unclear whether mixing occurs. In the impact simulations in Appendix~\ref{sec:appendix:Modelling_mixing_in_SPH}, we get stable or increasing values with increasing resolution for the simulations with iron/rock and iron/ice bodies, while the simulations with rock/ice bodies show a clear decreasing trend. This decreasing trend is also present in our Jupiter impact simulations (see Sections~\ref{sec:Oblique_Impact}~-~\ref{sec:Intermediate_Impact}).

Figure~\ref{fig:Mbeta_vs_CMB_density_contrast} shows the value of $M_{mix,\beta}$, scaled to the total mass of core material, at the end of the impact simulations from Tables~\ref{tab:Jupiter_impact_parameters_late},~\ref{tab:Mixing_Runs_Pure} and~\ref{tab:Mixing_Runs} that were run to \SI{35.4}{\hour} vs.~the core-mantle boundary (CMB) density contrast. The impacts between iron/rock and iron/ice bodies show, with little scatter, a trend that $M_{mix,\beta}$ decreases with increasing CMB density contrast. The impacts between rock/ice bodies and the head-on and core-hit impacts on Jupiter deviate from this trend towards less mixing, while the oblique impact on Jupiter lies on the trend line between the iron/rock and iron/ice results, but with larger scatter. 

We can therefore conclude that the impacts between iron/rock and iron/ice bodies mix core material into the mantle, while the collisions between rock/ice bodies and those on Jupiter are less expected to mix. The oblique impact on Jupiter is borderline, but since $M_{mix,\beta}$ decreases with increasing resolution, we also don't expect much mixing in this case. 

\subsubsection{Improving thermodynamic consistency in SPH}
Besides inherent issues of SPH with modeling mechanical mixing, the primary reason for suppressed mixing in SPH is the method’s inherently inconsistent treatment of thermodynamics. In SPH, each particle is assigned a specific material, and its density is calculated based on the surrounding particles -- either through a smoothing operation or by applying the mass conservation equation. Thermodynamic values are then derived using this density and the EOS of the particle’s assigned material. Mixing could be significantly improved by adopting an alternative approach: instead of relying solely on the particle’s assigned material, the material content of all particles within the kernel could be used to evaluate an additive volume law for determining thermodynamic variables. Such an approach would also eliminate the need for an interface correction when handling material boundaries. Furthermore, this could be extended to allow particles to change their material composition. An implementation of this method in SPH seems promising and should be addressed in future work. 

\section{Conclusions}\label{sec:Conclusions}
In this paper, we explored the possibility that Jupiter's fuzzy core is the result of a giant impact. To assess the likelihood of a high-energy collision during Jupiter's growth, we first analyzed the distribution of impact conditions for giant impacts on Jupiter using 3D N-body simulations, allowing us to identify the most probable collision scenario. We then performed high-resolution SPH impact simulations to investigate how a massive collision affects Jupiter’s pre-existing compact core. Following the impact simulations, we employed the MESA stellar evolution code to track the long-term evolution of the post-impact internal structure. Since the immediate post-impact state may be unstable over long time scales, we examined whether there can be significant mixing, core erosion, or further structural reorganization over time. Understanding these processes is critical for determining whether a giant impact can create and sustain the observed dilute core structure of Jupiter. We also discuss key aspects of modeling material mixing in SPH simulations and the challenges in simulating mixing. Our key conclusions can be summarized as follows:

\begin{itemize}
\item A head-on impact onto young Jupiter is dynamically very unlikely. The last giant impact is expected to occur at an oblique angle of $\theta=\SI{45}{\degree}$.
\item SPH simulations show that such an oblique impact does not disrupt Jupiter's core enough to form a dilute core.
\item Although head-on impacts can temporarily disrupt a compact heavy-element core, the heavy elements quickly settle into a compact core. None of our simulations can lead to a structure that resembles a fuzzy core.
\item Thermal evolution simulations show that when considering reasonable interior temperatures for Jupiter post-impact, giant impacts do not lead to an extended dilute core as inferred by interior models. Convective mixing tends to homogenize the planetary envelope, leaving a distinct compact core. 
\item Jupiter's fuzzy core has not formed via a giant impact. The formation of a dilute core is likely to be related to Jupiter's formation history. This also more naturally explains why Saturn also consists of a fuzzy core.
\end{itemize}

While our results, which are based on various methods, clearly suggest that Jupiter's fuzzy core was not formed by impacts, further investigations are welcome. For example, it would be interesting to consider a giant impact when Jupiter's primordial interior already consists of composition gradients due to its formation process. Investigations that include rotation and various compositions for both the target and impactor are also desirable. We also encourage further studies on mixing in impact simulations and more advanced thermodynamical modeling of mixtures. Future research should also include a more unified framework to connect SPH simulations with long-term thermal evolution simulations. Finally, we note that the origin of fuzzy cores in giant planets remains mysterious. Future studies should further investigate the role of impacts and convective mixing on the planetary evolution and final internal structure.

\acknowledgments
We thank the anonymous reviewer for valuable suggestions and comments that helped to substantially improve the paper. This work has been carried out within the framework of the National Centre of Competence in Research PlanetS supported by the Swiss National Science Foundation under grants 51NF40\_182901 and 51NF40\_205606. The authors acknowledge the financial support of the SNSF. We acknowledge access to Piz Daint and Alps.Eiger at the Swiss National Supercomputing Centre, Switzerland under the University of Zurich's share with the project ID UZH4. This work was supported by a grant from the Swiss National Supercomputing Centre (CSCS) under project ID S1285 on Piz Daint. N-body simulations were carried out on the Cray XC50 at the Center for Computational Astrophysics, National Astronomical Observatory of Japan.

\section*{Data Availability}
The data underlying Figures~\ref{fig:disk}~-~\ref{fig:Impact_Conditions}, \ref{fig:jupiter_impacts_mixing_timeseries}, \ref{fig:jupiter_profiles}~-~\ref{fig:mixing_resolution_study_2mat_dense_scaled}, \ref{fig:jupiter_demixing_timeseries} and \ref{fig:demixing_resolution_ratio_study} can be found in the GitHub repository \citet{meierJupiter_Fuzzy_Meier2025} with the current version deposited to Zenodo: \href{https://zenodo.org/records/15411990}{doi:10.5281/zenodo.15411990}. Raw initial condition and simulation result files will be shared on reasonable request to the corresponding author.

%

\vspace{5mm}
\facilities{Swiss National Supercomputing Centre (Piz Daint, Alps.Eiger)}


\software{pkdgrav3 \citep{potterPKDGRAV3TrillionParticle2017},
ballic \citep{reinhardtNumericalAspectsGiant2017},
eoslib \citep{meierEOSlib2021,meierANEOSmaterial2021},
tipsy \citep{n-bodyshopTIPSYCodeDisplay2011},
numpy \citep{harrisArrayProgrammingNumPy2020},
scipy \citep{virtanenSciPy10Fundamental2020},
matplotlib \citep{hunterMatplotlib2DGraphics2007},
GNU parallel \citep{tangeGNUParallelCommandline2011},
ImageMagick \citep{imagemagickstudiollcImageMagick2024}
}

\bibliographystyle{aasjournal} 
\bibliography{main}

\begin{thebibliography}{}
\expandafter\ifx\csname natexlab\endcsname\relax\def\natexlab#1{#1}\fi
\providecommand{\url}[1]{\href{#1}{#1}}

\bibitem[{Alonso~Asensio {et~al.}(2023)Alonso~Asensio, Dalla~Vecchia, Potter, \& Stadel}]{alonsoasensioMeshfreeHydrodynamicsPkdgrav32023}
Alonso~Asensio, I., Dalla~Vecchia, C., Potter, D., \& Stadel, J. 2023, Monthly Notices of the Royal Astronomical Society, 519, 300.
\newblock \url{https://doi.org/10.1093/mnras/stac3447}

\bibitem[{Asphaug \& Reufer(2014)}]{asphaugMercuryOtherIronrich2014}
Asphaug, E., \& Reufer, A. 2014, Nature Geoscience, 7, 564.
\newblock \url{https://www.nature.com/articles/ngeo2189}

\bibitem[{Ballantyne {et~al.}(2024)Ballantyne, Asphaug, Denton, Emsenhuber, \& Jutzi}]{ballantyneSputnikPlanitiaImpactor2024}
Ballantyne, H.~A., Asphaug, E., Denton, C.~A., Emsenhuber, A., \& Jutzi, M. 2024, Nature Astronomy, 8, 748.
\newblock \url{https://www.nature.com/articles/s41550-024-02248-1}

\bibitem[{Ballantyne {et~al.}(2023)Ballantyne, Jutzi, Golabek, Mishra, Cheng, Rozel, \& Tackley}]{ballantyneInvestigatingFeasibilityImpactinduced2023}
Ballantyne, H.~A., Jutzi, M., Golabek, G.~J., {et~al.} 2023, Icarus, 392, 115395.
\newblock \url{https://www.sciencedirect.com/science/article/pii/S0019103522004870}

\bibitem[{Batygin \& Morbidelli(2023)}]{batyginFormationRockySuperearths2023}
Batygin, K., \& Morbidelli, A. 2023, Nature Astronomy, 7, 330.
\newblock \url{https://www.nature.com/articles/s41550-022-01850-5}

\bibitem[{Becker {et~al.}(2014)Becker, Lorenzen, Fortney, Nettelmann, Sch{\"o}ttler, \& Redmer}]{beckerINITIOEQUATIONSSTATE2014}
Becker, A., Lorenzen, W., Fortney, J.~J., {et~al.} 2014, The Astrophysical Journal Supplement Series, 215, 21.
\newblock \url{https://iopscience.iop.org/article/10.1088/0067-0049/215/2/21}

\bibitem[{Bolton {et~al.}(2017)Bolton, Adriani, Adumitroaie, Allison, Anderson, Atreya, Bloxham, Brown, Connerney, DeJong, Folkner, Gautier, Grassi, Gulkis, Guillot, Hansen, Hubbard, Iess, Ingersoll, Janssen, Jorgensen, Kaspi, Levin, Li, Lunine, Miguel, Mura, Orton, Owen, Ravine, Smith, Steffes, Stone, Stevenson, Thorne, Waite, Durante, Ebert, Greathouse, Hue, Parisi, Szalay, \& Wilson}]{boltonJupitersInteriorDeep2017}
Bolton, S.~J., Adriani, A., Adumitroaie, V., {et~al.} 2017, Science, 356, 821.
\newblock \url{https://www.science.org/doi/10.1126/science.aal2108}

\bibitem[{Canup(2004)}]{canupSimulationsLateLunarforming2004}
Canup, R.~M. 2004, Icarus, 168, 433.
\newblock \url{http://www.sciencedirect.com/science/article/pii/S0019103503002999}

\bibitem[{Chabrier \& Debras(2021)}]{chabrierNewEquationState2021}
Chabrier, G., \& Debras, F. 2021, The Astrophysical Journal, 917, 4.
\newblock \url{https://dx.doi.org/10.3847/1538-4357/abfc48}

\bibitem[{Chabrier {et~al.}(2019)Chabrier, Mazevet, \& Soubiran}]{chabrierNewEquationState2019}
Chabrier, G., Mazevet, S., \& Soubiran, F. 2019, The Astrophysical Journal, 872, 51.
\newblock \url{https://ui.adsabs.harvard.edu/abs/2019ApJ...872...51C}

\bibitem[{Chau {et~al.}(2018)Chau, Reinhardt, Helled, \& Stadel}]{chauFormingMercuryGiant2018}
Chau, A., Reinhardt, C., Helled, R., \& Stadel, J. 2018, The Astrophysical Journal, 865, 35.
\newblock \url{https://dx.doi.org/10.3847/1538-4357/aad8b0}

\bibitem[{Cresswell \& Nelson(2008)}]{cresswellThreedimensionalSimulationsMultiple2008}
Cresswell, P., \& Nelson, R.~P. 2008, Astronomy \& Astrophysics, 482, 677.
\newblock \url{https://www.aanda.org/articles/aa/abs/2008/17/aa9178-07/aa9178-07.html}

\bibitem[{Cumming {et~al.}(2018)Cumming, Helled, \& Venturini}]{cummingPrimordialEntropyJupiter2018}
Cumming, A., Helled, R., \& Venturini, J. 2018, Monthly Notices of the Royal Astronomical Society, 477, 4817.
\newblock \url{https://doi.org/10.1093/mnras/sty1000}

\bibitem[{Debras \& Chabrier(2019)}]{debrasNewModelsJupiter2019}
Debras, F., \& Chabrier, G. 2019, The Astrophysical Journal, 872, 100.
\newblock \url{https://ui.adsabs.harvard.edu/abs/2019ApJ...872..100D}

\bibitem[{Dehnen \& Aly(2012)}]{dehnenImprovingConvergenceSmoothed2012}
Dehnen, W., \& Aly, H. 2012, Monthly Notices of the Royal Astronomical Society, 425, 1068.
\newblock \url{https://doi.org/10.1111/j.1365-2966.2012.21439.x}

\bibitem[{Deng {et~al.}(2019{\natexlab{a}})Deng, Ballmer, Reinhardt, Meier, Mayer, Stadel, \& Benitez}]{dengPrimordialEarthMantle2019}
Deng, H., Ballmer, M.~D., Reinhardt, C., {et~al.} 2019{\natexlab{a}}, The Astrophysical Journal, 887, 211.
\newblock \url{https://doi.org/10.3847%2F1538-4357%2Fab50b9}

\bibitem[{Deng {et~al.}(2019{\natexlab{b}})Deng, Reinhardt, Benitez, Mayer, Stadel, \& Barr}]{dengEnhancedMixingGiant2019}
Deng, H., Reinhardt, C., Benitez, F., {et~al.} 2019{\natexlab{b}}, The Astrophysical Journal, 870, 127.
\newblock \url{https://dx.doi.org/10.3847/1538-4357/aaf399}

\bibitem[{Durante {et~al.}(2020)Durante, Parisi, Serra, Zannoni, Notaro, Racioppa, Buccino, Lari, Gomez~Casajus, Iess, Folkner, Tommei, Tortora, \& Bolton}]{duranteJupiterGravityField2020}
Durante, D., Parisi, M., Serra, D., {et~al.} 2020, Geophysical Research Letters, 47, e2019GL086572.
\newblock \url{https://onlinelibrary.wiley.com/doi/abs/10.1029/2019GL086572}

\bibitem[{Emsenhuber {et~al.}(2018)Emsenhuber, Jutzi, \& Benz}]{emsenhuberSPHCalculationsMarsscale2018}
Emsenhuber, A., Jutzi, M., \& Benz, W. 2018, Icarus, 301, 247.
\newblock \url{https://linkinghub.elsevier.com/retrieve/pii/S0019103517302397}

\bibitem[{Folkner {et~al.}(2017)Folkner, Iess, Anderson, Asmar, Buccino, Durante, Feldman, Gomez~Casajus, Gregnanin, Milani, Parisi, Park, Serra, Tommei, Tortora, Zannoni, Bolton, Connerney, \& Levin}]{folknerJupiterGravityField2017}
Folkner, W.~M., Iess, L., Anderson, J.~D., {et~al.} 2017, Geophysical Research Letters, 44, 4694.
\newblock \url{https://onlinelibrary.wiley.com/doi/abs/10.1002/2017GL073140}

\bibitem[{Fuentes(2025)}]{fuentes3DSimulationsSemiconvection2025}
Fuentes, J.~R. 2025, The Astrophysical Journal, 982, 44.
\newblock \url{https://dx.doi.org/10.3847/1538-4357/adb8ec}

\bibitem[{Fuentes {et~al.}(2022)Fuentes, Cumming, \& Anders}]{fuentesLayerFormationStably2022}
Fuentes, J.~R., Cumming, A., \& Anders, E.~H. 2022, Physical Review Fluids, 7, 124501.
\newblock \url{https://link.aps.org/doi/10.1103/PhysRevFluids.7.124501}

\bibitem[{Fuentes {et~al.}(2024)Fuentes, Hindman, Fraser, \& Anders}]{fuentesEvolutionSemiconvectiveStaircases2024}
Fuentes, J.~R., Hindman, B.~W., Fraser, A.~E., \& Anders, E.~H. 2024, The Astrophysical Journal Letters, 975, L1.
\newblock \url{https://dx.doi.org/10.3847/2041-8213/ad84dc}

\bibitem[{Fuller(2014)}]{fullerSaturnRingSeismology2014}
Fuller, J. 2014, Icarus, 242, 283.
\newblock \url{https://www.sciencedirect.com/science/article/pii/S0019103514004205}

\bibitem[{Garaud(2018)}]{garaudDoubleDiffusiveConvectionLow2018}
Garaud, P. 2018, Annual Review of Fluid Mechanics, 50, 275.
\newblock \url{https://www.annualreviews.org/content/journals/10.1146/annurev-fluid-122316-045234}

\bibitem[{Guillot {et~al.}(2004)Guillot, Stevenson, Hubbard, \& Saumon}]{guillotInteriorJupiter2004}
Guillot, T., Stevenson, D.~J., Hubbard, W.~B., \& Saumon, D. 2004, The Interior of {{Jupiter}}, Vol.~1 (Cambridge University Press), 35--57.
\newblock \url{https://ui.adsabs.harvard.edu/abs/2004jpsm.book...35G}

\bibitem[{Harris {et~al.}(2020)Harris, Millman, {van der Walt}, Gommers, Virtanen, Cournapeau, Wieser, Taylor, Berg, Smith, Kern, Picus, Hoyer, {van Kerkwijk}, Brett, Haldane, {del R{\'i}o}, Wiebe, Peterson, {G{\'e}rard-Marchant}, Sheppard, Reddy, Weckesser, Abbasi, Gohlke, \& Oliphant}]{harrisArrayProgrammingNumPy2020}
Harris, C.~R., Millman, K.~J., {van der Walt}, S.~J., {et~al.} 2020, Nature, 585, 357.
\newblock \url{https://www.nature.com/articles/s41586-020-2649-2}

\bibitem[{Helled \& Stevenson(2017)}]{helledFuzzinessGiantPlanets2017}
Helled, R., \& Stevenson, D. 2017, The Astrophysical Journal Letters, 840, L4.
\newblock \url{https://dx.doi.org/10.3847/2041-8213/aa6d08}

\bibitem[{Helled \& Stevenson(2024)}]{helledFuzzyCoresJupiter2024}
Helled, R., \& Stevenson, D.~J. 2024, AGU Advances, 5, e2024AV001171.
\newblock \url{https://onlinelibrary.wiley.com/doi/abs/10.1029/2024AV001171}

\bibitem[{Helled {et~al.}(2022)Helled, Stevenson, Lunine, Bolton, Nettelmann, Atreya, Guillot, Militzer, Miguel, \& Hubbard}]{helledRevelationsJupiterFormation2022}
Helled, R., Stevenson, D.~J., Lunine, J.~I., {et~al.} 2022, Icarus, 378, 114937.
\newblock \url{https://www.sciencedirect.com/science/article/pii/S0019103522000586}

\bibitem[{Hopkins(2015)}]{hopkinsNewClassAccurate2015}
Hopkins, P.~F. 2015, Monthly Notices of the Royal Astronomical Society, 450, 53.
\newblock \url{http://academic.oup.com/mnras/article/450/1/53/992679/A-new-class-of-accurate-meshfree-hydrodynamic}

\bibitem[{Howard {et~al.}(2023)Howard, Guillot, Bazot, Miguel, Stevenson, Galanti, Kaspi, Hubbard, Militzer, Helled, Nettelmann, Idini, \& Bolton}]{howardJupitersInteriorJuno2023}
Howard, S., Guillot, T., Bazot, M., {et~al.} 2023, Astronomy \& Astrophysics, 672, A33.
\newblock \url{https://www.aanda.org/articles/aa/abs/2023/04/aa45625-22/aa45625-22.html}

\bibitem[{Hunter(2007)}]{hunterMatplotlib2DGraphics2007}
Hunter, J.~D. 2007, Computing in Science Engineering, 9, 90.
\newblock \url{https://ieeexplore.ieee.org/document/4160265}

\bibitem[{Iess {et~al.}(2018)Iess, Folkner, Durante, Parisi, Kaspi, Galanti, Guillot, Hubbard, Stevenson, Anderson, Buccino, Casajus, Milani, Park, Racioppa, Serra, Tortora, Zannoni, Cao, Helled, Lunine, Miguel, Militzer, Wahl, Connerney, Levin, \& Bolton}]{iessMeasurementJupiterAsymmetric2018}
Iess, L., Folkner, W.~M., Durante, D., {et~al.} 2018, Nature, 555, 220.
\newblock \url{https://www.nature.com/articles/nature25776}

\bibitem[{{ImageMagick Studio LLC}(2024)}]{imagemagickstudiollcImageMagick2024}
{ImageMagick Studio LLC}. 2024, ImageMagick.
\newblock \url{https://imagemagick.org}

\bibitem[{Jermyn {et~al.}(2023)Jermyn, Bauer, Schwab, Farmer, Ball, Bellinger, Dotter, Joyce, Marchant, Mombarg, Wolf, Sunny~Wong, Cinquegrana, Farrell, Smolec, Thoul, Cantiello, Herwig, Toloza, Bildsten, Townsend, \& Timmes}]{jermynModulesExperimentsStellar2023}
Jermyn, A.~S., Bauer, E.~B., Schwab, J., {et~al.} 2023, The Astrophysical Journal Supplement Series, 265, 15.
\newblock \url{https://dx.doi.org/10.3847/1538-4365/acae8d}

\bibitem[{Kanagawa {et~al.}(2018)Kanagawa, Tanaka, \& Szuszkiewicz}]{kanagawaRadialMigrationGapopening2018}
Kanagawa, K.~D., Tanaka, H., \& Szuszkiewicz, E. 2018, The Astrophysical Journal, 861, 140.
\newblock \url{https://dx.doi.org/10.3847/1538-4357/aac8d9}

\bibitem[{Kegerreis {et~al.}(2022)Kegerreis, {Ruiz-Bonilla}, Eke, Massey, Sandnes, \& Teodoro}]{kegerreisImmediateOriginMoon2022}
Kegerreis, J.~A., {Ruiz-Bonilla}, S., Eke, V.~R., {et~al.} 2022, The Astrophysical Journal Letters, 937, L40.
\newblock \url{https://dx.doi.org/10.3847/2041-8213/ac8d96}

\bibitem[{Kippenhahn {et~al.}(2013)Kippenhahn, Weigert, \& Weiss}]{kippenhahnStellarStructureEvolution2013}
Kippenhahn, R., Weigert, A., \& Weiss, A. 2013, Stellar {{Structure}} and {{Evolution}} (Springer-Verlag Berlin Heidelberg), doi:10.1007/978-3-642-30304-3.
\newblock \url{https://ui.adsabs.harvard.edu/abs/2013sse..book.....K}

\bibitem[{Knierim \& Helled(2024)}]{knierimConvectiveMixingGas2024}
Knierim, H., \& Helled, R. 2024, The Astrophysical Journal, 977, 227.
\newblock \url{https://dx.doi.org/10.3847/1538-4357/ad8dd0}

\bibitem[{Leconte \& Chabrier(2012)}]{leconteNewVisionGiant2012}
Leconte, J., \& Chabrier, G. 2012, Astronomy \& Astrophysics, 540, A20.
\newblock \url{https://www.aanda.org/articles/aa/abs/2012/04/aa17595-11/aa17595-11.html}

\bibitem[{Ledoux(1947)}]{ledouxStellarModelsConvection1947}
Ledoux, P. 1947, The Astrophysical Journal, 105, 305.
\newblock \url{https://ui.adsabs.harvard.edu/abs/1947ApJ...105..305L}

\bibitem[{Li {et~al.}(2010)Li, Agnor, \& Lin}]{liEMBRYOIMPACTSGAS2010}
Li, S.~L., Agnor, C., \& Lin, D. N.~C. 2010, The Astrophysical Journal, 720, 1161.
\newblock \url{https://dx.doi.org/10.1088/0004-637X/720/2/1161}

\bibitem[{Liu {et~al.}(2019)Liu, Hori, M{\"u}ller, Zheng, Helled, Lin, \& Isella}]{liuFormationJupitersDiluted2019}
Liu, S.-F., Hori, Y., M{\"u}ller, S., {et~al.} 2019, Nature, 572, 355.
\newblock \url{https://www.nature.com/articles/s41586-019-1470-2}

\bibitem[{Lodders(2010)}]{loddersSolarSystemAbundances2010}
Lodders, K. 2010, in Principles and {{Perspectives}} in {{Cosmochemistry}}, ed. A.~Goswami \& B.~E. Reddy (Berlin, Heidelberg: Springer), 379--417

\bibitem[{Lucy(1977)}]{lucyNumericalApproachTesting1977}
Lucy, L.~B. 1977, The Astronomical Journal, 82, 1013.
\newblock \url{http://adsabs.harvard.edu/abs/1977AJ.....82.1013L}

\bibitem[{Mankovich \& Fuller(2021)}]{mankovichDiffuseCoreSaturn2021}
Mankovich, C.~R., \& Fuller, J. 2021, Nature Astronomy, 5, 1103.
\newblock \url{https://www.nature.com/articles/s41550-021-01448-3}

\bibitem[{Matzkevich {et~al.}(2024)Matzkevich, Reinhardt, Meier, Stadel, \& Helled}]{matzkevichOutcomeCollisionsGaseous2024}
Matzkevich, Y., Reinhardt, C., Meier, T., Stadel, J., \& Helled, R. 2024, Astronomy \& Astrophysics, 691, A184.
\newblock \url{https://www.aanda.org/articles/aa/abs/2024/11/aa50900-24/aa50900-24.html}

\bibitem[{Meier \& Reinhardt(2021{\natexlab{a}})}]{meierEOSlib2021}
Meier, T., \& Reinhardt, C. 2021{\natexlab{a}}, {{EOSlib}},  Zenodo, doi:10.5281/zenodo.4662637.
\newblock \url{https://zenodo.org/record/4662637}

\bibitem[{Meier \& Reinhardt(2021{\natexlab{b}})}]{meierANEOSmaterial2021}
---. 2021{\natexlab{b}}, {{ANEOSmaterial}},  Zenodo, doi:10.5281/zenodo.4662606.
\newblock \url{https://zenodo.org/record/4662606}

\bibitem[{Meier {et~al.}(2021)Meier, Reinhardt, \& Stadel}]{meierEOSResolutionConspiracy2021}
Meier, T., Reinhardt, C., \& Stadel, J.~G. 2021, Monthly Notices of the Royal Astronomical Society, 505, 1806.
\newblock \url{https://academic.oup.com/mnras/article/505/2/1806/6279686}

\bibitem[{Meier {et~al.}(2024)Meier, Reinhardt, Timpe, Stadel, \& Moore}]{meierSystematicSurveyMoonforming2024}
Meier, T., Reinhardt, C., Timpe, M., Stadel, J., \& Moore, B. 2024, The Astrophysical Journal, 978, 11.
\newblock \url{https://dx.doi.org/10.3847/1538-4357/ad9248}

\bibitem[{Meier {et~al.}(2025)Meier, Shibata, \& M{\"u}ller}]{meierJupiter_Fuzzy_Meier2025}
Meier, T., Shibata, S., \& M{\"u}ller, S. 2025, Jupiter\_{{Fuzzy}}\_{{Meier}},  Zenodo, doi:10.5281/zenodo.15356900.
\newblock \url{https://zenodo.org/records/15356900}

\bibitem[{Melosh(2007)}]{meloshHydrocodeEquationState2007}
Melosh, H.~J. 2007, Meteoritics \& Planetary Science, 42, 2079.
\newblock \url{http://doi.wiley.com/10.1111/j.1945-5100.2007.tb01009.x}

\bibitem[{Miguel {et~al.}(2022)Miguel, Bazot, Guillot, Howard, Galanti, Kaspi, Hubbard, Militzer, Helled, Atreya, Connerney, Durante, Kulowski, Lunine, Stevenson, \& Bolton}]{miguelJupiterInhomogeneousEnvelope2022}
Miguel, Y., Bazot, M., Guillot, T., {et~al.} 2022, Astronomy \& Astrophysics, 662, A18.
\newblock \url{https://www.aanda.org/articles/aa/abs/2022/06/aa43207-22/aa43207-22.html}

\bibitem[{Militzer \& Hubbard(2024)}]{militzerStudyJupitersInterior2024}
Militzer, B., \& Hubbard, W.~B. 2024, Icarus, 411, 115955.
\newblock \url{https://www.sciencedirect.com/science/article/pii/S0019103524000137}

\bibitem[{Militzer {et~al.}(2022)Militzer, Hubbard, Wahl, Lunine, Galanti, Kaspi, Miguel, Guillot, Moore, Parisi, Connerney, Helled, Cao, Mankovich, Stevenson, Park, Wong, Atreya, Anderson, \& Bolton}]{militzerJunoSpacecraftMeasurements2022}
Militzer, B., Hubbard, W.~B., Wahl, S., {et~al.} 2022, The Planetary Science Journal, 3, 185.
\newblock \url{https://ui.adsabs.harvard.edu/abs/2022PSJ.....3..185M}

\bibitem[{Moll {et~al.}(2017)Moll, Garaud, Mankovich, \& Fortney}]{mollDoublediffusiveErosionCore2017}
Moll, R., Garaud, P., Mankovich, C., \& Fortney, J.~J. 2017, The Astrophysical Journal, 849, 24.
\newblock \url{https://dx.doi.org/10.3847/1538-4357/aa8d74}

\bibitem[{Monaghan(1992)}]{monaghanSmoothedParticleHydrodynamics1992}
Monaghan, J.~J. 1992, Annual Review of Astronomy and Astrophysics, 30, 543.
\newblock \url{http://adsabs.harvard.edu/abs/1992ARA%26A..30..543M}

\bibitem[{Moran(1950)}]{moranNotesContinuousStochastic1950}
Moran, P. A.~P. 1950, Biometrika, 37, 17.
\newblock \url{https://www.jstor.org/stable/2332142}

\bibitem[{Mordasini(2020)}]{mordasiniPlanetaryEvolutionAtmospheric2020}
Mordasini, C. 2020, Astronomy and Astrophysics, 638, A52.
\newblock \url{https://ui.adsabs.harvard.edu/abs/2020A&A...638A..52M}

\bibitem[{More {et~al.}(1988)More, Warren, Young, \& Zimmerman}]{moreNewQuotidianEquation1988}
More, R.~M., Warren, K.~H., Young, D.~A., \& Zimmerman, G.~B. 1988, The Physics of Fluids, 31, 3059.
\newblock \url{https://aip.scitation.org/doi/10.1063/1.866963}

\bibitem[{M{\"u}ller {et~al.}(2020{\natexlab{a}})M{\"u}ller, {Ben-Yami}, \& Helled}]{mullerTheoreticalObservationalUncertainties2020}
M{\"u}ller, S., {Ben-Yami}, M., \& Helled, R. 2020{\natexlab{a}}, The Astrophysical Journal, 903, 147.
\newblock \url{https://dx.doi.org/10.3847/1538-4357/abba19}

\bibitem[{M{\"u}ller \& Helled(2024)}]{mullerCanJupitersAtmospheric2024}
M{\"u}ller, S., \& Helled, R. 2024, The Astrophysical Journal, 967, 7.
\newblock \url{https://dx.doi.org/10.3847/1538-4357/ad3738}

\bibitem[{M{\"u}ller {et~al.}(2020{\natexlab{b}})M{\"u}ller, Helled, \& Cumming}]{mullerChallengeFormingFuzzy2020}
M{\"u}ller, S., Helled, R., \& Cumming, A. 2020{\natexlab{b}}, Astronomy \& Astrophysics, 638, A121.
\newblock \url{https://www.aanda.org/articles/aa/abs/2020/06/aa37376-19/aa37376-19.html}

\bibitem[{{N-Body Shop}(2011)}]{n-bodyshopTIPSYCodeDisplay2011}
{N-Body Shop}. 2011, Astrophysics Source Code Library, ascl:1111.015.
\newblock \url{https://ui.adsabs.harvard.edu/abs/2011ascl.soft11015N}

\bibitem[{Nettelmann {et~al.}(2021)Nettelmann, Movshovitz, Ni, Fortney, Galanti, Kaspi, Helled, Mankovich, \& Bolton}]{nettelmannTheoryFiguresSeventh2021}
Nettelmann, N., Movshovitz, N., Ni, D., {et~al.} 2021, The Planetary Science Journal, 2, 241.
\newblock \url{https://iopscience.iop.org/article/10.3847/PSJ/ac390a/meta}

\bibitem[{Paxton {et~al.}(2010)Paxton, Bildsten, Dotter, Herwig, Lesaffre, \& Timmes}]{paxtonMODULESEXPERIMENTSStelLAR2010}
Paxton, B., Bildsten, L., Dotter, A., {et~al.} 2010, The Astrophysical Journal Supplement Series, 192, 3.
\newblock \url{https://dx.doi.org/10.1088/0067-0049/192/1/3}

\bibitem[{Paxton {et~al.}(2013)Paxton, Cantiello, Arras, Bildsten, Brown, Dotter, Mankovich, Montgomery, Stello, Timmes, \& Townsend}]{paxtonMODULESEXPERIMENTSStelLAR2013}
Paxton, B., Cantiello, M., Arras, P., {et~al.} 2013, The Astrophysical Journal Supplement Series, 208, 4.
\newblock \url{https://dx.doi.org/10.1088/0067-0049/208/1/4}

\bibitem[{Paxton {et~al.}(2015)Paxton, Marchant, Schwab, Bauer, Bildsten, Cantiello, Dessart, Farmer, Hu, Langer, Townsend, Townsley, \& Timmes}]{paxtonMODULESEXPERIMENTSStelLAR2015}
Paxton, B., Marchant, P., Schwab, J., {et~al.} 2015, The Astrophysical Journal Supplement Series, 220, 15.
\newblock \url{https://dx.doi.org/10.1088/0067-0049/220/1/15}

\bibitem[{Paxton {et~al.}(2018)Paxton, Schwab, Bauer, Bildsten, Blinnikov, Duffell, Farmer, Goldberg, Marchant, Sorokina, Thoul, Townsend, \& Timmes}]{paxtonModulesExperimentsStellar2018}
Paxton, B., Schwab, J., Bauer, E.~B., {et~al.} 2018, The Astrophysical Journal Supplement Series, 234, 34.
\newblock \url{https://dx.doi.org/10.3847/1538-4365/aaa5a8}

\bibitem[{Paxton {et~al.}(2019)Paxton, Smolec, Schwab, Gautschy, Bildsten, Cantiello, Dotter, Farmer, Goldberg, Jermyn, Kanbur, Marchant, Thoul, Townsend, Wolf, Zhang, \& Timmes}]{paxtonModulesExperimentsStellar2019}
Paxton, B., Smolec, R., Schwab, J., {et~al.} 2019, The Astrophysical Journal Supplement Series, 243, 10.
\newblock \url{https://dx.doi.org/10.3847/1538-4365/ab2241}

\bibitem[{Potter {et~al.}(2017)Potter, Stadel, \& Teyssier}]{potterPKDGRAV3TrillionParticle2017}
Potter, D., Stadel, J., \& Teyssier, R. 2017, Computational Astrophysics and Cosmology, 4, 2.
\newblock \url{https://comp-astrophys-cosmol.springeropen.com/articles/10.1186/s40668-017-0021-1}

\bibitem[{Price(2012)}]{priceSmoothedParticleHydrodynamics2012}
Price, D.~J. 2012, Journal of Computational Physics, 231, 759.
\newblock \url{https://www.sciencedirect.com/science/article/pii/S0021999110006753}

\bibitem[{Radko {et~al.}(2014)Radko, Bulters, Flanagan, \& Campin}]{radkoDoubleDiffusiveRecipesPart2014}
Radko, T., Bulters, A., Flanagan, J.~D., \& Campin, J.-M. 2014, Journal of Physical Oceanography, doi:10.1175/JPO-D-13-0155.1.
\newblock \url{https://journals.ametsoc.org/view/journals/phoc/44/5/jpo-d-13-0155.1.xml}

\bibitem[{Rein \& Spiegel(2015)}]{reinIAS15FastAdaptive2015}
Rein, H., \& Spiegel, D.~S. 2015, Monthly Notices of the Royal Astronomical Society, 446, 1424.
\newblock \url{https://ui.adsabs.harvard.edu/abs/2015MNRAS.446.1424R}

\bibitem[{Reinhardt {et~al.}(2020)Reinhardt, Chau, Stadel, \& Helled}]{reinhardtBifurcationHistoryUranus2020}
Reinhardt, C., Chau, A., Stadel, J., \& Helled, R. 2020, Monthly Notices of the Royal Astronomical Society, 492, 5336.
\newblock \url{https://academic.oup.com/mnras/article/492/4/5336/5637902}

\bibitem[{Reinhardt {et~al.}(2022)Reinhardt, Meier, Stadel, Otegi, \& Helled}]{reinhardtFormingIronrichPlanets2022}
Reinhardt, C., Meier, T., Stadel, J.~G., Otegi, J.~F., \& Helled, R. 2022, Monthly Notices of the Royal Astronomical Society, 517, 3132.
\newblock \url{https://doi.org/10.1093/mnras/stac1853}

\bibitem[{Reinhardt \& Stadel(2017)}]{reinhardtNumericalAspectsGiant2017}
Reinhardt, C., \& Stadel, J. 2017, Monthly Notices of the Royal Astronomical Society, 467, doi:10.1093/mnras/stx322.
\newblock \url{http://adsabs.harvard.edu/abs/2017MNRAS.467.4252R}

\bibitem[{Robertson {et~al.}(2010)Robertson, Kravtsov, Gnedin, Abel, \& Rudd}]{robertsonComputationalEulerianHydrodynamics2010}
Robertson, B.~E., Kravtsov, A.~V., Gnedin, N.~Y., Abel, T., \& Rudd, D.~H. 2010, Monthly Notices of the Royal Astronomical Society, 401, 2463.
\newblock \url{http://www.scopus.com/inward/record.url?scp=74549179771&partnerID=8YFLogxK}

\bibitem[{{Ruiz-Bonilla} {et~al.}(2022){Ruiz-Bonilla}, Borrow, Eke, Kegerreis, Massey, Sandnes, \& Teodoro}]{ruiz-bonillaDealingDensityDiscontinuities2022}
{Ruiz-Bonilla}, S., Borrow, J., Eke, V.~R., {et~al.} 2022, Monthly Notices of the Royal Astronomical Society, 512, 4660.
\newblock \url{https://doi.org/10.1093/mnras/stac857}

\bibitem[{Sandnes {et~al.}(2025)Sandnes, Eke, Kegerreis, Massey, {Ruiz-Bonilla}, Schaller, \& Teodoro}]{sandnesREMIXSPHImproving2025}
Sandnes, T.~D., Eke, V.~R., Kegerreis, J.~A., {et~al.} 2025, Journal of Computational Physics, 113907.
\newblock \url{https://www.sciencedirect.com/science/article/pii/S0021999125001901}

\bibitem[{Sandnes {et~al.}(2024)Sandnes, Eke, Kegerreis, Massey, \& Teodoro}]{sandnesNoDiluteCore2024}
Sandnes, T.~D., Eke, V.~R., Kegerreis, J.~A., Massey, R.~J., \& Teodoro, L. F.~A. 2024, No Dilute Core Produced in Simulations of Giant Impacts onto {{Jupiter}},  arXiv, arXiv:2412.06094.
\newblock \url{http://arxiv.org/abs/2412.06094}

\bibitem[{Saumon {et~al.}(1995)Saumon, Chabrier, \& {van Horn}}]{saumonEquationStateLowMass1995}
Saumon, D., Chabrier, G., \& {van Horn}, H.~M. 1995, The Astrophysical Journal Supplement Series, 99, 713.
\newblock \url{http://adsabs.harvard.edu/doi/10.1086/192204}

\bibitem[{Shibata \& Helled(2024)}]{shibataFateRemnantSolid2024}
Shibata, S., \& Helled, R. 2024, Astronomy \& Astrophysics, 689, A26.
\newblock \url{https://www.aanda.org/articles/aa/abs/2024/09/aa49897-24/aa49897-24.html}

\bibitem[{Shibata {et~al.}(2023)Shibata, Helled, \& Kobayashi}]{shibataHeavyelementAccretionProtoJupiter2023}
Shibata, S., Helled, R., \& Kobayashi, H. 2023, Monthly Notices of the Royal Astronomical Society, 519, 1713.
\newblock \url{https://doi.org/10.1093/mnras/stac3568}

\bibitem[{Shuai {et~al.}(2024)Shuai, Sch{\"a}fer, Burger, \& Hui}]{shuaiMetalsilicateMixingPlanetesimal2024}
Shuai, K., Sch{\"a}fer, C.~M., Burger, C., \& Hui, H. 2024, Astronomy \& Astrophysics, 687, A194.
\newblock \url{https://www.aanda.org/articles/aa/abs/2024/07/aa47781-23/aa47781-23.html}

\bibitem[{Springel(2010{\natexlab{a}})}]{springelPurSiMuove2010}
Springel, V. 2010{\natexlab{a}}, Monthly Notices of the Royal Astronomical Society, 401, 791.
\newblock \url{https://doi.org/10.1111/j.1365-2966.2009.15715.x}

\bibitem[{Springel(2010{\natexlab{b}})}]{springelSmoothedParticleHydrodynamics2010}
---. 2010{\natexlab{b}}, Annual Review of Astronomy and Astrophysics, 48, 391.
\newblock \url{https://www.annualreviews.org/doi/10.1146/annurev-astro-081309-130914}

\bibitem[{Springel \& Hernquist(2002)}]{springelCosmologicalSmoothedParticle2002}
Springel, V., \& Hernquist, L. 2002, Monthly Notices of the Royal Astronomical Society, 333, 649.
\newblock \url{https://academic.oup.com/mnras/article/333/3/649/1002394}

\bibitem[{Stevenson {et~al.}(2022)Stevenson, Bodenheimer, Lissauer, \& D'Angelo}]{stevensonMixingCondensableConstituents2022}
Stevenson, D.~J., Bodenheimer, P., Lissauer, J.~J., \& D'Angelo, G. 2022, The Planetary Science Journal, 3, 74.
\newblock \url{https://iopscience.iop.org/article/10.3847/PSJ/ac5c44/meta}

\bibitem[{Stewart(2020)}]{stewartEquationStateModel2020a}
Stewart, S.~T. 2020, Equation of {{State Model Iron ANEOS}}: {{Documentation}} and {{Comparisons}} ({{Version SLVTv0}}.{{2G1}}),  Zenodo, doi:10.5281/zenodo.3866507.
\newblock \url{https://zenodo.org/record/3866507}

\bibitem[{Stewart {et~al.}(2019)Stewart, Davies, Duncan, Lock, Root, Townsend, Kraus, Caracas, \& Jacobsen}]{stewartEquationStateModel2019}
Stewart, S.~T., Davies, E.~J., Duncan, M.~S., {et~al.} 2019, Equation of {{State Model Forsterite-ANEOS-SLVTv1}}.{{0G1}}: {{Documentation}} and {{Comparisons}},  Zenodo, doi:10.5281/zenodo.3478631.
\newblock \url{https://zenodo.org/record/3478631}

\bibitem[{Sur {et~al.}(2024)Sur, Su, Tejada~Arevalo, Chen, \& Burrows}]{surAPPLEEvolutionCode2024}
Sur, A., Su, Y., Tejada~Arevalo, R., Chen, Y.-X., \& Burrows, A. 2024, The Astrophysical Journal, 971, 104.
\newblock \url{https://dx.doi.org/10.3847/1538-4357/ad57c3}

\bibitem[{Sur {et~al.}(2025)Sur, Tejada~Arevalo, Su, \& Burrows}]{surSimultaneousEvolutionaryFits2025}
Sur, A., Tejada~Arevalo, R., Su, Y., \& Burrows, A. 2025, The Astrophysical Journal Letters, 980, L5.
\newblock \url{https://dx.doi.org/10.3847/2041-8213/adad62}

\bibitem[{Tange(2011)}]{tangeGNUParallelCommandline2011}
Tange, O. 2011, ;login: The USENIX Magazine, 36, 42.
\newblock \url{http://www.gnu.org/s/parallel}

\bibitem[{Tejada~Arevalo {et~al.}(2025)Tejada~Arevalo, Sur, Su, \& Burrows}]{tejadaarevaloJupiterEvolutionaryModels2025}
Tejada~Arevalo, R., Sur, A., Su, Y., \& Burrows, A. 2025, The Astrophysical Journal, 979, 243.
\newblock \url{https://dx.doi.org/10.3847/1538-4357/ada030}

\bibitem[{Thompson \& Lauson(1974)}]{thompsonImprovementsCHARTRadiationhydrodynamic1974}
Thompson, S.~L., \& Lauson, H.~S. 1974, Improvements in the {{CHART D}} Radiation-Hydrodynamic Code {{III}}: Revised Analytic Equations of State, Tech. Rep. SC-RR--71-0714, Sandia Labs.
\newblock \url{http://inis.iaea.org/Search/search.aspx?orig_q=RN:6209386}

\bibitem[{Thompson {et~al.}(2019)Thompson, Lauson, Melosh, Collins, \& Stewart}]{thompsonMANEOS2019}
Thompson, S.~L., Lauson, H.~S., Melosh, H.~J., Collins, G.~S., \& Stewart, S.~T. 2019, M-{{ANEOS}},  Zenodo, doi:10.5281/zenodo.3525030.
\newblock \url{https://zenodo.org/record/3525030#.XvxXHygzaUk}

\bibitem[{Timpe {et~al.}(2023)Timpe, Reinhardt, Meier, Stadel, \& Moore}]{timpeSystematicSurveyMoonforming2023}
Timpe, M., Reinhardt, C., Meier, T., Stadel, J., \& Moore, B. 2023, The Astrophysical Journal, 959, 38.
\newblock \url{https://dx.doi.org/10.3847/1538-4357/acfc40}

\bibitem[{Tulekeyev {et~al.}(2024)Tulekeyev, Garaud, Idini, \& Fortney}]{tulekeyevConstraintsLongtermExistence2024}
Tulekeyev, A., Garaud, P., Idini, B., \& Fortney, J.~J. 2024, The Planetary Science Journal, 5, 190.
\newblock \url{https://iopscience.iop.org/article/10.3847/PSJ/ad6571/meta}

\bibitem[{Valletta \& Helled(2020)}]{vallettaGiantPlanetFormation2020}
Valletta, C., \& Helled, R. 2020, The Astrophysical Journal, 900, 133.
\newblock \url{https://dx.doi.org/10.3847/1538-4357/aba904}

\bibitem[{Vazan {et~al.}(2018)Vazan, Helled, \& Guillot}]{vazanJupiterEvolutionPrimordial2018}
Vazan, A., Helled, R., \& Guillot, T. 2018, Astronomy \& Astrophysics, 610, L14.
\newblock \url{https://www.aanda.org/articles/aa/abs/2018/02/aa32522-17/aa32522-17.html}

\bibitem[{Vazan {et~al.}(2015)Vazan, Helled, Kovetz, \& Podolak}]{vazanCONVECTIONMIXINGGIANT2015}
Vazan, A., Helled, R., Kovetz, A., \& Podolak, M. 2015, The Astrophysical Journal, 803, 32.
\newblock \url{https://dx.doi.org/10.1088/0004-637X/803/1/32}

\bibitem[{Vazan {et~al.}(2013)Vazan, Kovetz, Podolak, \& Helled}]{vazanEffectCompositionEvolution2013}
Vazan, A., Kovetz, A., Podolak, M., \& Helled, R. 2013, Monthly Notices of the Royal Astronomical Society, 434, 3283.
\newblock \url{https://doi.org/10.1093/mnras/stt1248}

\bibitem[{Venturini \& Helled(2020)}]{venturiniJupiterHeavyelementEnrichment2020}
Venturini, J., \& Helled, R. 2020, Astronomy \& Astrophysics, 634, A31.
\newblock \url{https://www.aanda.org/articles/aa/abs/2020/02/aa36591-19/aa36591-19.html}

\bibitem[{Virtanen {et~al.}(2020)Virtanen, Gommers, Oliphant, Haberland, Reddy, Cournapeau, Burovski, Peterson, Weckesser, Bright, {van der Walt}, Brett, Wilson, Millman, Mayorov, Nelson, Jones, Kern, Larson, Carey, Polat, Feng, Moore, VanderPlas, Laxalde, Perktold, Cimrman, Henriksen, Quintero, Harris, Archibald, Ribeiro, Pedregosa, \& {van Mulbregt}}]{virtanenSciPy10Fundamental2020}
Virtanen, P., Gommers, R., Oliphant, T.~E., {et~al.} 2020, Nature Methods, 17, 261.
\newblock \url{https://www.nature.com/articles/s41592-019-0686-2}

\bibitem[{Wahl {et~al.}(2017)Wahl, Hubbard, Militzer, Guillot, Miguel, Movshovitz, Kaspi, Helled, Reese, Galanti, Levin, Connerney, \& Bolton}]{wahlComparingJupiterInterior2017}
Wahl, S.~M., Hubbard, W.~B., Militzer, B., {et~al.} 2017, Geophysical Research Letters, 44, 4649.
\newblock \url{https://onlinelibrary.wiley.com/doi/abs/10.1002/2017GL073160}

\bibitem[{Woo {et~al.}(2023)Woo, Morbidelli, Grimm, Stadel, \& Brasser}]{wooTerrestrialPlanetFormation2023}
Woo, J. M.~Y., Morbidelli, A., Grimm, S.~L., Stadel, J., \& Brasser, R. 2023, Icarus, 396, 115497.
\newblock \url{https://www.sciencedirect.com/science/article/pii/S001910352300074X}

\bibitem[{Woo {et~al.}(2022)Woo, Reinhardt, Cilibrasi, Chau, Helled, \& Stadel}]{wooDidUranusRegular2022}
Woo, J. M.~Y., Reinhardt, C., Cilibrasi, M., {et~al.} 2022, Icarus, 375, 114842.
\newblock \url{https://ui.adsabs.harvard.edu/abs/2022Icar..37514842W}

\bibitem[{Wood {et~al.}(2013)Wood, Garaud, \& Stellmach}]{woodNEWMODELMIXING2013}
Wood, T.~S., Garaud, P., \& Stellmach, S. 2013, The Astrophysical Journal, 768, 157.
\newblock \url{https://dx.doi.org/10.1088/0004-637X/768/2/157}

\bibitem[{Woolfson(2007)}]{woolfsonPracticalSPHModels2007}
Woolfson, M.~M. 2007, Monthly Notices of the Royal Astronomical Society, 376, 1173.
\newblock \url{https://academic.oup.com/mnras/article/376/3/1173/1747046}

\bibitem[{Xie {et~al.}(2025)Xie, Howard, \& Mazzola}]{xieAccurateThermodynamicallyConsistent2025}
Xie, H., Howard, S., \& Mazzola, G. 2025, Accurate and Thermodynamically Consistent Hydrogen Equation of State for Planetary Modeling with Flow Matching,  arXiv, arXiv:2501.10594.
\newblock \url{http://arxiv.org/abs/2501.10594}

\end{thebibliography}

\appendix

\section{Mixing metrics}\label{sec:appendix:Mixing_metrics}
There exists no uniquely defined mixing metric, so in this appendix we propose several different expressions that define a mixing metric. We then highlight the properties, advantages and disadvantages of each. In all of this, we assume that there are only two types of materials in the simulation, which we call $A$ and $B$. In the case of the impacts on Jupiter, where three materials are present (rock, ice and hydrogen-helium) we identify both rock and ice particles as $A$, while the hydrogen-helium particles are designated $B$. In some of the simulations with bodies consisting of only heavy elements (see Appendix~\ref{sec:appendix:Mixing_and_de-mixing_in_SPH:mixing} and Table~\ref{tab:Mixing_Runs}) we use a "core" of the same material as the "mantle". We then designate the "core" as material $A$ and the "mantle" as material $B$.

\subsection{Fractional density}\label{sec:appendix:Mixing_metrics:Fractional_Density}
The most straightforward approach is to calculate for each particle the fractional contribution of particles of opposite material to the SPH density estimator

\begin{equation}
\alpha_i = \frac{\sum_j\kappa_{ij}m_jW_{ij}}{\sum_jm_jW_{ij}}\,,
\end{equation}

\noindent where the summation is over all neighbors of particle $i$ ($N_{Neighbor}=400$ in this paper), $W_{ij}$ is the SPH kernel weight and $\kappa_{ij}=1$ when the materials of particle $i$ and $j$ are different and $\kappa_{ij}=0$ otherwise. This is equivalent to $(1-\bar{w}_i)$ defined in Equation 2 of \citet{sandnesNoDiluteCore2024} with the difference that we do not weight the contributions with $m_j/\rho_j$ to get a method independent metric (i.e., it does not depend on how the density is obtained, e.g., smoothed density with interface correction or evolved density with density diffusion). The total mixed mass in a snapshot can then be defined as the sum of all particle masses weighted with this metric:

\begin{equation}
M_{mix,\alpha}=\sum_i \alpha_i m_i\,.
\end{equation}

The maximum value of $M_{mix,\alpha}$ for a given mass $M_A$ of material $A$ (assuming that there are many more particles of material $B$) is reached when each $A$ particle has only neighbors of material $B$ and tends to the value $2M_A$ if the resolution goes to infinity. If one wants the maximum value $2M_A$ to be reached when the particles of material $A$ and $B$ are perfectly mixed in a 50:50 ratio, the value can be slightly modified as follows

\begin{equation}
\beta_i = 1.0 - 2.0 \left\vert\alpha_i-0.5\right\vert\,,
\end{equation}

\noindent with which we can again define the mixed mass in a snapshot as 

\begin{equation}
M_{mix,\beta}=\sum_i \beta_i m_i\,.
\end{equation}

\subsection{Spatial autocorrelation}
One drawback of mixing metrics based on the fractional density is that even fully segregated states (e.g. the initial state where a layer of particles $A$ exists under a layer of particles $B$ with a clear separation surface) results in a significant value for $M_{mix,\alpha/\beta}$ due to the particles at the interface getting a fractional density contribution from the other material. This can be remedied by using Moran's $I$ \citep{moranNotesContinuousStochastic1950} defined as

\begin{equation}
I_i=\frac{N}{W}\frac{\sum_j\sum_k\omega_{jk}(x_j-\bar{x})(x_k-\bar{x})}{\sum_j(x_j-\bar{x})^2}\,,
\end{equation}

\noindent where $W=\sum_j\sum_k\omega_{jk}$, $x_i$ is either $0$ or $1$ depending on the material ($A$ or $B$) and $\bar{x}$ is the average value of all $x_j$ in the kernel. We choose the spatial weights $\omega_{jk}$ such that $\omega_{jk}=1$ if particle $k$ is the closest neighbor to particle $j$ and $\omega_{jk}=0$ in all other cases. The expected value of Moran's $I$ under the null hypothesis of no spatial autocorrelation is

\begin{equation}
E(I)=-\frac{1}{N-1}\,.
\end{equation}

We thus can define a new metric and total mixed mass as

\begin{equation}
\gamma_i = \begin{cases}1&I_i<E(I)\\0&I_i\geq E(I)\end{cases}\quad M_{mix,\gamma}=\sum_i \gamma_i m_i\,,
\end{equation}

\noindent which results in values several orders of magnitude smaller than $M_{mix,\alpha/\beta}$ in segregated states.

Using only Moran's $I$ will severely overestimate the mixed mass, as even small deviations from correlated states can cause $I$ to be below $E(I)$. Thus, one can combine Moran's $I$ with a fractional density based metric by defining

\begin{equation}
\delta_{i,\beta_0} = \begin{cases}1&I_i<E(I) \mbox{ and } \beta_i > \beta_0\\0&\mbox{otherwise}\end{cases}\,,
\end{equation}

\noindent where we only accept a particle as mixed if both $I_i<E(I)$ and its fractional density metric is in a certain range ($\beta_0/2 < \alpha_i < 1-\beta_0/2$) where $\beta_0$ is a free parameter \citep{shuaiMetalsilicateMixingPlanetesimal2024}. In this case, the total mixed mass is then defined as

\begin{equation}
M_{mix,\delta,\beta_0}=\sum_i \delta_{i,\beta_0} m_i\,.
\end{equation}

A different way to combine the two metrics would be to modify $M_{mix,\beta}$ such that only the particles are considered that have $I_i<E(I)$:

\begin{equation}
M_{mix,\beta\gamma}=\sum_i \beta_i\gamma_i m_i\,.
\end{equation}

\subsection{Interface area estimation}
A third approach to find a quantification of mixing in SPH is to estimate the contact area between particles of different materials. To estimate the contact area, we calculate the Meshless finite-mass (MFM) area estimate (for a derivation see \citet{hopkinsNewClassAccurate2015} or \citet{alonsoasensioMeshfreeHydrodynamicsPkdgrav32023}) for each particle pair. The area estimate is given as

\begin{equation}
\mathbf{A}_{ij}^\alpha = V_i\tilde{\psi}_j^\alpha(\mathbf{x}_i) - V_j\tilde{\psi}_i^\alpha(\mathbf{x}_j)\,,
\end{equation}

\noindent where $V_i$ (the effective volume of particle $i$) and $\psi_i$ (the fraction of the volume element associated with particle $i$) are defined as

\begin{align}
\psi_i(\mathbf{x})&=\frac{1}{\omega(\mathbf{x})}W(\mathbf{x}-\mathbf{x}_i, h(\mathbf{x}))\\
\omega(\mathbf{x})&=\sum_jW(\mathbf{x}-\mathbf{x}_i, h(\mathbf{x}))\\
V_i&=\sum_j\psi_i(\mathbf{x}_j)\,.
\end{align}

\noindent $\tilde{\psi}_j^\alpha$ is given as

\begin{equation}
\tilde{\psi}_j^\alpha(\mathbf{x}_i)=\sum_\beta\mathbf{B}_i^{\alpha\beta}(\mathbf{x}_j-\mathbf{x}_i)^\beta\psi_j(\mathbf{x}_i)\,,
\end{equation}

\noindent where the matrix $\mathbf{B}_i=\mathbf{E}_i^{-1}$ is given by inverting this matrix

\begin{equation}
\mathbf{E}_i^{\alpha\beta}=\sum_j(\mathbf{x}_j-\mathbf{x}_i)^\alpha(\mathbf{x}_j-\mathbf{x}_i)^\beta\psi_j(\mathbf{x}_i)\,.
\end{equation}

We then sum up the terms where the two particles differ in material

\begin{equation}
A_i = \sum_j\begin{cases}\left\vert\mathbf{A}_{ij}^\alpha\right\vert & i \mbox{ and } j \mbox{ have different material}\\0& i \mbox{ and } j \mbox{ have same the material}\end{cases}\,.
\end{equation}

By summing this area over all particles, we get the total interface area:

\begin{equation}
A_I = \sum_i A_i\,,
\end{equation}

\noindent but every contribution is counted twice. For states where the true value of the interface area is known (e.g. for the initial conditions), the interface area is overestimated by \SIrange{20}{50}{\percent} after removing the double counting.

We can also define a mass based mixing metric similar to the metrics above, where we sum up the mass of all particles with a nonzero area value:

\begin{equation}
\epsilon_i = \begin{cases}1&A_i> 0.0\\0&\mbox{otherwise}\end{cases}\quad M_{mix,\epsilon}=\sum_i \epsilon_i m_i\,.
\end{equation}

\subsection{Mixing metric benchmark}\label{sec:appendix:Mixing_metric_benchmark}
We compare the different mixing metrics defined above by applying them to a set of benchmark cases at different resolutions. We use the following cases based on a body-centered cubic grid (BCC):

\begin{itemize}
\item Case 1: Fully segregated sphere with a core of material A and a mantle of material B, with a 1:2 mass ratio.
\item Case 2: Sphere of material B, with a core of 1:1 mixed material such that all the cell centers are material A and the corners material B.
\item Case 3: Sphere of material B with a core of 1:1 mixed material but the material labels are randomly assigned.
\item Case 4: Sphere of mixed material, with labels A and B randomly assigned, with a mixing ratio (A:B) of 1:2.
\item Case 5-8: Cases 1-4 again, but instead of a perfect BCC grid, a uniformly distributed random offset of $\{-0.5a,0.5a\}$, where $a$ is the grid spacing, is applied to all particles in all directions before the materials are assigned.
\end{itemize}

\begin{figure*}[ht!]
\centering
\includegraphics[width=\linewidth]{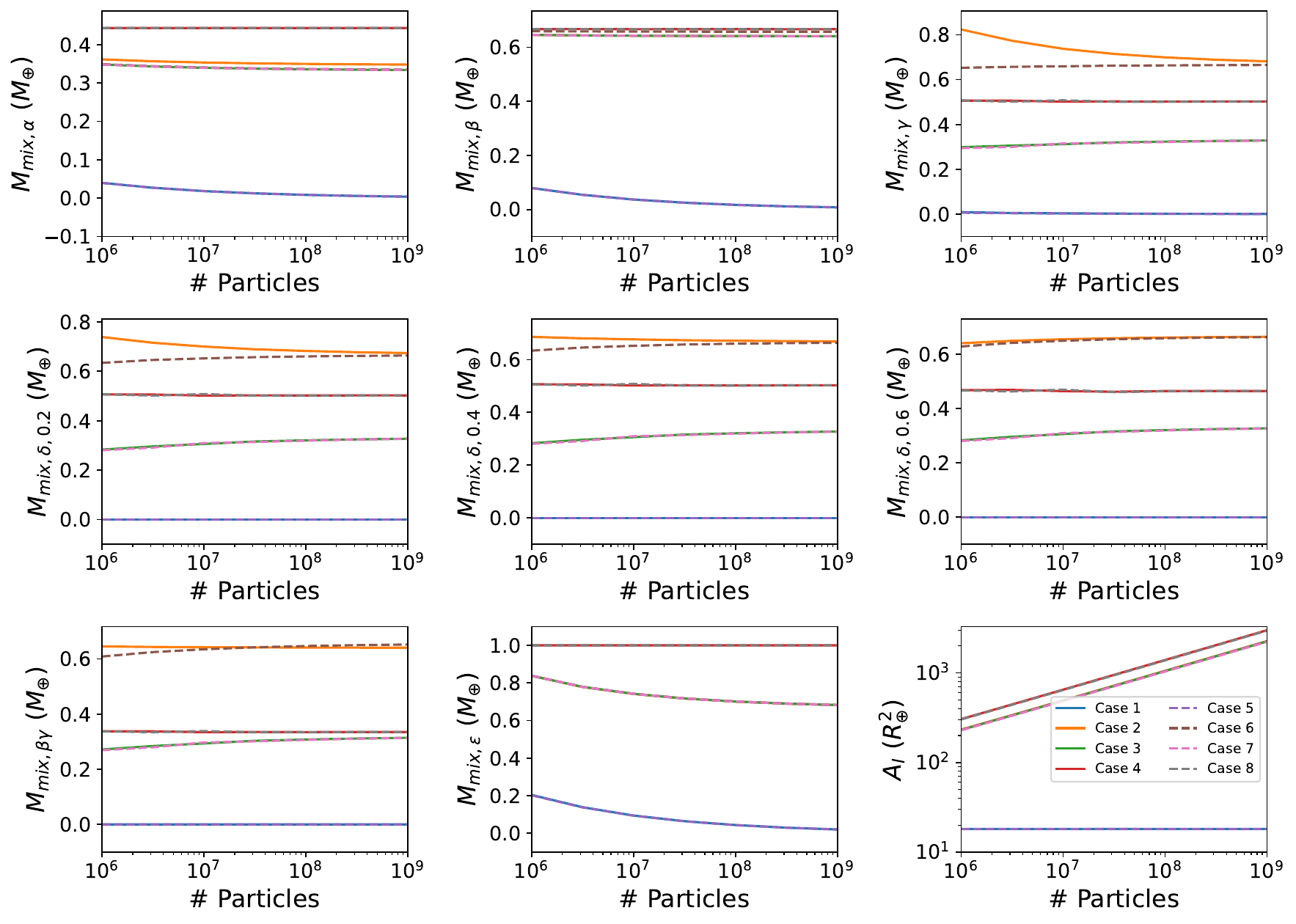}
\caption{Resolution study of the mixing metrics defined in Appendix~\ref{sec:appendix:Mixing_metrics} applied to the 8 test cases defined in Appendix~\ref{sec:appendix:Mixing_metric_benchmark}. We show the results for three different values of $\beta_0\in\{0.2,0.4,0.6\}$ for the metric $M_{mix,\delta,\beta_0}$. The resolution of the test cases ranges from \SI{e6}{} to \SI{e9}{} particles. The results are discussed in Appendix~\ref{sec:appendix:Mixing_metric_benchmark}.}
\label{fig:mixing_resolution_study}
\end{figure*}

Figure~\ref{fig:mixing_resolution_study} shows the results of applying all 7 mixing metrics defined above (we chose three different values of $\beta_0\in\{0.2,0.4,0.6\}$ for the metric $M_{mix,\delta,\beta_0}$) to the 8 test cases defined above at resolutions between \SI{e6}{} and \SI{e9}{} particles. We can make the following observations:

\begin{itemize}
\item $M_{mix,\alpha}$ measures how close the material is to being completely diluted. In the fully segregated cases (1 and 5), the significant, but relatively small value ($\sim\SI{0.05}{\Mearth}$) further decreases with $O(N^{-\frac{1}{3}})$. In the other cases, the value is nearly constant, but we get a significant difference between the value of case 2 and the value of cases 3, 6 and 7, even though they are all close to \SI{0.33}{\Mearth}. As discussed in Section~\ref{sec:appendix:Mixing_metrics:Fractional_Density}, this metric values low relative mass fractions, perfect 1:1 mixing is thus not the maximum value. The cases where the mean mixing ration is 1:2 (cases 4 and 8) have a significantly higher value than the cases with a 1:1 mean ratio in the mixed region (cases 2, 3, 6 and 7).
\item $M_{mix,\beta}$ measures how close the material is to being perfectly mixed. In the fully segregated cases (1 and 5), the value shows the same behavior as $M_{mix,\alpha}$, but it starts with a value that is roughly twice as large. The mixed cases all show very similar values of $\sim\SI{0.65}{\Mearth}$. Significant is that cases 2 and 3 show the same value, whereas for $M_{mix,\alpha}$, they differ slightly. Also, the difference between the cases with 1:2 and 1:1 mean mixing ratio is much smaller.
\item $M_{mix,\gamma}$ measures how close the particles are to perfect spatial anti-correlation. The fully segregated cases 1 and 5 show very small values which also decrease with $O(N^{-\frac{1}{3}})$. Case 2 shows a significant dependence on the resolution, because a layer of roughly one kernel radius beyond the boundary of the core is recognized as mixed. This is not the case if the particle positions are noisy. In random arrangements, many particles are not recognized as mixed, reducing the value by a factor of $\sim 0.5$, compared to the cases where the particles are arranged such that the materials are well distributed (cases 3 and 7 vs 2 and 6). In these cases there is no difference between the perfect alignment and the noisy cases.
\item $M_{mix,\delta,\beta_0}$ contains a free parameter $\beta_0$ that has to be chosen carefully. The values of this metric follow closely the values of $M_{mix,\gamma}$, as $\beta>0.6$ for almost all particles that have $\gamma=1$, but having a large $\beta_0$ suppresses the overestimation in the case 2 at low resolutions. In the perfectly segregated cases this metric results in perfect zero values. The value that this metric reports depends more strongly on $\beta_0$ if the particles are less well mixed than in these benchmark cases.
\item $M_{mix,\beta\gamma}$ measures how close the material is to being perfectly mixed while also being distributed somewhat mixed. This metric results in $\sim\SI{0.65}{\Mearth}$ for the cases with 1:1 mixing ratio in which the materials are well distributed (cases 2 and 6) and $\sim\SI{0.33}{\Mearth}$ for those with random distribution. In the perfectly segregated cases this metric results in perfect zero values.
\item $M_{mix,\epsilon}$ measures, without a self-evident threshold value for the area per particle, the sum of the masses of those particles that have at least one particle of the other material in its kernel. This is represented in the resulting value, as for the cases 4 and 8, all particles are recognized as "mixed", whereas in the other mixed cases, all particles inside the mixed region as well as a layer of one kernel radius around it is counted towards the mixed mass. In the fully segregated cases (1 and 5), a layer of 2 kernel radii is counted as mixed. The kernel radius scales with $O(N^{-\frac{1}{3}})$ which is thus also the scaling of this metric.
\item $A_I$ measures the contact area between materials, which may be of interest in the context of chemical processes. In the fully segregated cases (1 and 5), this metric results in $\sim3$ times the correct value for the interface area (\SI{50}{\percent} overestimation after considering the double counting). In the mixed cases, the area shows a strong dependency on the resolution (increases with $O(N^\frac{1}{3})$), as expected, with cases 4 and 8 having a slightly higher value than cases 2, 3, 6 and 7. There is no significant difference between the perfectly aligned cases and those with noise. Single particles in very dilute regions can have disproportionate contributions.
\end{itemize}

Given these properties, we choose $M_{mix,\beta}$ $M_{mix,\delta,0.2}$ and $M_{mix,\beta\gamma}$ for the analysis of the SPH simulations in the main part of the paper.

\section{Modelling mixing in SPH}\label{sec:appendix:Modelling_mixing_in_SPH}
Here we investigate whether the material separation observed in the impact simulations in Section~\ref{sec:Impact_Simulations} is due to the nature of SPH or the specific setup of the cases we considered. We therefore ran a set of collision simulations with different bodies containing iron, rock, and ice (see Tables~\ref{tab:Mixing_Runs_Pure} and~\ref{tab:Mixing_Runs}) at various resolutions. We included additional cases without impacts to investigate the behavior of bodies with different structures and compositions. We considered an initially mixed core for both Jupiter-like and terrestrial bodies 
(see Table~\ref{tab:demixing_parameters}). 

\subsection{Mixing in heavy-element bodies}\label{sec:appendix:Mixing_and_de-mixing_in_SPH:mixing}
\begin{figure*}[ht!]
\centering
\includegraphics[width=\linewidth]{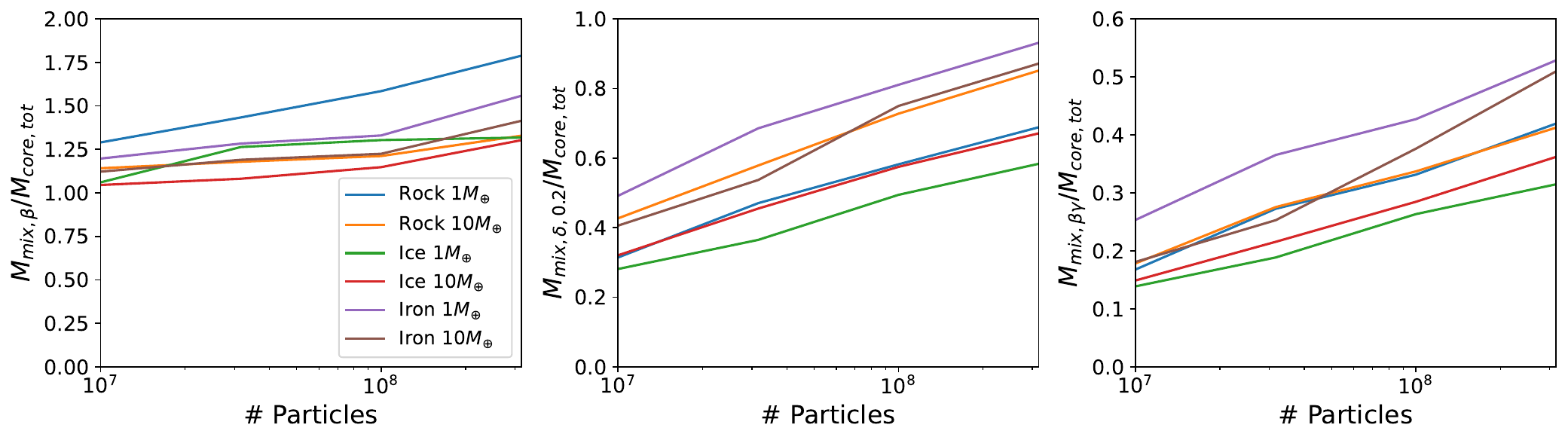}
\caption{Resolution study of the mixing metrics as defined in Appendix~\ref{sec:appendix:Mixing_metrics} at the end (\SI{35.4}{\hour} after the impact) for the simulations with heavy-element bodies (see Table~\ref{tab:Mixing_Runs_Pure}). As there is only one material in each simulation, we mark the central third of both the target and impactor as "core" and the remaining material as "mantle". The mixing metric values are rescaled to the total mass of "core" material in the simulation. The resulting values of all three metrics show a significant increasing trend with increasing resolution.}
\label{fig:mixing_resolution_study_pure_dense_scaled}
\end{figure*}

\begin{figure*}[ht!]
\centering
\includegraphics[width=\linewidth]{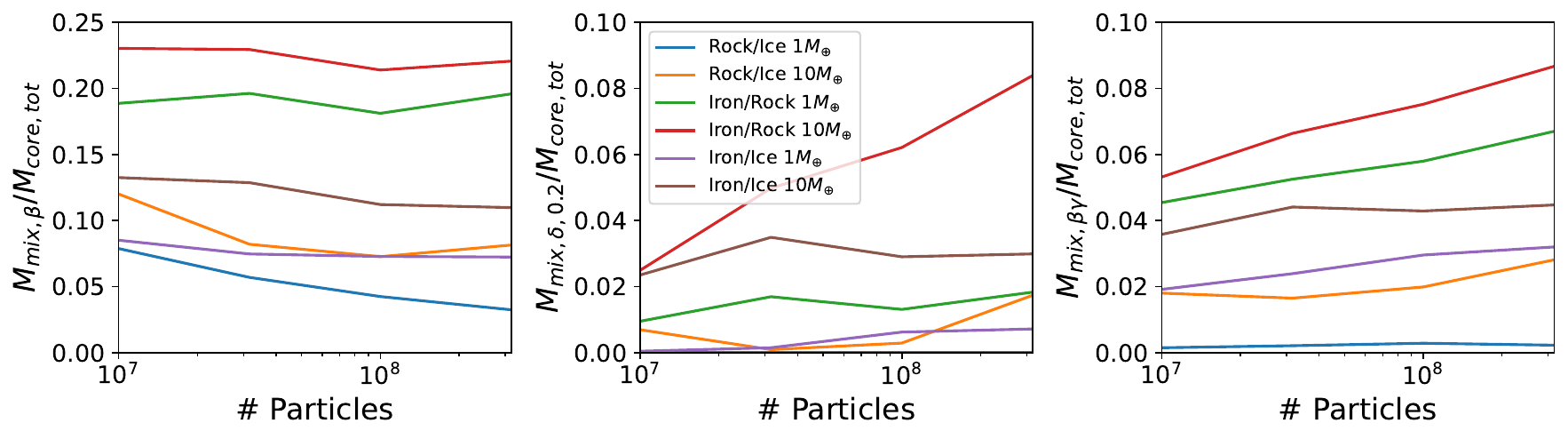}
\caption{Resolution study of the mixing metrics as defined in Appendix~\ref{sec:appendix:Mixing_metrics} at the end (\SI{35.4}{\hour} after the impact) for the simulations of heavy-element bodies in Table~\ref{tab:Mixing_Runs}. The mixing metric values are rescaled to the total mass of core material in the simulation. The impacts with an iron/rock target of \SI{10}{\Mearth} results in the largest value for all metrics, while the impacts with a rock/ice target of \SI{1}{\Mearth} results in the lowest values (especially for $M_{mix,\delta,0.2}$ where this configuration results in miniscule values).}
\label{fig:mixing_resolution_study_2mat_dense_scaled}
\end{figure*}

Here we investigate the collisions between a \SI{1}{\Mearth} or \SI{10}{\Mearth} target and an impactor of half that mass, each with a 2:1 mantle to core mass ratio. Both the cores and the mantles consist of either ice, rock, or iron, in all stable combinations (i.e. density decreases radially). This means that we have simulations where both the core and the mantle are of the same material (see Table~\ref{tab:Mixing_Runs_Pure}) and simulations where the core and the mantle have different compositions (see Table~\ref{tab:Mixing_Runs}). The impacts occur at the mutual escape velocity of the two bodies with an impact angle of $\theta=\SI{30}{\degree}$. This leads to strong turbulent flow patterns and a differentially rotating final body, which enhances mixing. We run these simulations at resolutions of $10^{7}$, $10^{7.5}$, $10^{8}$ and $10^{8.5}$ particles.

The simulations where the core and the mantle have the same composition provide an upper bound for the values of the mixing metrics that can be obtained in this type of collisions, because there is no material interface that could prevent mixing. Figure~\ref{fig:mixing_resolution_study_pure_dense_scaled} shows the mixing metrics at the end of the simulation (\SI{35.4}{\hour} after the impact) scaled to the total "core" mass in the simulations for the different resolutions. We find a strong increasing trend in all metrics with increasing resolution, where higher particle numbers can resolve finer flow patterns. Compared to the maximum values achieved with a random distribution ($M_{mix,\beta}=1.92M_{core,tot}$, $M_{mix,\delta,0.2}=0.99M_{core,tot}$ and $M_{mix,\beta\gamma}=0.93M_{core,tot}$, see Figure~\ref{fig:mixing_resolution_study}) we reach between \SI{60}{\percent} and \SI{85}{\percent} for $M_{mix,\beta}$ and $M_{mix,\delta,0.2}$ and between \SI{35}{\percent} and \SI{60}{\percent} for $M_{mix,\beta\gamma}$. This is a rather good result given that in this type of collision some parts of the target can remain undisturbed and that the values are still increasing at the end of the simulation (after \SI{35.4}{\hour}) due to ongoing differential rotation. The value of $M_{mix,\beta\gamma}$ before the collision decreases with increasing resolution as expected from the benchmarks in Appendix~\ref{sec:appendix:Mixing_metric_benchmark}.

We next consider simulations where the core and mantle have different compositions (see Table~\ref{tab:Mixing_Runs}). Figure~\ref{fig:mixing_resolution_study_2mat_dense_scaled} shows the scaled mixing metrics at the end of the simulation as a function of the resolution. Here we observe much less mixing than in the pure-material collisions and get a compact core. However, significant core material can be suspended in the mantle, especially in the iron/rock combination. The highest values of the mixing metrics are achieved in collisions with a \SI{10}{\Mearth} iron/rock target where we get up to \SI{10}{\percent} of the maximum achievable value (see Figure~\ref{fig:mixing_resolution_study}). In general, for each material combination, collisions involving the \SI{10}{\Mearth} target produce higher mixing metric values than the corresponding collisions with the \SI{1}{\Mearth} target. Overall, the results of the \SI{10}{\Mearth} collisions are ordered as follows: Iron/rock produces the highest values, iron/ice give intermediate values and rock/ice result in the smallest values. The rock/ice results also show a downward trend in $M_{mix,\beta}$ with increasing resolution. We note that the iron/rock and iron/ice results do not follow this trend which is also not observed in the $M_{mix,\delta,0.2}$ and $M_{mix,\beta\gamma}$ metrics.

\subsection{De-mixing in Jupiter-like bodies}\label{sec:Premixed_initial_conditions}
\begin{figure*}[ht!]
\centering
\includegraphics[width=\linewidth]{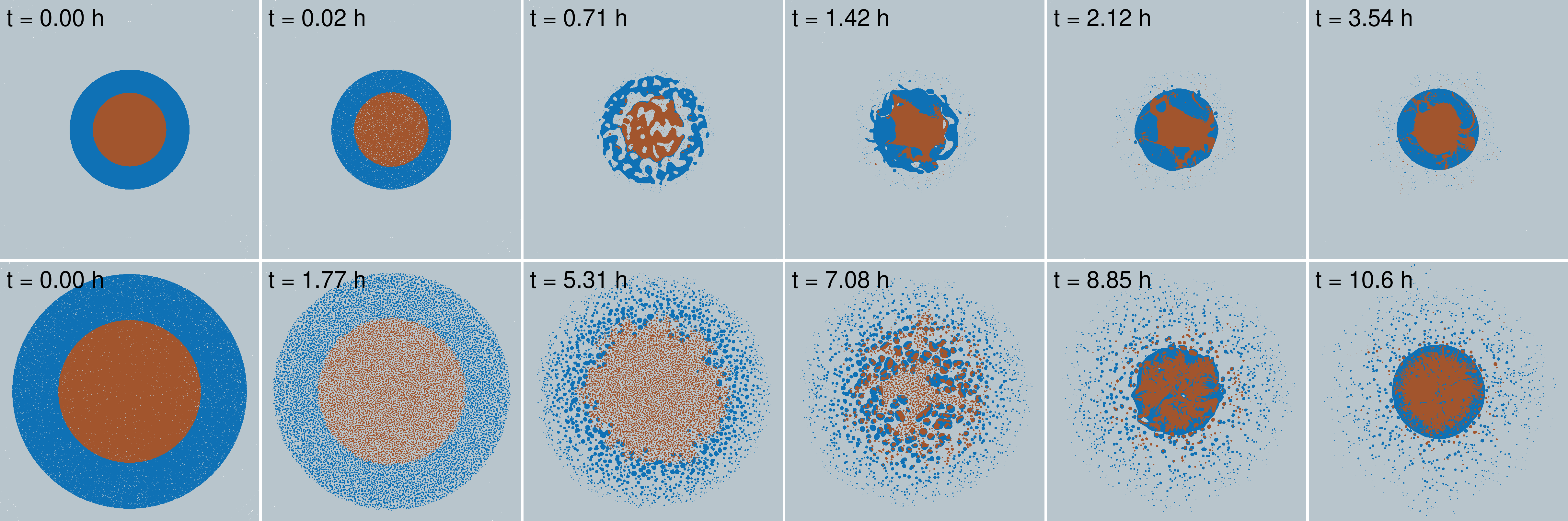}
\caption{Thin slices through the simulation starting with a 50:50 mixed core with \SI{e9}{} particles (run 5.4.2, top) and the simulation starting with a 18:82 mixed core (run 5.4.4, bottom) with \texttt{tipsy} \citep{n-bodyshopTIPSYCodeDisplay2011}. The side length of the domain shown is \SI{12.5}{\Rearth}. Grey pixels contain only hydrogen-helium particles, blue pixels contain at least one ice particle and brown pixels contain at least one rock particle (rock content supersedes ice content). In the 50:50 simulation, the heavy particles nucleate immediately after the beginning of the simulation, and then quickly fall into the center, expelling most hydrogen-helium particles. In the 18:82 simulation, the heavy particles also immediately nucleate, but then only over time, the pockets of heavy material slowly grow and fall into the center. The process is much slower than in the case of a 50:50 mixture.}
\label{fig:Bubbling_50_18}
\end{figure*}

\begin{figure*}[ht!]
\centering
\includegraphics[width=\linewidth]{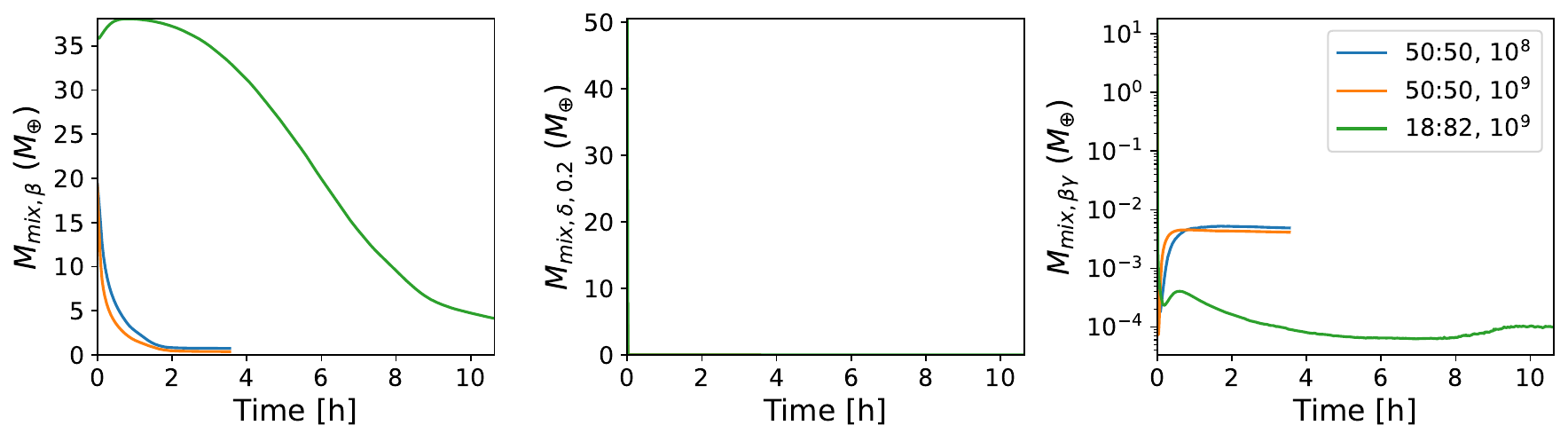}
\caption{Timeseries of the mixing metrics as defined in Appendix~\ref{sec:appendix:Mixing_metrics} applied to the runs starting with initially mixed Jupiter-like bodies in Table~\ref{tab:demixing_parameters}. $M_{mix,\delta,0.2}$ falls to exactly \SI{0.0}{\Mearth} shortly after the start of the simulation, while $M_{mix,\beta\gamma}$ also drops very fast, but does not reach exactly \SI{0.0}{\Mearth}.}
\label{fig:jupiter_demixing_timeseries}
\end{figure*}
 
In the impact simulations in Section~\ref{sec:Impact_Simulations} we do not observe well-mixed states. In order to better understand the meaning of the inferred values we can compare them to simulations with an initially well mixed core that is left to evolve without an impact. We explore two scenarios: the first starts with a \SI{20}{\Mearth} core that consists of a 50:50 mixture between the heavy elements and H-He (a total of \SI{10}{\Mearth} heavy material) and the second starts with a \SI{100}{\Mearth} core consisting of a 18:82 mixture between the heavy elements and H-He (a total of \SI{18}{\Mearth} heavy elements) similar to the profile proposed in \citet{militzerJunoSpacecraftMeasurements2022}. We create mixture tables for 50:50 and 18:82 mixtures between Ice:H-He and Rock:H-He using the additive volume law. These mixture tables are then used in \texttt{ballic} to create initial models in hydrostatic equilibrium. The particles of these mixed materials are then replaced with one of the two constituent materials, for each particle randomly chosen according to the mixing ratio. This ensures that once the body is transferred to the SPH simulation, they do not experience any sudden re-equilibration motion at the beginning of the simulation. Figure~\ref{fig:Bubbling_50_18} shows several snapshots of \SI{e9}{} particle simulations (runs 5.4.2 and 5.4.4) of these two scenarios. We find that in both cases the heavy elements quickly de-mix which leads to the formation of a compact core on a timescale of hours. However, the de-mixing timescale depends on the initial mixing ratio, as in the simulation with 18:82 ratio it takes much longer for the heavy elements to reach the center.

Figure~\ref{fig:jupiter_demixing_timeseries} shows the timeseries of the three mixing metrics $M_{mix,\beta}$, $M_{mix,\delta,0.2}$ and $M_{mix,\beta\gamma}$ for simulations of the first scenario at \SI{e8}{} and \SI{e9}{} particles and the second scenario at \SI{e9}{} particles. Note that the values of the mixing metrics in the two scenarios correspond to different amounts of heavy elements. We find that in the case of the 50:50 mixture, the heavy elements settle within \SI{2}{\hour}, whereas in the case of the 18:82 mixture they settle within about \SI{10}{\hour}, and that also at the end of the simulation the metric $M_{mix,\beta}$ has not stabilized. The heavy elements in the simulation with \SI{e9}{} particles settle faster than in the \SI{e8}{} particle simulation. We also observe this trend in the de-mixing simulations with heavy-element bodies in Section~\ref{sec:Mixing_and_de-mixing_in_SPH:Demixing}. Although the mixing metric $M_{mix,\delta,0.2}$ starts with \SI{9.24}{\Mearth}, \SI{9.59}{\Mearth} and \SI{50.5}{\Mearth} for the three simulations respectively, they immediately fall to \SI{0.0}{\Mearth} in less than \SI{1}{\minute} in all cases. $M_{mix,\beta\gamma}$ also starts with reasonable values of \SI{8.87}{\Mearth}, \SI{9.20}{\Mearth} and \SI{18.2}{\Mearth}, respectively, and falls by at least four orders of magnitude in the same time, but does not reach an absolute zero. 

Overall, the values of the mixing metrics at the end of the settling process are significantly smaller compared to the values observed in the impact simulations in Sections~\ref{sec:Oblique_Impact},~\ref{sec:Head-on_Impact} and~\ref{sec:Intermediate_Impact}. This is especially true for $M_{mix,\delta,0.2}$, where any value above \SI{0.0}{\Mearth} seems to indicate that mixing occurs. 

\subsection{De-mixing in heavy-element bodies}\label{sec:Mixing_and_de-mixing_in_SPH:Demixing}
\begin{figure*}[ht!]
\centering
\includegraphics[width=\linewidth]{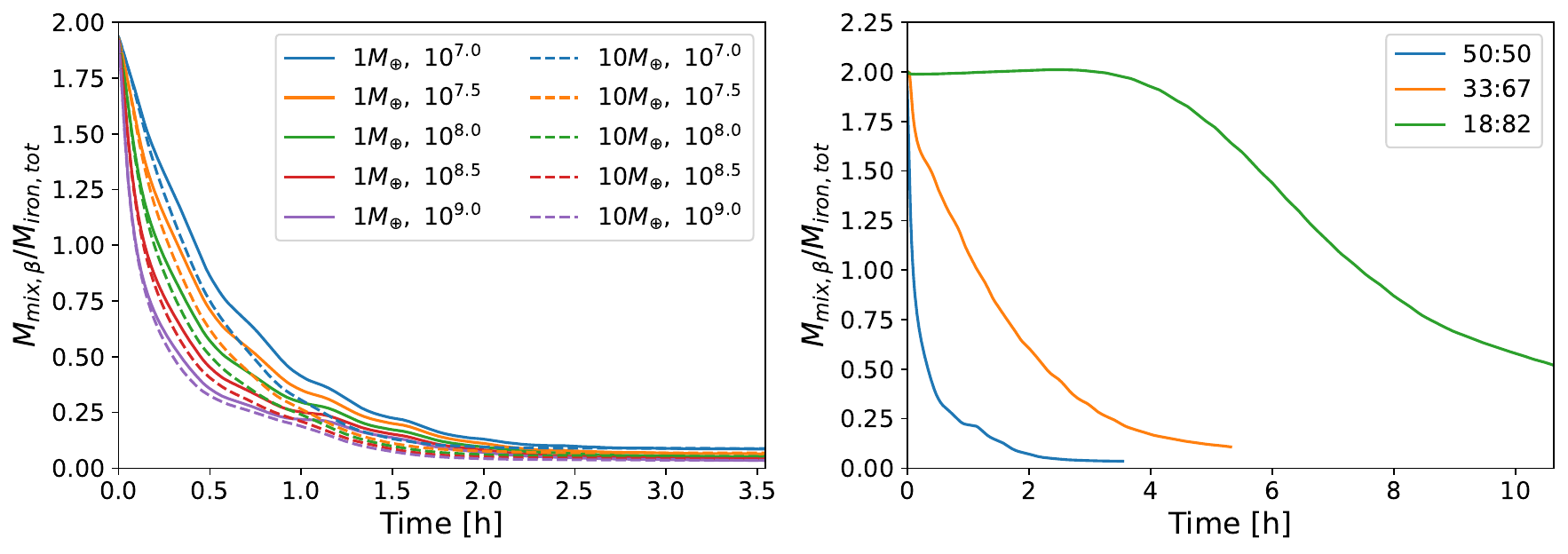}
\caption{Timeseries of the mixing metric $M_{mix,\beta}$ as defined in Appendix~\ref{sec:appendix:Mixing_metrics} for the simulations starting with initially mixed iron/rock cores (see Table~\ref{tab:demixing_parameters}). Left: Results for the simulations with \SI{1}{\Mearth} and \SI{10}{\Mearth} bodies with 50:50 mixing ratio at different resolutions. Right: Results for the simulations with \SI{1}{\Mearth} bodies at a resolution of $10^{9.0}$ and different mixing ratios. The settling is faster the higher the resolution, the heavier the body and the higher the initial mixing ratio.}
\label{fig:demixing_resolution_ratio_study}
\end{figure*}

We next run settling simulations with initially premixed bodies without H-He. The procedure to create the initial conditions is exactly the same as before: with EOS tables for 50:50, 33:67 and 18:82 mixtures for Rock:Iron calculated with the additive volume law, we build bodies in hydrostatic equilibrium with a core of iron/rock mixture and a mantle of rock at different masses, resolutions and mixing rations (see Table~\ref{tab:demixing_parameters}). The left panel of Figure~\ref{fig:demixing_resolution_ratio_study} shows the mixing metric $M_{mix,\beta}$ vs.~time for simulations of \SI{1}{\Mearth} and \SI{10}{\Mearth} bodies starting with a 50:50 mixed core at resolutions between $10^{7.0}$ and $10^{9.0}$. For both masses, higher resolution leads to faster settling. The value of $M_{mix,\beta}$ decreases with increasing resolution. This is because the result for the core mantle boundary depends on the used resolution. While in all the cases we infer a nearly perfectly segregated core with only trace amounts of iron still suspended in the mantle, the contribution of the core mantle boundary to $M_{mix,\beta}$ is different. For the resolutions we considered, the iron in the \SI{10}{\Mearth} simulation settles faster than in the \SI{1}{\Mearth} simulation. The right panel of Figure~\ref{fig:demixing_resolution_ratio_study} shows the mixing metric $M_{mix,\beta}$ vs.~time for simulations with a \SI{1}{\Mearth} body at a resolution of $10^{9.0}$ with mixing ratios for the core of 50:50, 33:67 and 18:82. At lower mixing ratios, the settling process takes much more time, with the 18:82 simulation taking \SI{3}{\hour} before $M_{mix,\beta}$ starts to decrease. This is consistent with the results with Jupiter-like bodies in Section~\ref{sec:Premixed_initial_conditions}. In all simulations in Figure~\ref{fig:demixing_resolution_ratio_study}, the values of the other two mixing metrics ($M_{mix,\delta,0.2}$ and $M_{mix,\beta\gamma}$) immediately drop by multiple orders of magnitude in the first minute of the simulation.

\section{Extended data tables}\label{sec:appendix:Extended_data_tables}
In this Appendix, all SPH simulation runs used in Sections~\ref{sec:Impact_Simulations} and~\ref{sec:Mixing_and_de-mixing_in_SPH} are defined. Listed are the run number, the composition and resolution of both the target and the impactor (if applicable) and the impact velocity and impact parameter (if applicable).

\begin{table*}[ht!]
\centering
\begin{tabular}{|c|c|c|c|c|c|c|c|c|c|c|}
\toprule
& \multicolumn{4}{c|}{Target} & \multicolumn{4}{c|}{Impactor} & \multicolumn{2}{c|}{Impact} \\
Run & Rock & Ice & H-He & $N$ & Rock & Ice & H-He & $N$ & $v$ & $b$\\
\midrule
3.1.1 & \SI{2.70}{\Mearth} & \SI{7.30}{\Mearth} & \SI{296.7}{\Mearth} & \SI{25e6}{} & \SI{2.16}{\Mearth} & \SI{5.84}{\Mearth} & \SI{2.0}{\Mearth} & \SI{815100}{} & \SI{46.0}{\kilo\meter\per\second} & \SI{0.0}{} \\
3.1.2 & \SI{2.70}{\Mearth} & \SI{7.30}{\Mearth} & \SI{296.7}{\Mearth} & \SI{25e6}{} & \SI{2.16}{\Mearth} & \SI{5.84}{\Mearth} & \SI{2.0}{\Mearth} & \SI{815100}{} & \SI{46.0}{\kilo\meter\per\second} & \SI{0.7}{} \\
\midrule
3.2.1 & \SI{4.59}{\Mearth} & \SI{12.41}{\Mearth} & \SI{298.7}{\Mearth} & \SI{25e6}{} & \SI{1.00}{\Mearth} & -- & -- & \SI{79185}{} & \SI{46.0}{\kilo\meter\per\second} & \SI{0.0}{} \\
3.2.2 & \SI{4.59}{\Mearth} & \SI{12.41}{\Mearth} & \SI{298.7}{\Mearth} & \SI{25e6}{} & \SI{1.00}{\Mearth} & -- & -- & \SI{79185}{} & \SI{46.0}{\kilo\meter\per\second} & \SI{0.7}{} \\
\midrule
5.1.3 & \SI{2.70}{\Mearth} & \SI{7.30}{\Mearth} & \SI{296.7}{\Mearth} & \SI{96.84e6}{} & \SI{2.16}{\Mearth} & \SI{5.84}{\Mearth} & \SI{2.0}{\Mearth} & \SI{31.57e6}{} & \SI{46.0}{\kilo\meter\per\second} & \SI{0.0}{} \\
5.1.4 & \SI{2.70}{\Mearth} & \SI{7.30}{\Mearth} & \SI{296.7}{\Mearth} & \SI{96.84e6}{} & \SI{2.16}{\Mearth} & \SI{5.84}{\Mearth} & \SI{2.0}{\Mearth} & \SI{31.57e6}{} & \SI{46.0}{\kilo\meter\per\second} & \SI{0.7}{} \\
5.1.8 & \SI{2.70}{\Mearth} & \SI{7.30}{\Mearth} & \SI{296.7}{\Mearth} & \SI{96.84e6}{} & \SI{2.16}{\Mearth} & \SI{5.84}{\Mearth} & \SI{2.0}{\Mearth} & \SI{31.57e6}{} & \SI{46.0}{\kilo\meter\per\second} & \SI{0.2}{} \\
\midrule
5.1.5$^\#$ & \SI{2.70}{\Mearth} & \SI{7.30}{\Mearth} & \SI{296.7}{\Mearth} & \SI{96.84e6}{} & \SI{2.16}{\Mearth} & \SI{5.84}{\Mearth} & \SI{2.0}{\Mearth} & \SI{31.57e6}{} & \SI{46.0}{\kilo\meter\per\second} & \SI{0.0}{} \\
5.1.6$^\#$ & \SI{2.70}{\Mearth} & \SI{7.30}{\Mearth} & \SI{296.7}{\Mearth} & \SI{96.84e6}{} & \SI{2.16}{\Mearth} & \SI{5.84}{\Mearth} & \SI{2.0}{\Mearth} & \SI{31.57e6}{} & \SI{46.0}{\kilo\meter\per\second} & \SI{0.7}{} \\
5.1.7$^\#$ & \SI{2.70}{\Mearth} & \SI{7.30}{\Mearth} & \SI{296.7}{\Mearth} & \SI{96.84e6}{} & \SI{2.16}{\Mearth} & \SI{5.84}{\Mearth} & \SI{2.0}{\Mearth} & \SI{31.57e6}{} & \SI{46.0}{\kilo\meter\per\second} & \SI{0.2}{} \\
\midrule
5.1.1 & \SI{2.70}{\Mearth} & \SI{7.30}{\Mearth} & \SI{296.7}{\Mearth} & \SI{968.4e6}{} & \SI{2.16}{\Mearth} & \SI{5.84}{\Mearth} & \SI{2.0}{\Mearth} & \SI{315.7e6}{} & \SI{46.0}{\kilo\meter\per\second} & \SI{0.0}{} \\
5.1.2 & \SI{2.70}{\Mearth} & \SI{7.30}{\Mearth} & \SI{296.7}{\Mearth} & \SI{968.4e6}{} & \SI{2.16}{\Mearth} & \SI{5.84}{\Mearth} & \SI{2.0}{\Mearth} & \SI{315.7e6}{} & \SI{46.0}{\kilo\meter\per\second} & \SI{0.7}{} \\
5.1.9$^\dagger$ & \SI{2.70}{\Mearth} & \SI{7.30}{\Mearth} & \SI{296.7}{\Mearth} & \SI{968.4e6}{} & \SI{2.16}{\Mearth} & \SI{5.84}{\Mearth} & \SI{2.0}{\Mearth} & \SI{315.7e6}{} & \SI{46.0}{\kilo\meter\per\second} & \SI{0.2}{} \\
\midrule
5.1.10$^\dagger$ & \SI{2.70}{\Mearth} & \SI{7.30}{\Mearth} & \SI{296.7}{\Mearth} & \SI{2.060e9}{} & \SI{2.16}{\Mearth} & \SI{5.84}{\Mearth} & \SI{2.0}{\Mearth} & \SI{67.00e6}{} & \SI{46.0}{\kilo\meter\per\second} & \SI{0.0}{} \\
\midrule
5.5.1$^{*\dagger}$ & \SI{2.70}{\Mearth} & \SI{7.30}{\Mearth} & \SI{296.7}{\Mearth} & \SI{968.4e6}{} & \SI{2.16}{\Mearth} & \SI{5.84}{\Mearth} & \SI{2.0}{\Mearth} & \SI{315.7e6}{} & \SI{46.0}{\kilo\meter\per\second} & \SI{0.0}{} \\
5.5.2$^{*\dagger}$ & \SI{2.70}{\Mearth} & \SI{7.30}{\Mearth} & \SI{296.7}{\Mearth} & \SI{968.4e6}{} & \SI{2.16}{\Mearth} & \SI{5.84}{\Mearth} & \SI{2.0}{\Mearth} & \SI{315.7e6}{} & \SI{46.0}{\kilo\meter\per\second} & \SI{0.7}{} \\
5.5.3$^{*\dagger}$ & \SI{2.70}{\Mearth} & \SI{7.30}{\Mearth} & \SI{296.7}{\Mearth} & \SI{968.4e6}{} & \SI{2.16}{\Mearth} & \SI{5.84}{\Mearth} & \SI{2.0}{\Mearth} & \SI{315.7e6}{} & \SI{46.0}{\kilo\meter\per\second} & \SI{0.2}{} \\
\midrule
5.5.4$^*$ & \SI{2.70}{\Mearth} & \SI{7.30}{\Mearth} & \SI{296.7}{\Mearth} & \SI{96.84e6}{} & \SI{2.16}{\Mearth} & \SI{5.84}{\Mearth} & \SI{2.0}{\Mearth} & \SI{31.57e6}{} & \SI{46.0}{\kilo\meter\per\second} & \SI{0.0}{} \\
5.5.5$^*$ & \SI{2.70}{\Mearth} & \SI{7.30}{\Mearth} & \SI{296.7}{\Mearth} & \SI{96.84e6}{} & \SI{2.16}{\Mearth} & \SI{5.84}{\Mearth} & \SI{2.0}{\Mearth} & \SI{31.57e6}{} & \SI{46.0}{\kilo\meter\per\second} & \SI{0.7}{} \\
5.5.6$^*$ & \SI{2.70}{\Mearth} & \SI{7.30}{\Mearth} & \SI{296.7}{\Mearth} & \SI{96.84e6}{} & \SI{2.16}{\Mearth} & \SI{5.84}{\Mearth} & \SI{2.0}{\Mearth} & \SI{31.57e6}{} & \SI{46.0}{\kilo\meter\per\second} & \SI{0.2}{} \\
\bottomrule
\end{tabular}
\caption{Parameters of the SPH collision simulations on Jupiter. The surface temperature of all targets and impactors is set to \SI{300}{\kelvin} and the simulation time is \SI{35.4}{\hour}. Exceptions: Simulations marked with $^\dagger$ are not run for the full simulation time, simulations marked with $^*$ use the SCvH equation of state to model the hydrogen-helium mixture, simulations marked with $^\#$ have a higher surface temperature of \SI{1200}{\kelvin}.}
\label{tab:Jupiter_impact_parameters_late}
\end{table*}

\begin{table*}[ht!]
\centering
\begin{tabular}{|c|c|c|c|c|c|c|c|}
\toprule
& \multicolumn{3}{c|}{Target} & \multicolumn{2}{c|}{Impactor} & \multicolumn{2}{c|}{Impact} \\
Run & Core & Envelope & $N$ & Mass & $N$ & $v$ & $b$\\
\midrule
5.2.1 & \SI{10}{\Mearth} Rock & \SI{296.7}{\Mearth} H-He & \SI{96.8e6}{} & \SI{10}{\Mearth} Rock & \SI{3.2e6}{} & \SI{46.0}{\kilo\meter\per\second} & \SI{0.0}{} \\
5.2.2 & \SI{10}{\Mearth} Rock & \SI{296.7}{\Mearth} H-He & \SI{96.8e6}{} & \SI{10}{\Mearth} Rock & \SI{3.2e6}{} & \SI{46.0}{\kilo\meter\per\second} & \SI{0.7}{} \\
5.2.3 & \SI{10}{\Mearth} Rock & \SI{296.7}{\Mearth} H-He & \SI{96.8e6}{} & \SI{10}{\Mearth} Ice & \SI{3.2e6}{} & \SI{46.0}{\kilo\meter\per\second} & \SI{0.0}{} \\
5.2.4 & \SI{10}{\Mearth} Rock & \SI{296.7}{\Mearth} H-He & \SI{96.8e6}{} & \SI{10}{\Mearth} Ice & \SI{3.2e6}{} & \SI{46.0}{\kilo\meter\per\second} & \SI{0.7}{} \\
5.2.5$^\dagger$ & \SI{10}{\Mearth} Rock & \SI{296.7}{\Mearth} H-He & \SI{96.8e6}{} & \SI{10}{\Mearth} Helium & \SI{3.2e6}{} & \SI{46.0}{\kilo\meter\per\second} & \SI{0.0}{} \\
5.2.6$^\dagger$ & \SI{10}{\Mearth} Rock & \SI{296.7}{\Mearth} H-He & \SI{96.8e6}{} & \SI{10}{\Mearth} Helium & \SI{3.2e6}{} & \SI{46.0}{\kilo\meter\per\second} & \SI{0.7}{} \\
\midrule
5.2.7 & \SI{10}{\Mearth} Ice & \SI{296.7}{\Mearth} H-He & \SI{96.8e6}{} & \SI{10}{\Mearth} Rock & \SI{3.2e6}{} & \SI{46.0}{\kilo\meter\per\second} & \SI{0.0}{} \\
5.2.8 & \SI{10}{\Mearth} Ice & \SI{296.7}{\Mearth} H-He & \SI{96.8e6}{} & \SI{10}{\Mearth} Rock & \SI{3.2e6}{} & \SI{46.0}{\kilo\meter\per\second} & \SI{0.7}{} \\
5.2.9 & \SI{10}{\Mearth} Ice & \SI{296.7}{\Mearth} H-He & \SI{96.8e6}{} & \SI{10}{\Mearth} Ice & \SI{3.2e6}{} & \SI{46.0}{\kilo\meter\per\second} & \SI{0.0}{} \\
5.2.10 & \SI{10}{\Mearth} Ice & \SI{296.7}{\Mearth} H-He & \SI{96.8e6}{} & \SI{10}{\Mearth} Ice & \SI{3.2e6}{} & \SI{46.0}{\kilo\meter\per\second} & \SI{0.7}{} \\
5.2.11$^\dagger$ & \SI{10}{\Mearth} Ice & \SI{296.7}{\Mearth} H-He & \SI{96.8e6}{} & \SI{10}{\Mearth} Helium & \SI{3.2e6}{} & \SI{46.0}{\kilo\meter\per\second} & \SI{0.0}{} \\
5.2.12$^\dagger$ & \SI{10}{\Mearth} Ice & \SI{296.7}{\Mearth} H-He & \SI{96.8e6}{} & \SI{10}{\Mearth} Helium & \SI{3.2e6}{} & \SI{46.0}{\kilo\meter\per\second} & \SI{0.7}{} \\
\midrule
5.2.13 & \SI{10}{\Mearth} Helium & \SI{296.7}{\Mearth} H-He & \SI{96.8e6}{} & \SI{10}{\Mearth} Rock & \SI{3.2e6}{} & \SI{46.0}{\kilo\meter\per\second} & \SI{0.0}{} \\
5.2.14 & \SI{10}{\Mearth} Helium & \SI{296.7}{\Mearth} H-He & \SI{96.8e6}{} & \SI{10}{\Mearth} Rock & \SI{3.2e6}{} & \SI{46.0}{\kilo\meter\per\second} & \SI{0.7}{} \\
5.2.15 & \SI{10}{\Mearth} Helium & \SI{296.7}{\Mearth} H-He & \SI{96.8e6}{} & \SI{10}{\Mearth} Ice & \SI{3.2e6}{} & \SI{46.0}{\kilo\meter\per\second} & \SI{0.0}{} \\
5.2.16 & \SI{10}{\Mearth} Helium & \SI{296.7}{\Mearth} H-He & \SI{96.8e6}{} & \SI{10}{\Mearth} Ice & \SI{3.2e6}{} & \SI{46.0}{\kilo\meter\per\second} & \SI{0.7}{} \\
5.2.17$^\dagger$ & \SI{10}{\Mearth} Helium & \SI{296.7}{\Mearth} H-He & \SI{96.8e6}{} & \SI{10}{\Mearth} Helium & \SI{3.2e6}{} & \SI{46.0}{\kilo\meter\per\second} & \SI{0.0}{} \\
5.2.18$^\dagger$ & \SI{10}{\Mearth} Helium & \SI{296.7}{\Mearth} H-He & \SI{96.8e6}{} & \SI{10}{\Mearth} Helium & \SI{3.2e6}{} & \SI{46.0}{\kilo\meter\per\second} & \SI{0.7}{} \\
\bottomrule
\end{tabular}
\caption{Parameters of the SPH simulations with a simplified Jupiter model. The surface temperature of all targets and impactors is set to \SI{300}{\kelvin} and the simulation time is \SI{35.4}{\hour}. Exceptions: Simulations marked with $^\dagger$ are not run for the full simulation time, because a body of \SI{10}{\Mearth} helium is not compact enough to penetrate Jupiter's envelope.}
\label{tab:Jupiter_simplified_impact_parameters}
\end{table*}

\begin{table*}[ht!]
\centering
\begin{tabular}{|c|c|c|c|c|c|c|c|c|c|c|}
\toprule
& \multicolumn{4}{c|}{Target} & \multicolumn{4}{c|}{Impactor} & \multicolumn{2}{c|}{Impact} \\
Run & Rock & Ice & H-He & $N$ & Rock & Ice & H-He & $N$ & $v$ & $b$\\
\midrule
1.1.1 & \SI{10}{\Mearth} & -- & \SI{10}{\Mearth} & \SI{20e6}{} & \SI{1}{\Mearth} & -- & -- & \SI{1e6}{} & \SI{20.0}{\kilo\meter\per\second} & \SI{0.0}{} \\
1.1.2 & \SI{10}{\Mearth} & -- & \SI{10}{\Mearth} & \SI{20e6}{} & \SI{1}{\Mearth} & -- & -- & \SI{1e6}{} & \SI{20.0}{\kilo\meter\per\second} & \SI{0.2}{} \\
1.1.3 & \SI{10}{\Mearth} & -- & \SI{10}{\Mearth} & \SI{20e6}{} & \SI{1}{\Mearth} & -- & -- & \SI{1e6}{} & \SI{40.0}{\kilo\meter\per\second} & \SI{0.0}{} \\
1.1.4 & \SI{10}{\Mearth} & -- & \SI{10}{\Mearth} & \SI{20e6}{} & \SI{1}{\Mearth} & -- & -- & \SI{1e6}{} & \SI{40.0}{\kilo\meter\per\second} & \SI{0.2}{} \\
\midrule
1.2.1 & \SI{10}{\Mearth} & -- & \SI{5}{\Mearth} & \SI{15e6}{} & \SI{1}{\Mearth} & -- & -- & \SI{1e6}{} & \SI{20.0}{\kilo\meter\per\second} & \SI{0.0}{} \\
1.2.2 & \SI{10}{\Mearth} & -- & \SI{5}{\Mearth} & \SI{15e6}{} & \SI{1}{\Mearth} & -- & -- & \SI{1e6}{} & \SI{20.0}{\kilo\meter\per\second} & \SI{0.2}{} \\
1.2.3 & \SI{10}{\Mearth} & -- & \SI{5}{\Mearth} & \SI{15e6}{} & \SI{1}{\Mearth} & -- & -- & \SI{1e6}{} & \SI{40.0}{\kilo\meter\per\second} & \SI{0.0}{} \\
1.2.4 & \SI{10}{\Mearth} & -- & \SI{5}{\Mearth} & \SI{15e6}{} & \SI{1}{\Mearth} & -- & -- & \SI{1e6}{} & \SI{40.0}{\kilo\meter\per\second} & \SI{0.2}{} \\
\midrule
1.3.1 & \SI{10}{\Mearth} & -- & \SI{10}{\Mearth} & \SI{20e6}{} & \SI{10}{\Mearth} & -- & -- & \SI{10e6}{} & \SI{20.0}{\kilo\meter\per\second} & \SI{0.0}{} \\
1.3.2 & \SI{10}{\Mearth} & -- & \SI{10}{\Mearth} & \SI{20e6}{} & \SI{10}{\Mearth} & -- & -- & \SI{10e6}{} & \SI{20.0}{\kilo\meter\per\second} & \SI{0.2}{} \\
1.3.3 & \SI{10}{\Mearth} & -- & \SI{10}{\Mearth} & \SI{20e6}{} & \SI{10}{\Mearth} & -- & -- & \SI{10e6}{} & \SI{40.0}{\kilo\meter\per\second} & \SI{0.0}{} \\
1.3.4 & \SI{10}{\Mearth} & -- & \SI{10}{\Mearth} & \SI{20e6}{} & \SI{10}{\Mearth} & -- & -- & \SI{10e6}{} & \SI{40.0}{\kilo\meter\per\second} & \SI{0.2}{} \\
\midrule
2.1.1 & -- & \SI{10}{\Mearth} & \SI{10}{\Mearth} & \SI{20e6}{} & -- & \SI{1}{\Mearth} & -- & \SI{1e6}{} & \SI{20.0}{\kilo\meter\per\second} & \SI{0.0}{} \\
2.1.2 & -- & \SI{10}{\Mearth} & \SI{10}{\Mearth} & \SI{20e6}{} & -- & \SI{1}{\Mearth} & -- & \SI{1e6}{} & \SI{20.0}{\kilo\meter\per\second} & \SI{0.2}{} \\
2.1.3 & -- & \SI{10}{\Mearth} & \SI{10}{\Mearth} & \SI{20e6}{} & -- & \SI{1}{\Mearth} & -- & \SI{1e6}{} & \SI{40.0}{\kilo\meter\per\second} & \SI{0.0}{} \\
2.1.4 & -- & \SI{10}{\Mearth} & \SI{10}{\Mearth} & \SI{20e6}{} & -- & \SI{1}{\Mearth} & -- & \SI{1e6}{} & \SI{40.0}{\kilo\meter\per\second} & \SI{0.2}{} \\
\midrule
2.2.1 & -- & \SI{10}{\Mearth} & \SI{5}{\Mearth} & \SI{15e6}{} & -- & \SI{1}{\Mearth} & -- & \SI{1e6}{} & \SI{20.0}{\kilo\meter\per\second} & \SI{0.0}{} \\
2.2.2 & -- & \SI{10}{\Mearth} & \SI{5}{\Mearth} & \SI{15e6}{} & -- & \SI{1}{\Mearth} & -- & \SI{1e6}{} & \SI{20.0}{\kilo\meter\per\second} & \SI{0.2}{} \\
2.2.3 & -- & \SI{10}{\Mearth} & \SI{5}{\Mearth} & \SI{15e6}{} & -- & \SI{1}{\Mearth} & -- & \SI{1e6}{} & \SI{40.0}{\kilo\meter\per\second} & \SI{0.0}{} \\
2.2.4 & -- & \SI{10}{\Mearth} & \SI{5}{\Mearth} & \SI{15e6}{} & -- & \SI{1}{\Mearth} & -- & \SI{1e6}{} & \SI{40.0}{\kilo\meter\per\second} & \SI{0.2}{} \\
\midrule
2.3.1 & -- & \SI{10}{\Mearth} & \SI{10}{\Mearth} & \SI{20e6}{} & -- & \SI{10}{\Mearth} & -- & \SI{10e6}{} & \SI{20.0}{\kilo\meter\per\second} & \SI{0.0}{} \\
2.3.2 & -- & \SI{10}{\Mearth} & \SI{10}{\Mearth} & \SI{20e6}{} & -- & \SI{10}{\Mearth} & -- & \SI{10e6}{} & \SI{20.0}{\kilo\meter\per\second} & \SI{0.2}{} \\
2.3.3 & -- & \SI{10}{\Mearth} & \SI{10}{\Mearth} & \SI{20e6}{} & -- & \SI{10}{\Mearth} & -- & \SI{10e6}{} & \SI{40.0}{\kilo\meter\per\second} & \SI{0.0}{} \\
2.3.4 & -- & \SI{10}{\Mearth} & \SI{10}{\Mearth} & \SI{20e6}{} & -- & \SI{10}{\Mearth} & -- & \SI{10e6}{} & \SI{40.0}{\kilo\meter\per\second} & \SI{0.2}{} \\
\bottomrule
\end{tabular}
\caption{Parameters of the SPH collision simulations with an early Jupiter model. The surface temperature of all targets and impactors is set to \SI{300}{\kelvin} and the simulation time is \SI{35.4}{\hour}.}
\label{tab:Jupiter_impact_parameters_early}
\end{table*}

\begin{table}[ht!]
\centering
\begin{tabular}{|c|c|c|c|}
\toprule
Run & $M_{mix,\beta}$ & $M_{mix,\delta,0.2}$ & $M_{mix,\beta\gamma}$\\
\midrule
1.1.1 & \SI{1.78e-1}{\Mearth} & \SI{5.21e-3}{\Mearth} & \SI{1.28e-2}{\Mearth} \\
1.1.2 & \SI{2.36e-01}{\Mearth} & \SI{4.32e-02}{\Mearth} & \SI{4.54e-02}{\Mearth} \\
1.1.3 & \SI{2.89e-01}{\Mearth} & \SI{5.35e-02}{\Mearth} & \SI{6.83e-02}{\Mearth} \\
1.1.4$^\dagger$ & \SI{3.92e-01}{\Mearth} & \SI{7.25e-02}{\Mearth} & \SI{7.60e-02}{\Mearth} \\
\midrule
1.2.1 & \SI{1.79e-01}{\Mearth} & \SI{1.37e-02}{\Mearth} & \SI{2.89e-02}{\Mearth} \\
1.2.2 & \SI{2.39e-01}{\Mearth} & \SI{4.12e-02}{\Mearth} & \SI{4.48e-02}{\Mearth} \\
1.2.3 & \SI{3.85e-01}{\Mearth} & \SI{7.99e-02}{\Mearth} & \SI{9.15e-02}{\Mearth} \\
1.2.4$^\dagger$ & \SI{4.60e-01}{\Mearth} & \SI{6.55e-02}{\Mearth} & \SI{9.37e-02}{\Mearth} \\
\midrule
1.3.1 & \SI{1.19e+00}{\Mearth} & \SI{3.20e-01}{\Mearth} & \SI{3.03e-01}{\Mearth} \\
1.3.2 & \SI{1.19e+00}{\Mearth} & \SI{4.71e-01}{\Mearth} & \SI{5.02e-01}{\Mearth} \\
1.3.3$^\dagger$ & \SI{2.44e+00}{\Mearth} & \SI{1.47e-01}{\Mearth} & \SI{9.31e-02}{\Mearth} \\
1.3.4$^\#$ & \SI{1.57e+00}{\Mearth} & \SI{3.93e-01}{\Mearth} & \SI{3.06e-01}{\Mearth} \\
\midrule
2.1.1 & \SI{3.24e-01}{\Mearth} & \SI{3.93e-02}{\Mearth} & \SI{5.55e-02}{\Mearth} \\
2.1.2 & \SI{3.75e-01}{\Mearth} & \SI{1.01e-01}{\Mearth} & \SI{9.84e-02}{\Mearth} \\
2.1.3 & \SI{4.67e-01}{\Mearth} & \SI{1.03e-01}{\Mearth} & \SI{1.14e-01}{\Mearth} \\
2.1.4$^\dagger$ & \SI{7.79e-01}{\Mearth} & \SI{2.58e-01}{\Mearth} & \SI{2.03e-01}{\Mearth} \\
\midrule
2.2.1 & \SI{2.83e-01}{\Mearth} & \SI{1.99e-02}{\Mearth} & \SI{5.70e-02}{\Mearth} \\
2.2.2 & \SI{4.40e-01}{\Mearth} & \SI{1.78e-01}{\Mearth} & \SI{1.31e-01}{\Mearth} \\
2.2.3 & \SI{7.58e-01}{\Mearth} & \SI{2.11e-01}{\Mearth} & \SI{1.67e-01}{\Mearth} \\
2.2.4$^\dagger$ & \SI{9.15e-01}{\Mearth} & \SI{2.60e-01}{\Mearth} & \SI{1.88e-01}{\Mearth} \\
\midrule
2.3.1 & \SI{1.94e+00}{\Mearth} & \SI{6.24e-01}{\Mearth} & \SI{4.45e-01}{\Mearth} \\
2.3.2 & \SI{2.13e+00}{\Mearth} & \SI{1.48e+00}{\Mearth} & \SI{7.10e-01}{\Mearth} \\
2.3.3$^\dagger$ & \SI{2.50e+00}{\Mearth} & \SI{6.73e-02}{\Mearth} & \SI{6.27e-02}{\Mearth} \\
2.3.4$^\#$ & \SI{3.43e+00}{\Mearth} & \SI{1.73e-01}{\Mearth} & \SI{1.22e-01}{\Mearth} \\
\bottomrule
\end{tabular}
\caption{Value of the mixing metrics at the end of the simulations from Table~\ref{tab:Jupiter_impact_parameters_early}. For the discussion see Section~\ref{sec:Discussion_Collisions_on_early_Jupiter}. The runs marked with $^\dagger$ have a significant fraction of the heavy material not fallen back into Jupiter, while those marked with $^\#$ produce a secondary body that is flying away from Jupiter (potentially hit-and-run).}
\label{tab:Jupiter_impact_results_early}
\end{table}

\begin{table*}[ht!]
\centering
\begin{tabular}{|c|c|c|c|c|c|c|c|c|}
\toprule
& \multicolumn{3}{c|}{Target} & \multicolumn{3}{c|}{Impactor} & \multicolumn{2}{c|}{Impact} \\
Run & Core & Mantle & $N$ & Core & Mantle & $N$ & $v$ & $b$\\
\midrule
7.1.1 & \SI{0.333}{\Mearth} Rock & \SI{0.667}{\Mearth} Rock & \SI{6.67e6}{} & \SI{0.167}{\Mearth} Rock & \SI{0.333}{\Mearth} Rock & \SI{3.33e6}{} & \SI{9.72}{\kilo\meter\per\second} & \SI{0.5}{} \\
7.1.2 & \SI{0.333}{\Mearth} Rock & \SI{0.667}{\Mearth} Rock & \SI{21.0e6}{} & \SI{0.167}{\Mearth} Rock & \SI{0.333}{\Mearth} Rock & \SI{10.5e6}{} & \SI{9.72}{\kilo\meter\per\second} & \SI{0.5}{} \\
7.1.3 & \SI{0.333}{\Mearth} Rock & \SI{0.667}{\Mearth} Rock & \SI{66.7e6}{} & \SI{0.167}{\Mearth} Rock & \SI{0.333}{\Mearth} Rock & \SI{33.3e6}{} & \SI{9.72}{\kilo\meter\per\second} & \SI{0.5}{} \\
7.1.4 & \SI{0.333}{\Mearth} Rock & \SI{0.667}{\Mearth} Rock & \SI{210e6}{} & \SI{0.167}{\Mearth} Rock & \SI{0.333}{\Mearth} Rock & \SI{105e6}{} & \SI{9.72}{\kilo\meter\per\second} & \SI{0.5}{} \\
7.1.5 & \SI{3.33}{\Mearth} Rock & \SI{6.67}{\Mearth} Rock & \SI{6.67e6}{} & \SI{1.67}{\Mearth} Rock & \SI{3.33}{\Mearth} Rock & \SI{3.33e6}{} & \SI{22.5}{\kilo\meter\per\second} & \SI{0.5}{} \\
7.1.6 & \SI{3.33}{\Mearth} Rock & \SI{6.67}{\Mearth} Rock & \SI{21.0e6}{} & \SI{1.67}{\Mearth} Rock & \SI{3.33}{\Mearth} Rock & \SI{10.5e6}{} & \SI{22.5}{\kilo\meter\per\second} & \SI{0.5}{} \\
7.1.7 & \SI{3.33}{\Mearth} Rock & \SI{6.67}{\Mearth} Rock & \SI{66.7e6}{} & \SI{1.67}{\Mearth} Rock & \SI{3.33}{\Mearth} Rock & \SI{33.3e6}{} & \SI{22.5}{\kilo\meter\per\second} & \SI{0.5}{} \\
7.1.8 & \SI{3.33}{\Mearth} Rock & \SI{6.67}{\Mearth} Rock & \SI{210e6}{} & \SI{1.67}{\Mearth} Rock & \SI{3.33}{\Mearth} Rock & \SI{105e6}{} & \SI{22.5}{\kilo\meter\per\second} & \SI{0.5}{} \\
\midrule
7.2.1 & \SI{0.333}{\Mearth} Ice & \SI{0.667}{\Mearth} Ice & \SI{6.67e6}{} & \SI{0.167}{\Mearth} Ice & \SI{0.333}{\Mearth} Ice & \SI{3.33e6}{} & \SI{8.44}{\kilo\meter\per\second} & \SI{0.5}{} \\
7.2.2 & \SI{0.333}{\Mearth} Ice & \SI{0.667}{\Mearth} Ice & \SI{21.0e6}{} & \SI{0.167}{\Mearth} Ice & \SI{0.333}{\Mearth} Ice & \SI{10.5e6}{} & \SI{8.44}{\kilo\meter\per\second} & \SI{0.5}{} \\
7.2.3 & \SI{0.333}{\Mearth} Ice & \SI{0.667}{\Mearth} Ice & \SI{66.7e6}{} & \SI{0.167}{\Mearth} Ice & \SI{0.333}{\Mearth} Ice & \SI{33.3e6}{} & \SI{8.44}{\kilo\meter\per\second} & \SI{0.5}{} \\
7.2.4 & \SI{0.333}{\Mearth} Ice & \SI{0.667}{\Mearth} Ice & \SI{210e6}{} & \SI{0.167}{\Mearth} Ice & \SI{0.333}{\Mearth} Ice & \SI{105e6}{} & \SI{8.44}{\kilo\meter\per\second} & \SI{0.5}{} \\
7.2.5 & \SI{3.33}{\Mearth} Ice & \SI{6.67}{\Mearth} Ice & \SI{6.67e6}{} & \SI{1.67}{\Mearth} Ice & \SI{3.33}{\Mearth} Ice & \SI{3.33e6}{} & \SI{19.39}{\kilo\meter\per\second} & \SI{0.5}{} \\
7.2.6 & \SI{3.33}{\Mearth} Ice & \SI{6.67}{\Mearth} Ice & \SI{21.0e6}{} & \SI{1.67}{\Mearth} Ice & \SI{3.33}{\Mearth} Ice & \SI{10.5e6}{} & \SI{19.39}{\kilo\meter\per\second} & \SI{0.5}{} \\
7.2.7 & \SI{3.33}{\Mearth} Ice & \SI{6.67}{\Mearth} Ice & \SI{66.7e6}{} & \SI{1.67}{\Mearth} Ice & \SI{3.33}{\Mearth} Ice & \SI{33.3e6}{} & \SI{19.39}{\kilo\meter\per\second} & \SI{0.5}{} \\
7.2.8 & \SI{3.33}{\Mearth} Ice & \SI{6.67}{\Mearth} Ice & \SI{210e6}{} & \SI{1.67}{\Mearth} Ice & \SI{3.33}{\Mearth} Ice & \SI{105e6}{} & \SI{19.39}{\kilo\meter\per\second} & \SI{0.5}{} \\
\midrule
7.4.1 & \SI{0.333}{\Mearth} Iron & \SI{0.667}{\Mearth} Iron & \SI{6.67e6}{} & \SI{0.167}{\Mearth} Iron & \SI{0.333}{\Mearth} Iron & \SI{3.33e6}{} & \SI{11.5}{\kilo\meter\per\second} & \SI{0.5}{} \\
7.4.2 & \SI{0.333}{\Mearth} Iron & \SI{0.667}{\Mearth} Iron & \SI{21.0e6}{} & \SI{0.167}{\Mearth} Iron & \SI{0.333}{\Mearth} Iron & \SI{10.5e6}{} & \SI{11.5}{\kilo\meter\per\second} & \SI{0.5}{} \\
7.4.3 & \SI{0.333}{\Mearth} Iron & \SI{0.667}{\Mearth} Iron & \SI{66.7e6}{} & \SI{0.167}{\Mearth} Iron & \SI{0.333}{\Mearth} Iron & \SI{33.3e6}{} & \SI{11.5}{\kilo\meter\per\second} & \SI{0.5}{} \\
7.4.4 & \SI{0.333}{\Mearth} Iron & \SI{0.667}{\Mearth} Iron & \SI{210e6}{} & \SI{0.167}{\Mearth} Iron & \SI{0.333}{\Mearth} Iron & \SI{105e6}{} & \SI{11.5}{\kilo\meter\per\second} & \SI{0.5}{} \\
7.4.5 & \SI{3.33}{\Mearth} Iron & \SI{6.67}{\Mearth} Iron & \SI{6.67e6}{} & \SI{1.67}{\Mearth} Iron & \SI{3.33}{\Mearth} Iron & \SI{3.33e6}{} & \SI{26.7}{\kilo\meter\per\second} & \SI{0.5}{} \\
7.4.6 & \SI{3.33}{\Mearth} Iron & \SI{6.67}{\Mearth} Iron & \SI{21.0e6}{} & \SI{1.67}{\Mearth} Iron & \SI{3.33}{\Mearth} Iron & \SI{10.5e6}{} & \SI{26.7}{\kilo\meter\per\second} & \SI{0.5}{} \\
7.4.7 & \SI{3.33}{\Mearth} Iron & \SI{6.67}{\Mearth} Iron & \SI{66.7e6}{} & \SI{1.67}{\Mearth} Iron & \SI{3.33}{\Mearth} Iron & \SI{33.3e6}{} & \SI{26.7}{\kilo\meter\per\second} & \SI{0.5}{} \\
7.4.8 & \SI{3.33}{\Mearth} Iron & \SI{6.67}{\Mearth} Iron & \SI{210e6}{} & \SI{1.67}{\Mearth} Iron & \SI{3.33}{\Mearth} Iron & \SI{105e6}{} & \SI{26.7}{\kilo\meter\per\second} & \SI{0.5}{} \\
\bottomrule
\end{tabular}
\caption{Parameters of the SPH collision simulations with single material bodies of only heavy elements. The surface temperature of all targets and impactors is set to \SI{1000}{\kelvin} and the simulation time is \SI{35.4}{\hour}.}
\label{tab:Mixing_Runs_Pure}
\end{table*}

\begin{table*}[ht!]
\centering
\begin{tabular}{|c|c|c|c|c|c|c|c|c|}
\toprule
& \multicolumn{3}{c|}{Target} & \multicolumn{3}{c|}{Impactor} & \multicolumn{2}{c|}{Impact} \\
Run & Core & Mantle & $N$ & Core & Mantle & $N$ & $v$ & $b$\\
\midrule
7.3.1 & \SI{0.333}{\Mearth} Rock & \SI{0.667}{\Mearth} Ice & \SI{6.67e6}{} & \SI{0.167}{\Mearth} Rock & \SI{0.333}{\Mearth} Ice & \SI{3.33e6}{} & \SI{8.74}{\kilo\meter\per\second} & \SI{0.5}{} \\
7.3.2 & \SI{0.333}{\Mearth} Rock & \SI{0.667}{\Mearth} Ice & \SI{21.0e6}{} & \SI{0.167}{\Mearth} Rock & \SI{0.333}{\Mearth} Ice & \SI{10.5e6}{} & \SI{8.74}{\kilo\meter\per\second} & \SI{0.5}{} \\
7.3.3 & \SI{0.333}{\Mearth} Rock & \SI{0.667}{\Mearth} Ice & \SI{66.7e6}{} & \SI{0.167}{\Mearth} Rock & \SI{0.333}{\Mearth} Ice & \SI{33.3e6}{} & \SI{8.74}{\kilo\meter\per\second} & \SI{0.5}{} \\
7.3.4 & \SI{0.333}{\Mearth} Rock & \SI{0.667}{\Mearth} Ice & \SI{210e6}{} & \SI{0.167}{\Mearth} Rock & \SI{0.333}{\Mearth} Ice & \SI{105e6}{} & \SI{8.74}{\kilo\meter\per\second} & \SI{0.5}{} \\
7.3.5 & \SI{3.33}{\Mearth} Rock & \SI{6.67}{\Mearth} Ice & \SI{6.67e6}{} & \SI{1.67}{\Mearth} Rock & \SI{3.33}{\Mearth} Ice & \SI{3.33e6}{} & \SI{20.13}{\kilo\meter\per\second} & \SI{0.5}{} \\
7.3.6 & \SI{3.33}{\Mearth} Rock & \SI{6.67}{\Mearth} Ice & \SI{21.0e6}{} & \SI{1.67}{\Mearth} Rock & \SI{3.33}{\Mearth} Ice & \SI{10.5e6}{} & \SI{20.13}{\kilo\meter\per\second} & \SI{0.5}{} \\
7.3.7 & \SI{3.33}{\Mearth} Rock & \SI{6.67}{\Mearth} Ice & \SI{66.7e6}{} & \SI{1.67}{\Mearth} Rock & \SI{3.33}{\Mearth} Ice & \SI{33.3e6}{} & \SI{20.13}{\kilo\meter\per\second} & \SI{0.5}{} \\
7.3.8 & \SI{3.33}{\Mearth} Rock & \SI{6.67}{\Mearth} Ice & \SI{210e6}{} & \SI{1.67}{\Mearth} Rock & \SI{3.33}{\Mearth} Ice & \SI{105e6}{} & \SI{20.13}{\kilo\meter\per\second} & \SI{0.5}{} \\
\midrule
7.5.1 & \SI{0.333}{\Mearth} Iron & \SI{0.667}{\Mearth} Rock & \SI{6.67e6}{} & \SI{0.167}{\Mearth} Iron & \SI{0.333}{\Mearth} Rock & \SI{3.33e6}{} & \SI{10.12}{\kilo\meter\per\second} & \SI{0.5}{} \\
7.5.2 & \SI{0.333}{\Mearth} Iron & \SI{0.667}{\Mearth} Rock & \SI{21.0e6}{} & \SI{0.167}{\Mearth} Iron & \SI{0.333}{\Mearth} Rock & \SI{10.5e6}{} & \SI{10.12}{\kilo\meter\per\second} & \SI{0.5}{} \\
7.5.3 & \SI{0.333}{\Mearth} Iron & \SI{0.667}{\Mearth} Rock & \SI{66.7e6}{} & \SI{0.167}{\Mearth} Iron & \SI{0.333}{\Mearth} Rock & \SI{33.3e6}{} & \SI{10.12}{\kilo\meter\per\second} & \SI{0.5}{} \\
7.5.4 & \SI{0.333}{\Mearth} Iron & \SI{0.667}{\Mearth} Rock & \SI{210e6}{} & \SI{0.167}{\Mearth} Iron & \SI{0.333}{\Mearth} Rock & \SI{105e6}{} & \SI{10.12}{\kilo\meter\per\second} & \SI{0.5}{} \\
7.5.5 & \SI{3.33}{\Mearth} Iron & \SI{6.67}{\Mearth} Rock & \SI{6.67e6}{} & \SI{1.67}{\Mearth} Iron & \SI{3.33}{\Mearth} Rock & \SI{3.33e6}{} & \SI{23.45}{\kilo\meter\per\second} & \SI{0.5}{} \\
7.5.6 & \SI{3.33}{\Mearth} Iron & \SI{6.67}{\Mearth} Rock & \SI{21.0e6}{} & \SI{1.67}{\Mearth} Iron & \SI{3.33}{\Mearth} Rock & \SI{10.5e6}{} & \SI{23.45}{\kilo\meter\per\second} & \SI{0.5}{} \\
7.5.7 & \SI{3.33}{\Mearth} Iron & \SI{6.67}{\Mearth} Rock & \SI{66.7e6}{} & \SI{1.67}{\Mearth} Iron & \SI{3.33}{\Mearth} Rock & \SI{33.3e6}{} & \SI{23.45}{\kilo\meter\per\second} & \SI{0.5}{} \\
7.5.8 & \SI{3.33}{\Mearth} Iron & \SI{6.67}{\Mearth} Rock & \SI{210e6}{} & \SI{1.67}{\Mearth} Iron & \SI{3.33}{\Mearth} Rock & \SI{105e6}{} & \SI{23.45}{\kilo\meter\per\second} & \SI{0.5}{} \\
\midrule
7.6.1 & \SI{0.333}{\Mearth} Iron & \SI{0.667}{\Mearth} Ice & \SI{6.67e6}{} & \SI{0.167}{\Mearth} Iron & \SI{0.333}{\Mearth} Ice & \SI{3.33e6}{} & \SI{8.96}{\kilo\meter\per\second} & \SI{0.5}{} \\
7.6.2 & \SI{0.333}{\Mearth} Iron & \SI{0.667}{\Mearth} Ice & \SI{21.0e6}{} & \SI{0.167}{\Mearth} Iron & \SI{0.333}{\Mearth} Ice & \SI{10.5e6}{} & \SI{8.96}{\kilo\meter\per\second} & \SI{0.5}{} \\
7.6.3 & \SI{0.333}{\Mearth} Iron & \SI{0.667}{\Mearth} Ice & \SI{66.7e6}{} & \SI{0.167}{\Mearth} Iron & \SI{0.333}{\Mearth} Ice & \SI{33.3e6}{} & \SI{8.96}{\kilo\meter\per\second} & \SI{0.5}{} \\
7.6.4 & \SI{0.333}{\Mearth} Iron & \SI{0.667}{\Mearth} Ice & \SI{210e6}{} & \SI{0.167}{\Mearth} Iron & \SI{0.333}{\Mearth} Ice & \SI{105e6}{} & \SI{8.96}{\kilo\meter\per\second} & \SI{0.5}{} \\
7.6.5 & \SI{3.33}{\Mearth} Iron & \SI{6.67}{\Mearth} Ice & \SI{6.67e6}{} & \SI{1.67}{\Mearth} Iron & \SI{3.33}{\Mearth} Ice & \SI{3.33e6}{} & \SI{20.65}{\kilo\meter\per\second} & \SI{0.5}{} \\
7.6.6 & \SI{3.33}{\Mearth} Iron & \SI{6.67}{\Mearth} Ice & \SI{21.0e6}{} & \SI{1.67}{\Mearth} Iron & \SI{3.33}{\Mearth} Ice & \SI{10.5e6}{} & \SI{20.65}{\kilo\meter\per\second} & \SI{0.5}{} \\
7.6.7 & \SI{3.33}{\Mearth} Iron & \SI{6.67}{\Mearth} Ice & \SI{66.7e6}{} & \SI{1.67}{\Mearth} Iron & \SI{3.33}{\Mearth} Ice & \SI{33.3e6}{} & \SI{20.65}{\kilo\meter\per\second} & \SI{0.5}{} \\
7.6.8 & \SI{3.33}{\Mearth} Iron & \SI{6.67}{\Mearth} Ice & \SI{210e6}{} & \SI{1.67}{\Mearth} Iron & \SI{3.33}{\Mearth} Ice & \SI{105e6}{} & \SI{20.65}{\kilo\meter\per\second} & \SI{0.5}{} \\
\bottomrule
\end{tabular}
\caption{Parameters of the SPH collision simulations with two material bodies of only heavy elements. The surface temperature of all targets and impactors is set to \SI{1000}{\kelvin} and the simulation time is \SI{35.4}{\hour}.}
\label{tab:Mixing_Runs}
\end{table*}

\begin{table*}[ht!]
\centering
\begin{tabular}{|c|c|c|c|c|}
\toprule
Run & Core & Mantle & Envelope & $N$\\
\midrule
5.4.1 & \SI{5.40}{\Mearth} 50:50 Rock:H-He & \SI{14.60}{\Mearth} 50:50 Ice:H-He & \SI{286.7}{\Mearth} H-He & \SI{100e6}{} \\
5.4.2 & \SI{5.40}{\Mearth} 50:50 Rock:H-He & \SI{14.60}{\Mearth} 50:50 Ice:H-He & \SI{286.7}{\Mearth} H-He & \SI{1e9}{} \\
5.4.4$^\dagger$ & \SI{27.03}{\Mearth} 18:82 Rock:H-He & \SI{72.97}{\Mearth} 18:82 Ice:H-He & \SI{216.7}{\Mearth} H-He & \SI{1e9}{} \\
\midrule
7.7.1 & \SI{0.667}{\Mearth} 50:50 Iron:Rock & \SI{0.333}{\Mearth} Rock & -- & \SI{10e6}{} \\
7.7.2 & \SI{0.667}{\Mearth} 50:50 Iron:Rock & \SI{0.333}{\Mearth} Rock & -- & \SI{31.6e6}{} \\
7.7.3 & \SI{0.667}{\Mearth} 50:50 Iron:Rock & \SI{0.333}{\Mearth} Rock & -- & \SI{100e6}{} \\
7.7.4 & \SI{0.667}{\Mearth} 50:50 Iron:Rock & \SI{0.333}{\Mearth} Rock & -- & \SI{316e6}{} \\
7.7.5 & \SI{0.667}{\Mearth} 50:50 Iron:Rock & \SI{0.333}{\Mearth} Rock & -- & \SI{1e9}{} \\
7.7.6 & \SI{6.67}{\Mearth} 50:50 Iron:Rock & \SI{3.33}{\Mearth} Rock & -- & \SI{10e6}{} \\
7.7.7 & \SI{6.67}{\Mearth} 50:50 Iron:Rock & \SI{3.33}{\Mearth} Rock & -- & \SI{31.6e6}{} \\
7.7.8 & \SI{6.67}{\Mearth} 50:50 Iron:Rock & \SI{3.33}{\Mearth} Rock & -- & \SI{100e6}{} \\
7.7.9 & \SI{6.67}{\Mearth} 50:50 Iron:Rock & \SI{3.33}{\Mearth} Rock & -- & \SI{316e6}{} \\
7.7.10 & \SI{6.67}{\Mearth} 50:50 Iron:Rock & \SI{3.33}{\Mearth} Rock & -- & \SI{1e9}{} \\
\midrule
7.9.1$^*$ & \SI{0.99}{\Mearth} 33:67 Iron:Rock & \SI{0.01}{\Mearth} Rock & -- & \SI{1e9}{} \\
7.9.2$^\dagger$ & \SI{0.99}{\Mearth} 18:82 Iron:Rock & \SI{0.01}{\Mearth} Rock & -- & \SI{1e9}{} \\
\bottomrule
\end{tabular}
\caption{Parameters of the SPH simulations with initially mixed bodies. The surface temperature of all bodies with hydrogen-helium is set to \SI{300}{\kelvin} while those with only heavy elements have a surface temperature of \SI{1000}{\kelvin}. The simulation time is generally \SI{3.54}{\hour} but the run marked with $^*$ has a simulation time of \SI{5.31}{\hour} and the runs marked with $\dagger$ have a simulation time of \SI{10.62}{\hour}.}
\label{tab:demixing_parameters}
\end{table*}





\end{document}